\documentclass[11pt]{article}
\usepackage[utf8]{inputenc}
\usepackage{amsmath,amssymb,latexsym}
\usepackage[T1]{fontenc}
\usepackage{epsfig}
\usepackage{fullpage}
\usepackage{graphicx}
\usepackage{makeidx}
\usepackage{slashed}
\usepackage[dvipsnames]{xcolor}
\usepackage{subcaption}
\usepackage{feynmp}
\usepackage{cite}
\usepackage{hyperref}
\usepackage{cancel}
\usepackage{slashed}
\usepackage{mathtools}
\usepackage{bm}

\long\def\symbolfootnote[#1]#2{\begingroup
\def\thefootnote{\fnsymbol{footnote}}\footnote[#1]{#2}\endgroup}

\def\as{\alpha_s}

\newcommand{\Eq}{eq.~\eqref}
\newcommand{\eq}{eq.~\eqref}
\newcommand{\fig}[1]{fig.~\ref{fig:#1}}

\usepackage[textfont=sl]{caption}
\let\oldmarginpar\marginpar
\renewcommand\marginpar[1]{\-\oldmarginpar[\raggedleft\tiny #1]
{\raggedright\tiny #1}}

\allowdisplaybreaks

\begin{document}

\begin{flushright}
Nikhef/2021-001 
\end{flushright}

\vspace*{1.5cm}
\begin{center}  
{\Large\bf Next-to-leading power threshold corrections for finite order \\[1ex] and resummed colour-singlet cross sections}\\[10ex] 
 { 
 Melissa van Beekveld$^{\,a,b,c}$,
 Eric Laenen$^{\,c,d,e}$, Jort Sinninghe Damst\'{e}$^{\,c,d}$, Leonardo Vernazza$^{\,f}$}\\[1cm]
\end{center}
 {\it 
$^a$ {Rudolf Peierls Centre, Oxford University, 20 Parks Road, Oxford OX1
3PU, United Kingdom}\\
$^b$ {THEP, Radboud University,
 Heyendaalseweg 135, 6525~AJ Nijmegen, The Netherlands}\\
$^c$ {Nikhef, Science Park 105, 1098 XG Amsterdam, The Netherlands}\\
$^d$ {IoP, University of Amsterdam, Science Park 904, 1018 XE Amsterdam, The Netherlands}\\
 $^e$ {ITF, Utrecht University, Leuvenlaan 4, 3584 CE Utrecht, The Netherlands}\\
 $^f$ {Dipartimento di Fisica Teorica and Arnold-Regge Center, Universit`a di Torino, and INFN, Sezione di Torino, Via Pietro Giuria 1, I-10125 Torino, Italy}}
\vspace{1cm}

\begin{abstract}
\noindent We study next-to-leading-power (NLP) threshold corrections in colour-singlet production processes, with particular emphasis on Drell-Yan (DY) and single-Higgs production. We assess the quality of the partonic and hadronic threshold expansions for each process up to NNLO.  We determine  numerically the NLP leading-logarithmic (LL) resummed contribution in addition to the leading-power next-to-next-to-leading logarithmic (LP NNLL) resummed DY and Higgs cross sections, matched to NNLO. We find that the inclusion of NLP logarithms is numerically more relevant than increasing the precision to N$^3$LL at LP for these processes. We also perform an analytical and numerical comparison of LP NNLL + NLP LL resummation in soft-collinear effective theory and direct QCD, where we achieve excellent analytical and numerical agreement once the NLP LL terms are included in both formalisms. Our results underline the phenomenological importance of understanding the NLP structure of QCD cross sections. 
\end{abstract}

\vspace*{\fill}

\newpage
\reversemarginpar

%======================
%=======SECTION========
%======================
\section{Introduction}

The convergence of perturbative calculations of hadronic cross sections in Quantum Chromodynamics (QCD) is key to understanding the phenomenology of particle collisions at hadron colliders. While the perturbative approach relying on the smallness of QCD coupling constant $\alpha_s$ is often successful, its value in certain kinematic limits is conditional on the behaviour of logarithmic corrections. A well-known example is the presence of logarithms of the threshold variable $\xi$ in these corrections. These logarithms grow large as $\xi \rightarrow 0$, i.e. approaching the threshold boundary of phase space. One may generically express the structure of the dimensionless partonic coefficient function near threshold as
\begin{equation}
\label{threshold_simple}
    \Delta(\xi) = \sum_{n=0}^{\infty}\alpha_s^n \Bigg\{ \sum_{m=0}^{2n-1} c^{\rm LP}_{nm} \left(\frac{\ln^m \xi}{\xi}\right)_+ + d_n\delta(\xi)+ \sum_{m=0}^{2n-1} c^{\rm NLP}_{nm} \ln^m \xi + \dots \Bigg\}\,.
\end{equation}
The first term on the right is the class of leading-power (LP) contributions, while the second consists of constants localized at $\xi = 0$. The former contributions are well-studied and originate from the phase-space difference between virtual and real-emission diagrams after the KLN cancellation and mass factorisation of the soft and collinear singularities takes place. While singular in the $\xi\rightarrow 0$ limit, they are regularized by the plus prescription, which makes them integrable.  Long ago it was realized that the predictive power of the perturbative approach may be saved by resumming the logarithmic enhancements to all orders in $\alpha_s$, first to leading logarithmic (LL) order~\cite{Parisi:1980xd,Curci:1979am}, and soon after extended to subleading logarithmic accuracies~\cite{Sterman:1987aj, Catani:1989ne}. These initial results have been re-derived and generalized using either different techniques in direct QCD (dQCD)~\cite{Korchemsky:1993uz,Forte:2002ni,Contopanagos:1997nh} or the Soft-Collinear-Effective Theory (SCET) approach~\cite{Becher:2006nr}.

The third set of terms in eq.~\eqref{threshold_simple} is down by a single power in $\xi$ with respect to the first. These are the so-called next-to-leading power (NLP) contributions and are the main focus of this work. On the formal side, conjectures on the exponential structure of NLP contributions have already appeared in various places~\cite{Kramer:1996iq,Catani:2001ic,Laenen:2008ux,Moch:2009hr,Soar:2009yh,Grunberg:2009vs,deFlorian:2014vta,Presti:2014lqa,Ajjath:2020ulr}, and have been been proven at NLP LL accuracy for the DY and Higgs production cross sections using two methods. In ref.~\cite{Bahjat-Abbas:2019fqa} NLP LL resummation was achieved for general colour-singlet production processes using diagrammatic techniques in the dQCD framework. NLP resummation is also achieved in SCET, both for the DY~\cite{Beneke:2018gvs,Beneke:2019oqx} and Higgs~\cite{Beneke:2019mua} production processes~\footnote{Note that besides the threshold regime, resummation of NLP logarithm has been obtained in SCET also for thrust \cite{Moult:2018jjd}. Fixed-order NLP corrections for DY and single Higgs production have also been studied in SCET for the $N$-jettiness ~\cite{Moult:2016fqy, Boughezal:2016zws,Moult:2017jsg, Boughezal:2018mvf,Ebert:2018lzn} and the $q_T$ of the produced lepton pair/Higgs boson~\cite{Ebert:2018gsn,Cieri:2019tfv}. }. Note that these methods resum NLP LL terms that originate from next-to-soft gluon emissions only. A general resummation framework for soft-quark emissions is still under development (see e.g.~\cite{Moult:2019uhz,Liu:2020eqe,Beneke:2020ibj} for recent process in the SCET framework).  

Despite this interest in the behaviour of the NLP terms, relatively few studies focus on their numerical/phenomenological aspects. Early numerical studies on the threshold approximation of the Drell-Yan (DY) production process already noted the importance of subleading-power logarithms~\cite{Appell:1988ie,Magnea:1990qg}. Threshold expansions of the Higgs next-to-next-to-leading order (NNLO) and N$^3$LO coefficient functions are studied in~\cite{Anastasiou:2014lda,Anastasiou:2016cez,Anastasiou:2015vya}, showing that the convergence of this series is slow, thereby stressing the importance of NLP contributions. In ref.~\cite{Kramer:1996iq}, the LP resummation formalism was extended to include subleading contributions for the DY and Higgs production processes, and found that these impact the final result significantly. In ref.~\cite{Bonvini:2010tp,Bonvini:2014qga,Bonvini:2014joa,Bonvini:2016frm} it was found that the way one handles NLP contributions results in ambiguity of the LP resummed result, and that the numerical difference that follows from this ambiguity is of phenomenological relevance for both DY and Higgs~\footnote{We comment on the comparison with our work in section~\ref{sec:someremarks}. }. 
In~\cite{Basu:2007nu,vanBeekveld:2019cks}  numerical effects of subleading powers were studied for prompt photon production. 
All these studies point in the same direction: although not as divergent as the LP contributions, numerically NLP terms can become quite important.

Motivated by these considerations, we aim to answer three questions in this paper. First, what is the behaviour of the threshold expansion of the DY and Higgs partonic coefficient functions? Second, what is the numerical impact of NLP LL resummation in dQCD, and how does it compare to that of subleading logarithmic corrections at LP? And finally, to what extent do the dQCD and SCET NLP resummation formalisms agree, analytically and numerically? 

The outline of this paper is then as follows. In Section~\ref{sec:fo_threshold} we study the convergence of the fixed-order DY and Higgs expansions at threshold; for the DY threshold expansion this, to the best of our knowledge, has not been shown earlier (neither for the diagonal nor off-diagonal production channels). For the Higgs case in the dominant production channel, we confirm results that have appeared earlier~\cite{Anastasiou:2014lda,Herzog:2014wja}. In Section~\ref{sec:dQCDresum} we review and slightly improve LP and NLP resummation for these processes, and derive the resummation exponent (eq.~\eqref{eq:resumfunctions}) that is used for our numerical studies. These are presented in Section~\ref{sec:dQCD_results}, where we show the impact of the NLP LL resummation on top of the LP NNLL($^{\prime}$) resummed and matched DY and Higgs cross sections. We also present an assessment of the numerical importance for off-diagonal emissions, for which no resummation has been achieved yet~\footnote{Although important progress has been made, see e.g.~\cite{Presti:2014lqa,Moult:2019uhz,Beneke:2020ibj}}. In Section~\ref{sec:comparison} we perform a careful comparison of dQCD and SCET resummation to LP NNLL plus NLP LL accuracy. We summarize and conclude in Section~\ref{sec:discussion}, while certain technical aspects and collections of relevant perturbative coefficients can be found in three appendices. 

%======================
\section{Threshold expansions at fixed order}
\label{sec:fo_threshold}
In this section we study the convergence of the threshold expansion in fixed order calculations for the single Higgs and the DY processes, first at the parton level, and subsequently for the hadronic cross section. The partonic flux plays an important role in the threshold expansion of the latter. While earlier studies (see e.g. \cite{Appell:1988ie, Magnea:1990qg, vanNeerven:1991gh, Contopanagos:1993xh, Harlander:2002wh,Catani:2001cr, Kramer:1996iq, Bonvini:2010tp, Anastasiou:2014lda,Anastasiou:2016cez,Anastasiou:2015vya}) have addressed  partonic and/or hadronic threshold expansions for one or both of these processes, we provide here an extended analysis tailored to study NLP corrections. As such it also serves as an introduction to the numerical study of NLP effects in resummed cross sections in section~\ref{sec:dQCD_results}. 

\subsection{Threshold behaviour of the partonic coefficients for single Higgs production}
\noindent We start with the study of the behaviour of the partonic coefficient functions for single Higgs production at NLO \cite{Dawson:1990zj,Djouadi:1991tka,Spira:1995rr} and NNLO \cite{Anastasiou:2002yz,Ravindran:2003um,Harlander:2002wh}. First we introduce some relevant quantities and variables. The hadronic cross section for this process is given by 
\begin{equation} \label{higgs_z}
\sigma_{p p \rightarrow h + X}(m_h^2,S) =  \sigma^{\rm h}_0\sum_{i,j}\int_0^1 {\rm d}x_1 f_i(x_1,\mu)\int_0^1 {\rm d}x_2 f_j(x_2,\mu) \int_0^1{\rm d}z\,\Delta_{{\rm h},\,ij}(z,m_h^2/\mu^2)\delta\left(x_1x_2z-\tau\right)\, ,
\end{equation}
where  $\tau = m_h^2/S$, with $m_h$ the Higgs boson mass and $S$ the hadronic center-of-mass (CM) energy squared, $f_i(x,\mu)$ are the parton distribution functions (PDFs) and $\Delta_{{\rm h},ij}(z,m_h^2/\mu^2)$ is the partonic coefficient function. Note that the leading order (LO) partonic cross section $\sigma^h_0$ is factored out. In the infinite top-mass limit~\footnote{In this limit, the effect of the top quark mass is contained in a Wilson coefficient, whose lowest-order contribution is $-\as/(3\pi v)$~\cite{Anastasiou:2002yz} with $v^2 = 1/(\sqrt{2}G_F)$. We include this coefficient (squared) in eq.~\eqref{eq:higgsLO}.} it reads (see e.g.~ref.~\cite{Anastasiou:2002yz})
\begin{equation}
\label{eq:higgsLO}
\sigma^{\rm h}_0 = \frac{\sqrt{2}\,G_F}{72 (N_c^2-1)}\frac{\alpha_s^2}{\pi}\frac{m_h^2}{S}\,,
\end{equation}
in units of GeV$^{-2}$. The strong coupling is denoted by $\alpha_s\equiv\alpha_s(\mu)$, $N_c$ is the number of colours, and Fermi's constant $G_F$ has the value $1.16639\cdot 10^{-5}$~GeV$^{-2}$. Throughout this paper we employ the central member of the PDF4LHC15 NNLO 100 PDF set~\cite{Butterworth:2015oua} for proton-proton collisions with $\sqrt{S} = 13$ TeV, corresponding to $\alpha_s(M_Z^2) = 0.118$ with the $Z$-boson mass  $M_Z = 91.18$~GeV. In practice we use a polynomial fit of these PDFs (see appendix~ \ref{app:PDFs}), which is particularly convenient for obtaining the resummed results presented in section~\ref{sec:dQCD_results}. We choose the renormalisation and factorisation scale equal and denote both as $\mu$. The threshold variable is $1-z=1-m_{ h}^2/s$, with $\sqrt{s}$ the partonic CM energy. 
The hadronic cross section can also be expressed as
\begin{equation}
\label{eq:higgstotalxsec}
\sigma_{pp\hspace{.1em}\rightarrow\hspace{.1em} h+X} =\sigma_0^{\rm h}~\sum_{i,j}\int_{\tau}^1\frac{{\rm d}z}{z}~\mathcal{L}_{ij}\left(\frac{\tau}{z}\right)~\Delta_{{\rm h},ij}(z)\,,
\end{equation}
in terms of the parton flux 
\begin{equation}
\mathcal{L}_{ij}\left(\frac{\tau}{z}\right) = \int_{\tau/z}^1\frac{{\rm d}x}{x}\,f_i(x,\mu)\,f_j\left(\frac{\tau/z}{x},\mu\right)\,,
\end{equation}
and where we have suppressed scale dependence in both the flux and coefficient functions.
The perturbative expansion of the partonic coefficient functions $\Delta_{{\rm h},ij}(z)$ reads
\begin{equation}
\label{eq:pertexp}
\Delta_{{\rm h},ij}(z) = \Delta^{(0)}_{{\rm h},ij}(z) + \Delta^{(1)}_{{\rm h},ij}(z) +  \Delta^{(2)}_{{\rm h},ij}(z) + \dots\,,
\end{equation}
where $\Delta^{(n)}_{{\rm h},ij}(z)$ includes $\alpha_s^n$. The definition in \Eq{eq:higgstotalxsec} implies that $\Delta^{(0)}_{{\rm h},ij}(z)$ is normalized as
\begin{equation}
\Delta^{(0)}_{{\rm h},ij}(z) = \delta_{ig}\delta_{jg}\delta(1-z)\,, 
\end{equation}
since the Born-level process involves the gluon-gluon fusion channel.

Near $z=1$ the function $\Delta_{{\rm h},ij}(z)$ can be expanded as 
\begin{align}
\label{eq:defthreshold}
\Delta_{{\rm h},ij}(z) = \sum_{n=0}^{\infty}\left(\frac{\alpha_s}{\pi}\right)^n\bigg\{&c_n^{\delta}\delta(1-z)
 +\sum_{m=0}^{2n-1}\bigg[c_{nm}^{{\rm LP}}\left(\frac{\ln^m (1-z)}{1-z}\right)_+ +\, c_{nm}^{\rm NLP}\ln^m(1-z) \\ 
&\hspace{4cm}+c_{nm}^{\rm NNLP}(1-z)\ln^m(1-z)+ \dots\bigg]\bigg\}\,. \nonumber 
\end{align}
The $c_{nm}^{\rm LP}$ are the coefficients of the LP contributions (as in eq.~\eqref{threshold_simple}), which contain a factor of $(1-z)^{-1}$, while NLP contributions at $\mathcal{O}((1-z)^0)$ have coefficients $c_{nm}^{\rm NLP}$; in general N$^{i}$LP contributions contain a factor $(1-z)^{i-1}$. Explicit forms of the NLO and NNLO coefficients $\Delta^{(1)}_{{\rm h},ij}$ and $\Delta^{(2)}_{{\rm h},ij}$ are obtained from the functions  $\eta_{ij}^{(n)}(z)$ eq.~(45)-(57) in ref.~\cite{Anastasiou:2002yz}. We note that the factor of $1/z$ appearing in eq.~(15) of that reference is included in our partonic coefficient functions, i.e. we have~\footnote{We refer to appendix~\ref{AppNormalization} for the origin of the additional factor of $1/z$ with respect to the DY case (section~\ref{subsub:DY}).}
\begin{equation}
\label{eq:etadef}
  \Delta_{{\rm h},ij}^{(n)}(z) = \left(\frac{\alpha_s}{\pi}\right)^n\, \eta_{ij}^{(n)}(z)/z\,. 
\end{equation}
In our threshold expansion the factor $1/z$ is expanded too, an approach also taken in \cite{Harlander:2002wh}, following \cite{Catani:2001cr}. One may choose to keep that factor unexpanded. This would lead to somewhat different results,  a consequence of truncating the expansion \eq{eq:defthreshold}. 
Let us comment here further on the role of the extra factor of $1/z$. In general, one may define the partonic coefficient functions to contain additional powers of $1/z$, provided a compensating factor is introduced. For some power $p$ we would find instead
\begin{equation}
\sigma_{pp\hspace{.1em}\rightarrow\hspace{.1em} h+X} =\sigma_0^{\rm h}~\sum_{i,j}\int_{\tau}^1 {\rm d}z\, z^{p-1}~\mathcal{L}_{ij}\left(\frac{\tau}{z}\right)\left[\frac{\Delta_{{\rm h},ij}(z)}{z^p}\right]\,.
\end{equation}
If one expands the term in square brackets around threshold and truncate that series at some power in $(1-z)$, the above result is no longer equal to \eq{eq:higgstotalxsec}, and is thus sensitive to the $p$ additional inverse powers of $z$ that are expanded
\begin{equation}
    \frac{1}{z^p}\left[\frac{\ln(1-z)}{1-z}\right]_+ = \left[\frac{\ln(1-z)}{1-z}\right]_+ + p \ln(1-z) + \frac{1}{2}p(1+p)(1-z)\ln(1-z) +\cdots.
\end{equation}  
This makes results of threshold expansions somewhat ambiguous, an aspect also discussed at length in ref.~\cite{Herzog:2014wja} (see in particular the caption of fig.~2 in that reference). In our case we fix the power of $1/z$ by requiring the universality of LL NLP terms in the dominant channel of colour singlet production processes, as shown in \cite{DelDuca:2017twk,vanBeekveld:2019prq} at fixed order, as well as in \cite{Bahjat-Abbas:2019fqa} in a resummation context. In both DY and Higgs production the coefficients of the highest power of $\ln(1-z)$ at LP and NLP are, at each order in $\alpha_s$, identical but with opposite sign. This follows from both terms resulting from multiplying the same residual collinear singularity with an overall $\epsilon$-dependent power of $(1-z)$. This singularity, which is absorbed in the PDF by mass factorisation, is associated to the lowest-order splitting kernel, such that the logarithms in the finite part are indirectly generated by the splitting kernel too, as is well known. Expansion of the lowest order splitting function, $P_{qq}^{(1)}$ and $P_{gg}^{(1)}$ for DY and Higgs respectively, reveals that they indeed obey this relation between the LP and NLP terms
\begin{equation}
    P_{qq}^{(1)} = C_F\left(2\left[\frac{1}{1-z}\right]_+ -2 +\mathcal{O}(1-z)\right)\, , \hspace{10pt}     P_{gg}^{(1)} = C_A\left(2\left[\frac{1}{1-z}\right]_+ -2 +\mathcal{O}(1-z)\right)\, .
\end{equation}
The definition of the (partonic) cross section in \eq{eq:higgstotalxsec} satisfies the condition, as we show explicitly for the NLO coefficient function $\Delta_{{\rm h},ij}^{(1)}(z)$ below. The full one-loop coefficient reads \cite{Anastasiou:2002yz}
\begin{align}\label{eq:DeltaHiggs1}
\frac{\pi}{\alpha_s}\Delta^{(1)}_{{\rm h}, gg}(z) =&\, \frac{1}{z}\Bigg[\left(\frac{11}{2}+6\zeta_2\right)\delta\left(1-z\right) +12\left[\frac{\ln(1-z)}{1-z}\right]_+ - 12z(2-z+z^2)\ln(1-z) \nonumber\\
& -\frac{6(z^2-z+1)^2}{1-z}\ln(z)-\frac{11}{2}(1-z)^3\Bigg] \nonumber \\ 
 =&\, 12\left[\frac{\ln(1-z)}{1-z}\right]_+ + \left(\frac{11}{2}+6\zeta_2\right)\delta\left(1-z\right) - 12\ln(1-z) -1 +\mathcal{O}\left(1-z\right)\,,
\end{align}
where we have expanded around threshold in the second line, and where we observe the advertised relation between the coefficient of the LL at LP and NLP.

\begin{figure}[t!]
    \centering
    \mbox{
    \centering
    \begin{subfigure}{.492\textwidth}
    \centering
        \vspace*{-2.5pt}\includegraphics[width=\textwidth]{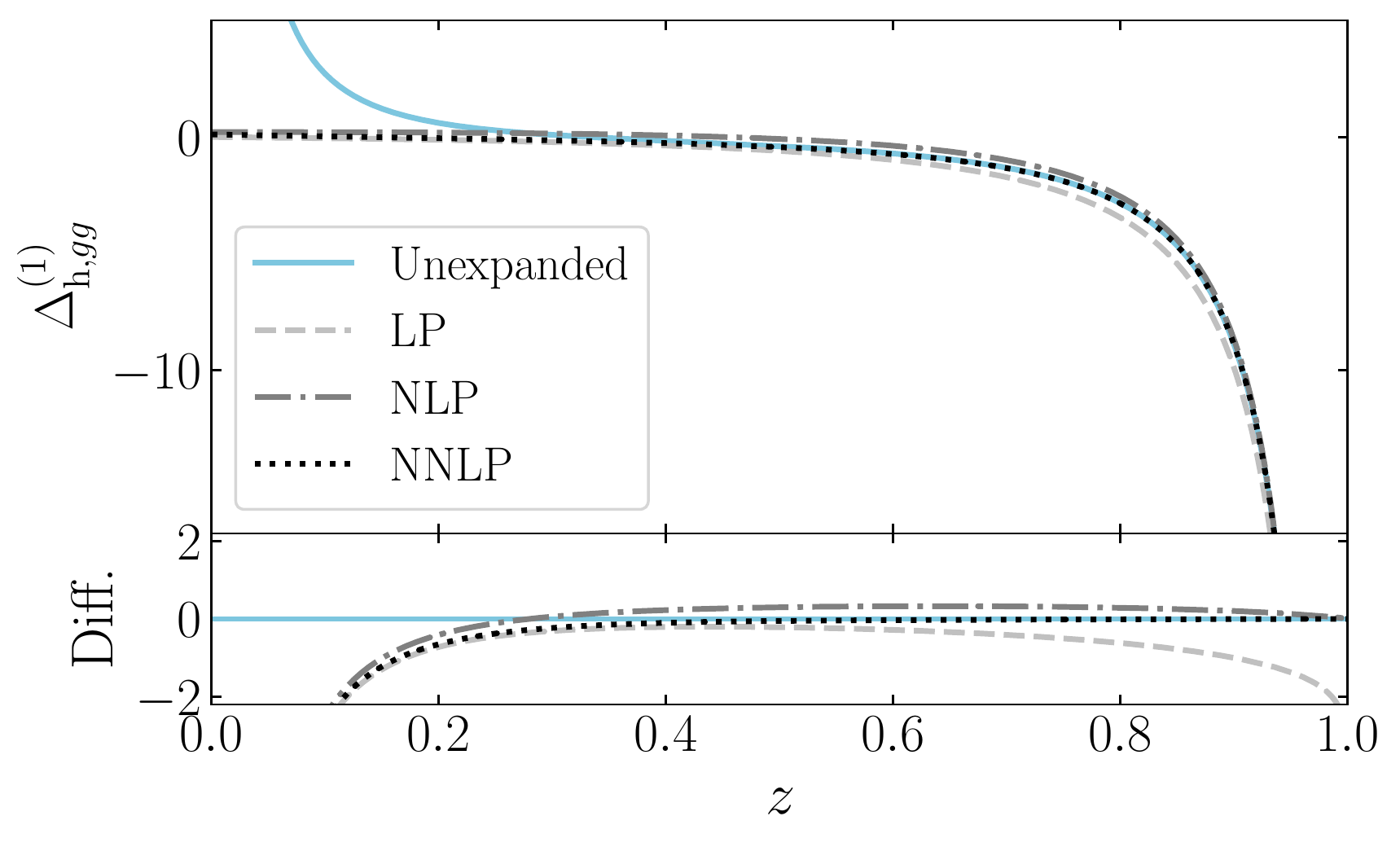}
        \caption{NLO $gg$-channel.}
        \label{fig:ggNLOhiggs}
    \end{subfigure}
    \begin{subfigure}{.498\textwidth}
        \includegraphics[width=\textwidth]{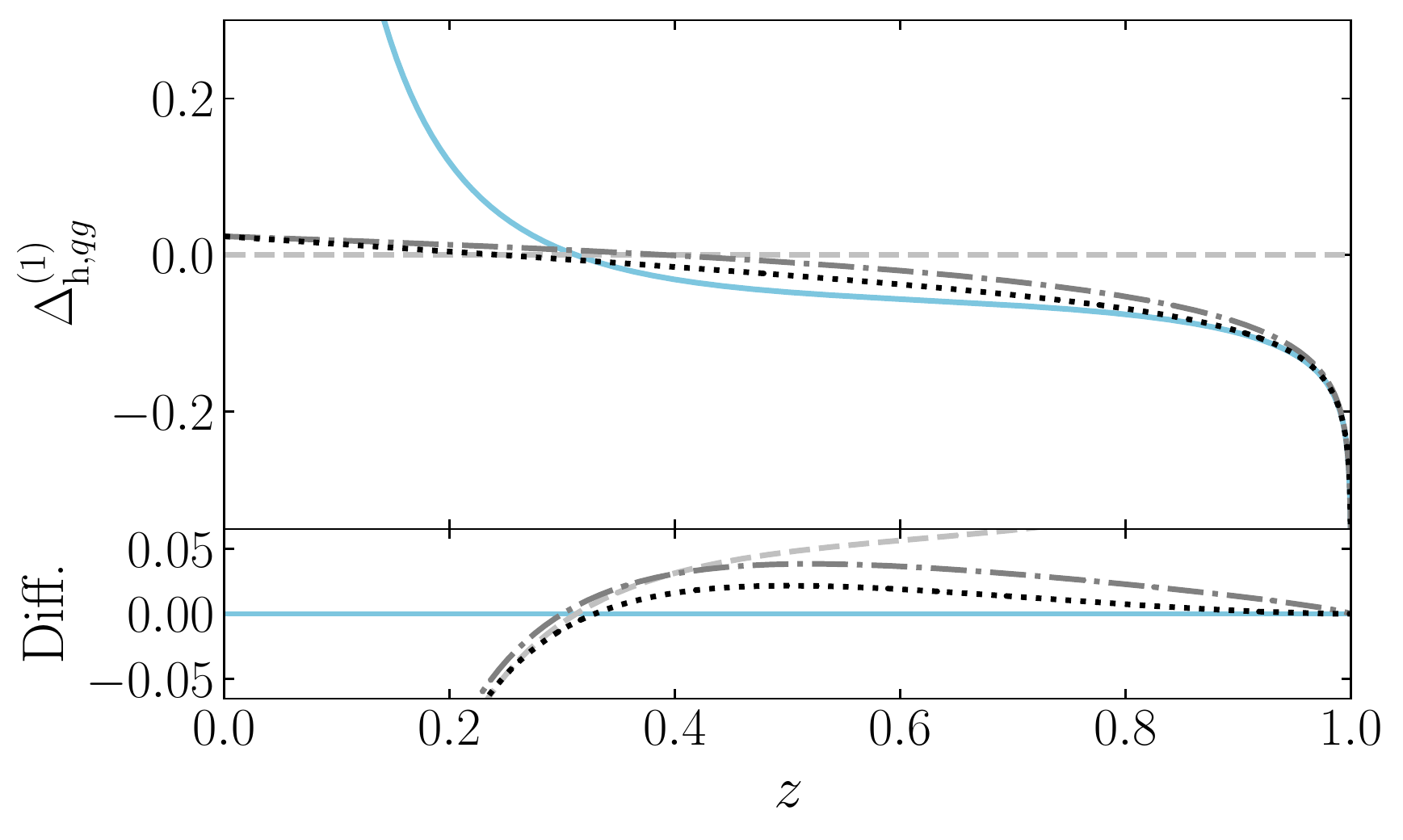}
        \caption{NLO $qg$-channel.}
        \label{fig:qgNLOhiggs}
    \end{subfigure}}
    \vspace{0.2cm}
    
    \hspace*{-1pt}\mbox{
    \centering
    \begin{subfigure}{.494\textwidth}
    \centering
        \includegraphics[width=\textwidth]{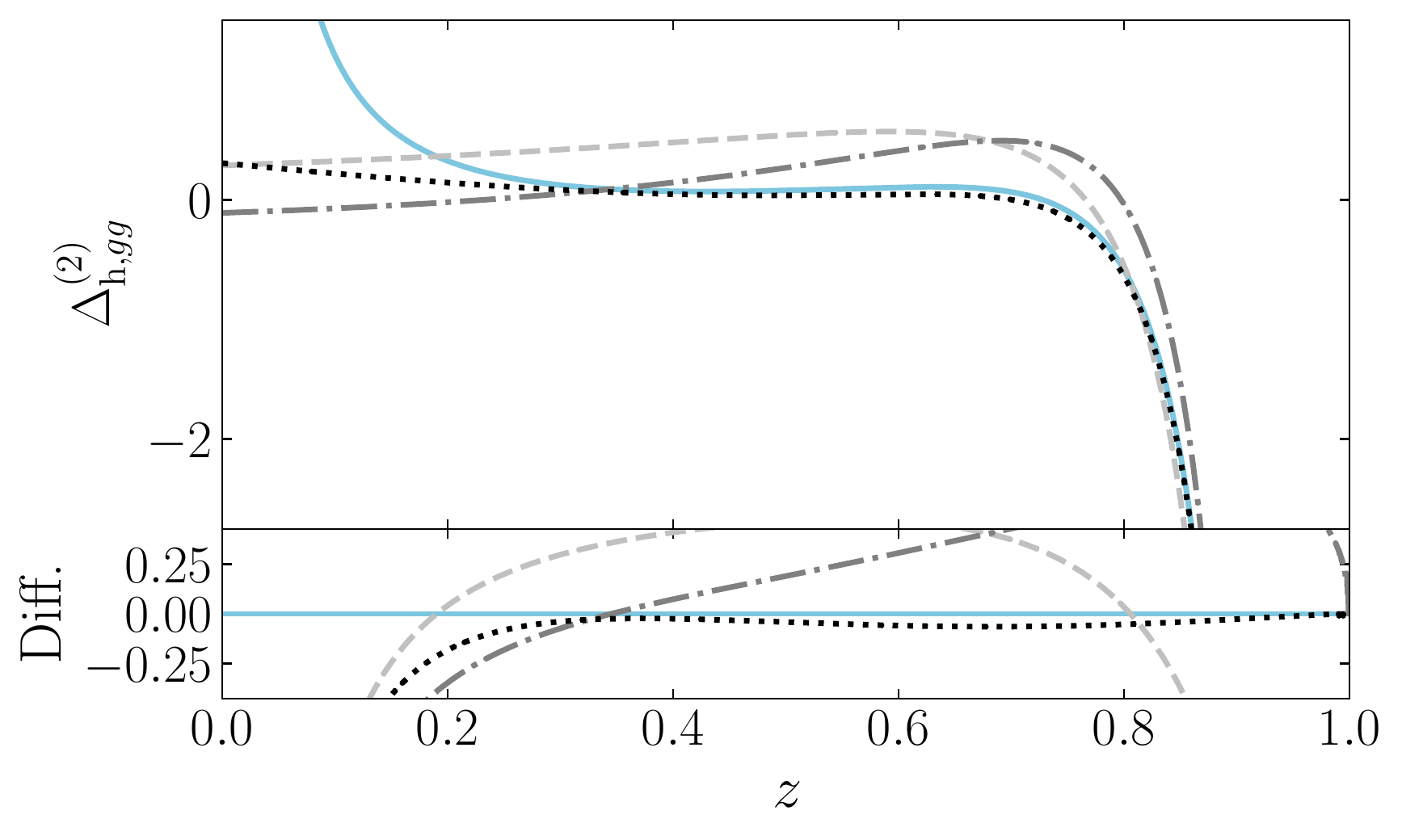}
        \caption{NNLO $gg$-channel }
        \label{fig:ggNNLOhiggs}
    \end{subfigure}
    \begin{subfigure}{.498\textwidth}
        \includegraphics[width=\textwidth]{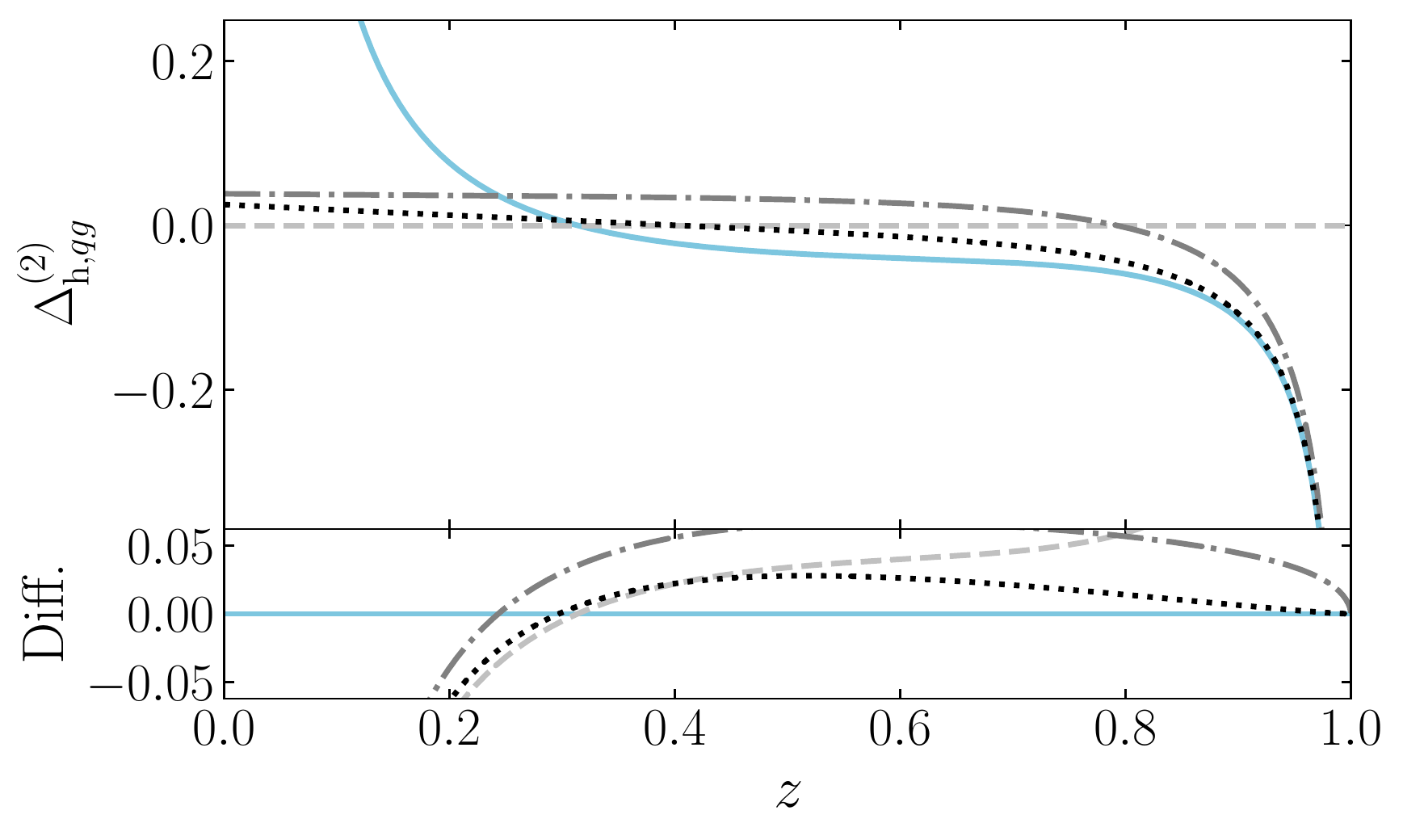}
        \caption{NNLO $qg$-channel.}
        \label{fig:qgNNLOhiggs}
    \end{subfigure}
    }
    \caption{Threshold expansions of the NLO (a,b) and NNLO (c,d) coefficients for single Higgs production. The unexpanded expressions are indicated by the coloured line, while the grey dashed, grey dash-dotted and black dotted lines show the LP, NLP and NNLP truncated expressions respectively. The bottom panes show the difference between the truncated and unexpanded expressions. The $y$-axis range of these difference plots is fixed at $20\%$ of the range of $\Delta_{\rm h}^{(1)}$ shown in the upper panes, for a uniformized comparison. We use eq.~(45), (46) and eq.~(48)-(50) in ref.~\cite{Anastasiou:2002yz}) for the unexpanded form of the partonic coefficients. Note that we have included the (trivial) LP approximation of the $qg$-channels, which vanishes, as these channels start contributing at NLP. }
    \label{fig:Higgsexpansion}
\end{figure}
We now examine how well the partonic coefficient functions are approximated by the threshold expansion at NLO and NNLO order,  for each partonic channel. Our default scale choice is $\mu = m_h$. In \fig{Higgsexpansion}a and b we show the threshold expansion of the two NLO coefficients $\Delta^{(1)}_{{\rm h},gg}$ and $\Delta^{(1)}_{{\rm h},qg}$. The third partonic NLO channel ($q\bar{q}$) only contributes at N$^4$LP, hence we leave it out of our discussion~\footnote{This is because this channel corresponds to the  $s$-channel process $q\bar{q}\rightarrow g \rightarrow gh$, which is proportional to $\frac{u^2+t^2}{s}$ with $s$, $t$ and $u$ Mandelstam variables. Parametrizing $t$ and $u$ as $t = -s(1-z)v$ and $u = -s(1-z)(1-v)$, this leads to a factor $(1-z)^2$ from the matrix element. An extra factor of $(1-z)$ follows from the phase space measure.}. We show the results without the $\delta(1-z)$ term, which is present in the $gg$-channel. In \fig{Higgsexpansion}c and d we show the equivalent NNLO results. 
Considering the $gg$-initiated channel first (\fig{ggNLOhiggs} and \fig{ggNNLOhiggs}), we observe that the LP threshold expansion of $\Delta^{(1)}_{{\rm h},gg}$ deviates considerably from the unexpanded result in the $z\rightarrow 1$ limit, which might be surprising at first sight. This is caused by the NLP terms:  $\ln^i(1-z)$ terms do not vanish as $z\rightarrow 1$, and are not captured by the LP truncation of the matrix element. Although subdominant to the LP contribution, they are not altogether negligible as $z\rightarrow 1$. The unexpanded NLO result is however well-described by the NLP approximation for $z\gtrsim 0.2$. None of the truncations captures the behaviour below $z\lesssim0.2$ well, due to the factor of $1/z$ in the partonic coefficient function. The threshold expansion of the NNLO coefficient function $\Delta^{(2)}_{{\rm h},gg}$ performs worse (\fig{ggNNLOhiggs}) than that of the NLO coefficient function: the LP approximation underestimates the unexpanded result in the large-$z$ domain, while the NLP approximation overestimates it. Convergence is seemingly only obtained at NNLP.  

Turning to the $qg$-channel (\fig{qgNLOhiggs}), whose LP approximation vanishes, we see that the NLP approximation overestimates the unexpanded result in the large-$z$ domain. Also here the small-$z$ domain is poorly described by the $z\rightarrow 1$ expansion of the full partonic coefficient function due to a $1/z$ factor. The NNLO $qg$ coefficient function (\fig{qgNNLOhiggs}) shows similar behaviour.

\subsection{Threshold behaviour of the partonic coefficient functions for Drell-Yan}
\label{subsub:DY}
Next we perform the same studies for the Drell-Yan process. The distribution in the squared invariant mass $Q^2$ is given by
\begin{equation}
\label{eq:DYtotalxsec}
\frac{{\rm d} \sigma_{{\rm DY}}}{{\rm d} Q^2} = \sigma_0^{\rm DY}\sum_{i,j}~\int_{\tau}^1\frac{{\rm d}z}{z}~\mathcal{L}_{ij}\left(\frac{\tau}{z}\right)~\Delta_{{\rm DY},ij}(z),
\end{equation}
with  $z=\frac{Q^2}{s}$.
The $\alpha_s$ expansion of $\Delta_{{\rm DY},ij}(z)$ is similar to \Eq{eq:pertexp}, and the threshold expansion of $\Delta_{{\rm DY},ij}(z)$ is as in \Eq{eq:defthreshold}. The coefficient $\Delta^{(0)}_{{\rm DY},ij}(z)$ is given by $\delta_{iq}\delta_{j\bar{q}}\delta(1-z)$ by extracting in \eq{eq:DYtotalxsec} the prefactor (see appendix~\ref{AppNormalization})
\begin{equation}
\label{eq:DYLOcoeff}
\sigma_0^{\rm DY} = \frac{4\pi \alpha_{\rm EM}^2}{3 Q^2 S}\frac{1}{N_c},
\end{equation}
where  $\alpha_{\rm EM}=1/127.94$ is the electromagnetic fine-structure constant at the scale $M_Z$.
To compare directly with single Higgs production we set $\mu=Q=m_h$ for the discussion in this subsection.\\
\begin{figure}[t!]
    \centering
    \mbox{
    \centering
    \begin{subfigure}{.482\textwidth}
    \centering
        \includegraphics[width=\textwidth]{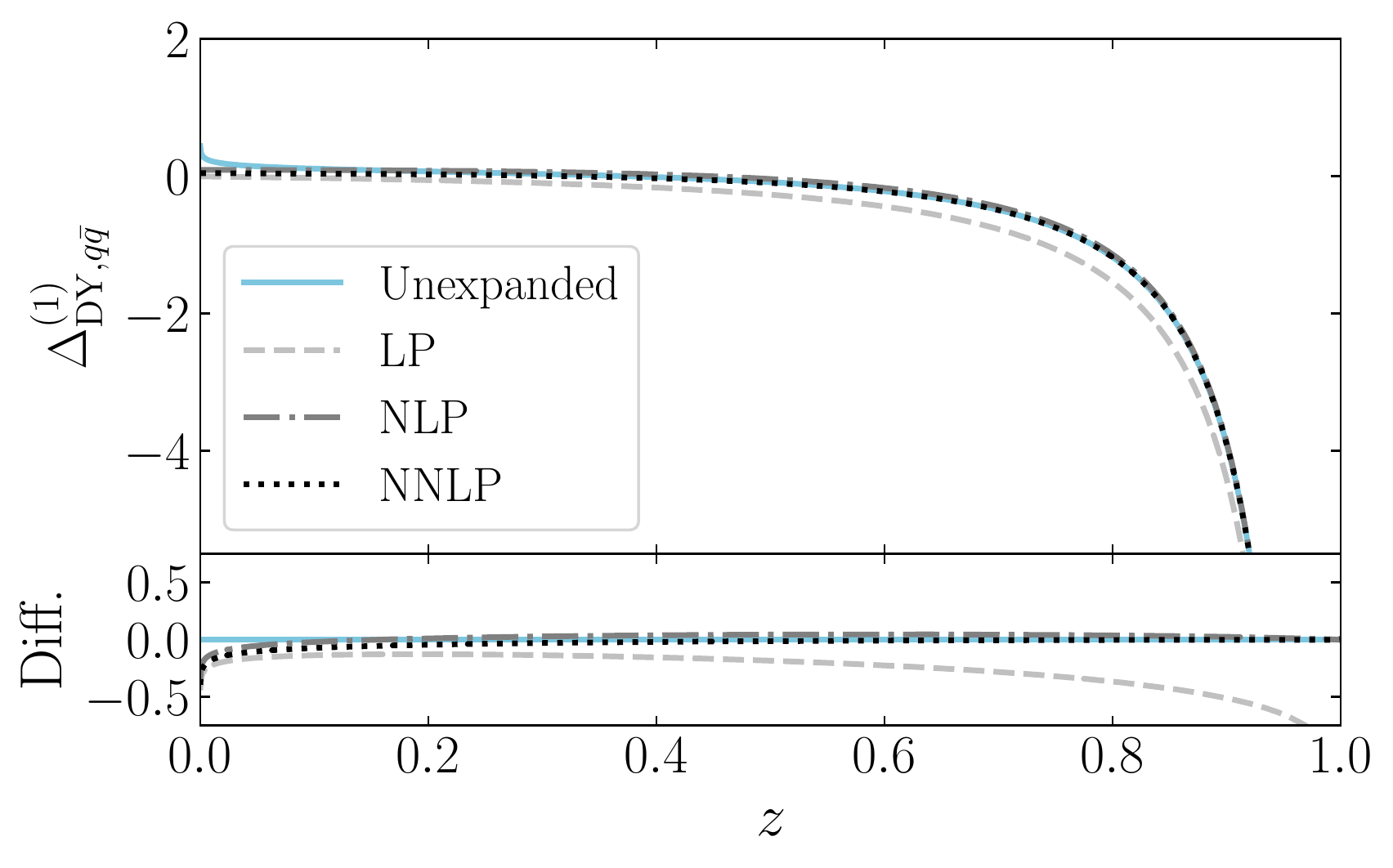}
       \caption{NLO $q\bar{q}$-channel. }
        \label{fig:qqbarNLODY}
    \end{subfigure}
    \begin{subfigure}{.502\textwidth}
        \includegraphics[width=\textwidth]{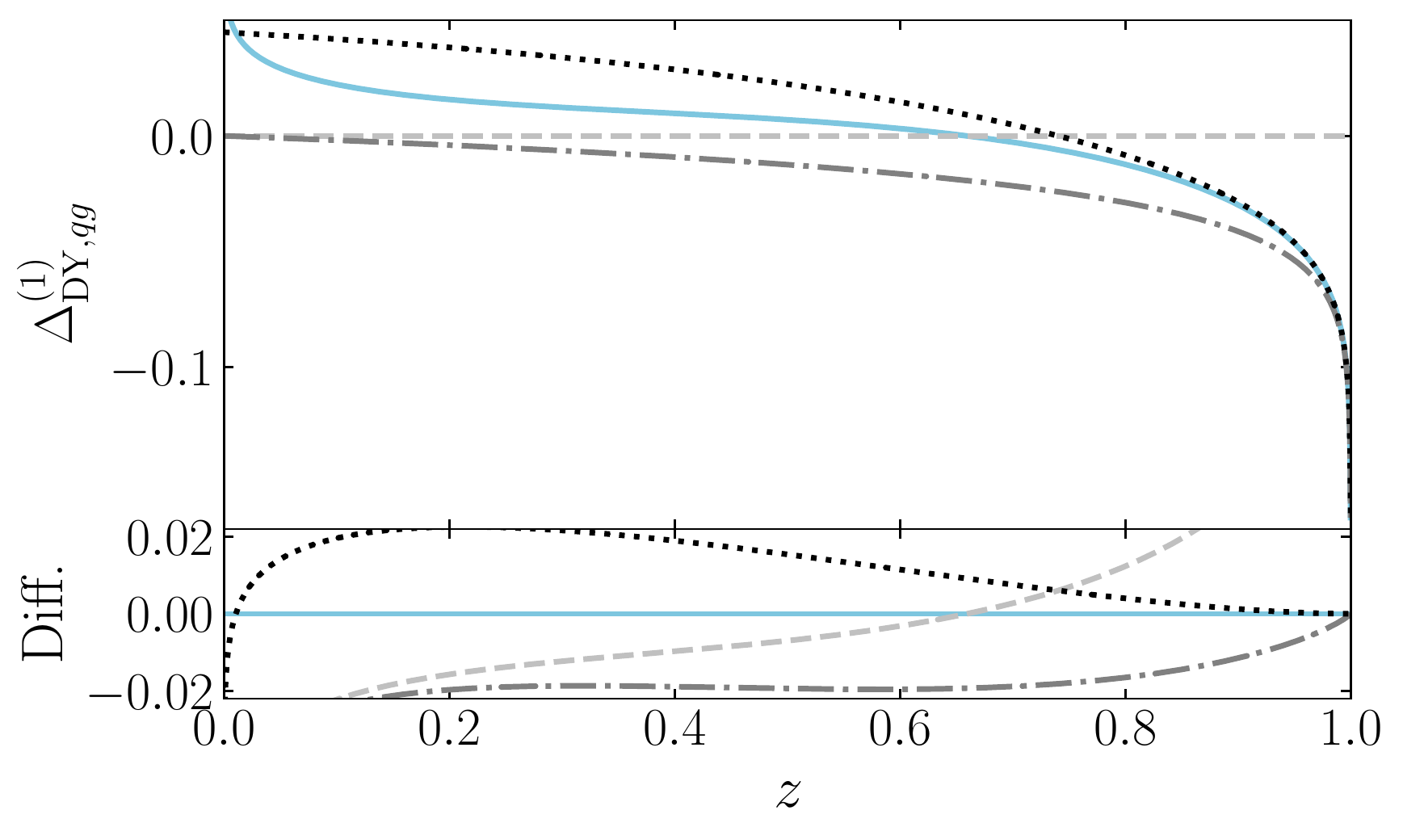}
       \caption{NLO $qg$-channel.}
        \label{fig:qgNLODY}
    \end{subfigure}
    }
    \vspace{0.2cm}
    \hspace{-7pt}
    \mbox{
    \centering
    \begin{subfigure}{.492\textwidth}
    \centering
        \vspace*{-3pt}\includegraphics[width=\textwidth]{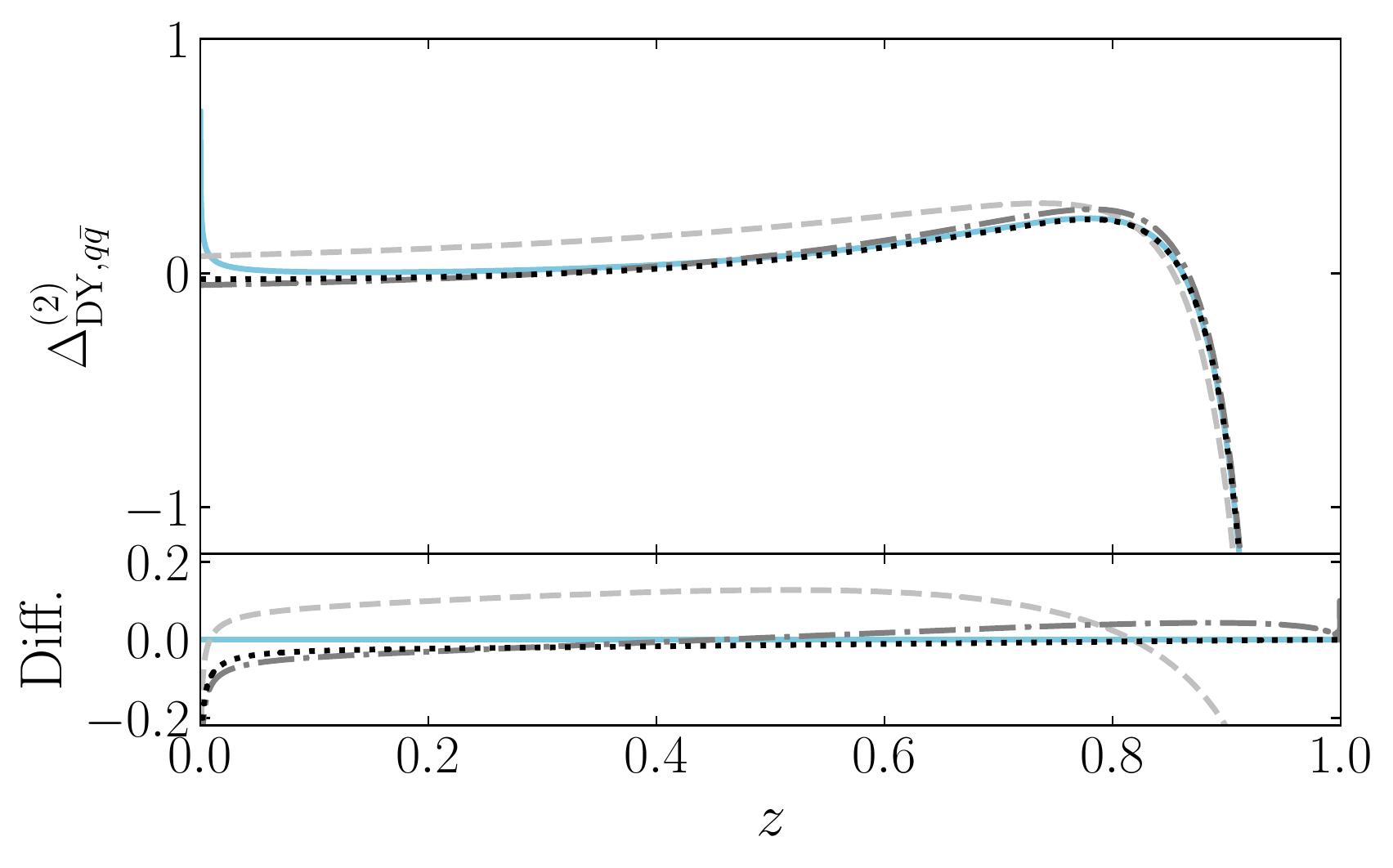}
       \caption{NNLO $q\bar{q}$-channel. }
        \label{fig:qqbarNNLODY}
    \end{subfigure}
    \hspace{1.5pt}\begin{subfigure}{.498\textwidth}
        \includegraphics[width=\textwidth]{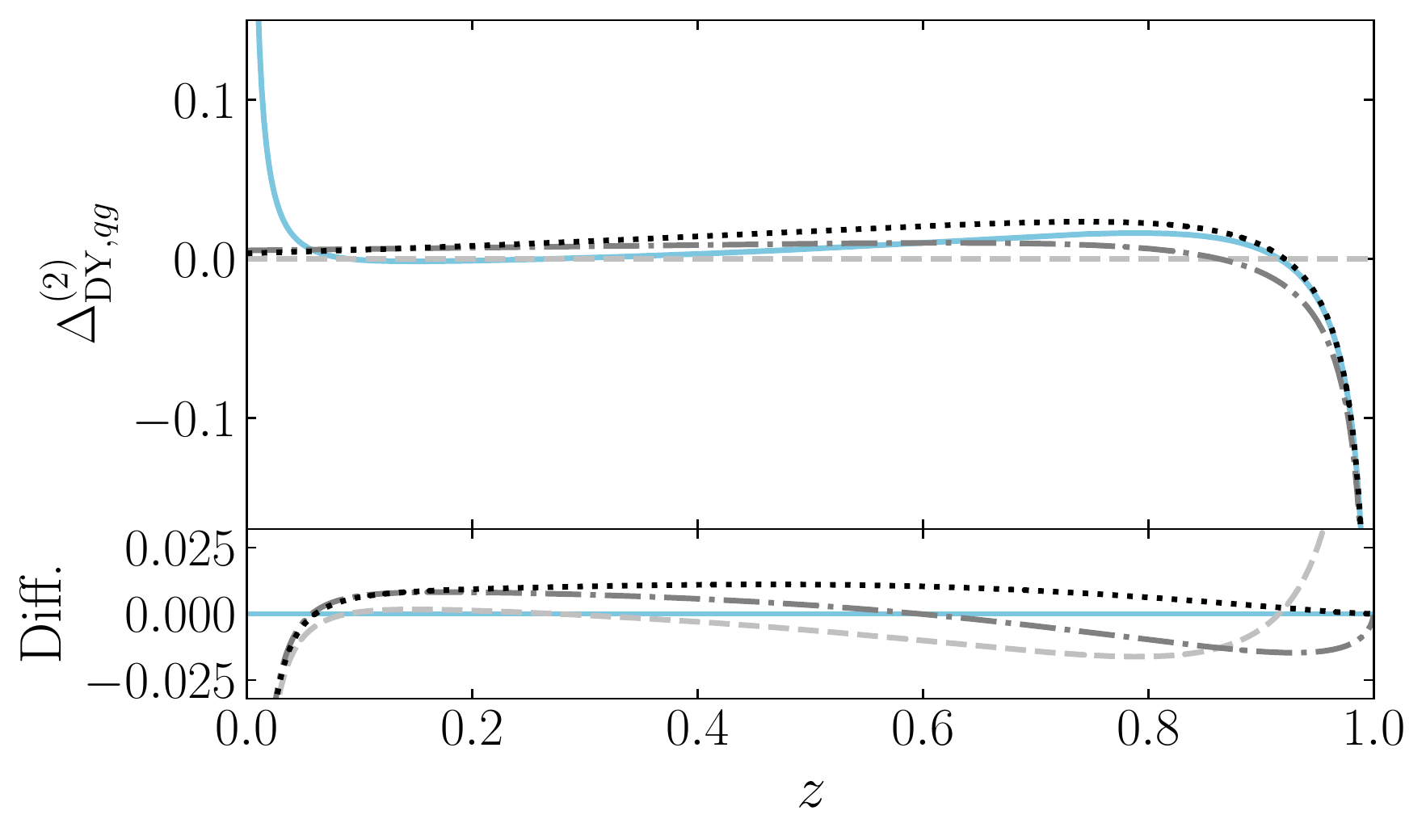}
       \caption{NNLO $qg$-channel. }
        \label{fig:qgNNLODY}
    \end{subfigure}
    }
    \caption{Expansions of the NLO (a,b) and NNLO (c,d) partonic coefficients for the DY production of an off-shell photon. The labeling is the same as in \fig{Higgsexpansion}. }\label{fig:DYexpansion} %(eq.~(B.2) of ref.~\cite{Hamberg:1990np} without the delta term). (eq.~(B.18) of ref.~\cite{Hamberg:1990np}, (eq.~(B.17) of ref.~\cite{Hamberg:1990np}), (eq.~(B.7) of ref.~\cite{Hamberg:1990np} without the constant contribution))
\end{figure}
\vspace{-1em}

In \fig{qqbarNLODY} and \fig{qqbarNNLODY}, we show the threshold expansions truncated up to NNLP of the NLO and NNLO $q\bar{q}$-channel, whose exact expressions are obtained from ref.~\cite{Hamberg:1990np}. Both perturbative orders are well described by the NLP approximation for a large range of $z$-values. The convergence of the threshold expansions of the $qg$-channel, shown in  \fig{qgNLODY} and \fig{qgNNLODY}, is worse than for the dominant $q\bar{q}$-channel. In contrast to the Higgs case, the NLP approximation underestimates the $\Delta_{{\rm DY},qg}^{(1)}$ and $\Delta_{{\rm DY},qg}^{(2)}$ coefficients near $z=1$. The NNLP expansion behaves slightly better, although convergence is again slow for this channel. As for the Higgs case, the partonic channels ($qq,gg$) that open up at NNLO do not contribute at NLP, as these channels require the emission of two soft quarks. 

Recapitulating, for both single Higgs production and DY we have seen that the \emph{partonic} threshold expansion works best for the dominant production channel. The NLP truncation overestimates the coefficient functions for intermediate and large values of $z$ in these dominant channels, but convergence is reached at NNLP. In the $qg$-channels, the NLP expansion overestimates the exact result for Higgs production for $z\rightarrow 1$, while it underestimates it for DY. At NNLP no substantial improvement is obtained in this channel. The contribution in the $z\rightarrow 0$ region, corresponding to the high-energy limit, is for Higgs production  more pronounced than for DY, for both production channels. This is due to the factor of $1/z$ that is part of the partonic coefficient function of the former process, which magnifies small-$z$ contributions. To what extent the various differences also manifest themselves in hadronic cross sections depends of course on the parton flux. This question is addressed in the next section.

\subsection{Threshold expansions of the Higgs and DY hadronic cross sections}
\begin{figure}[t]
    \centering
        \includegraphics[width=.475\textwidth]{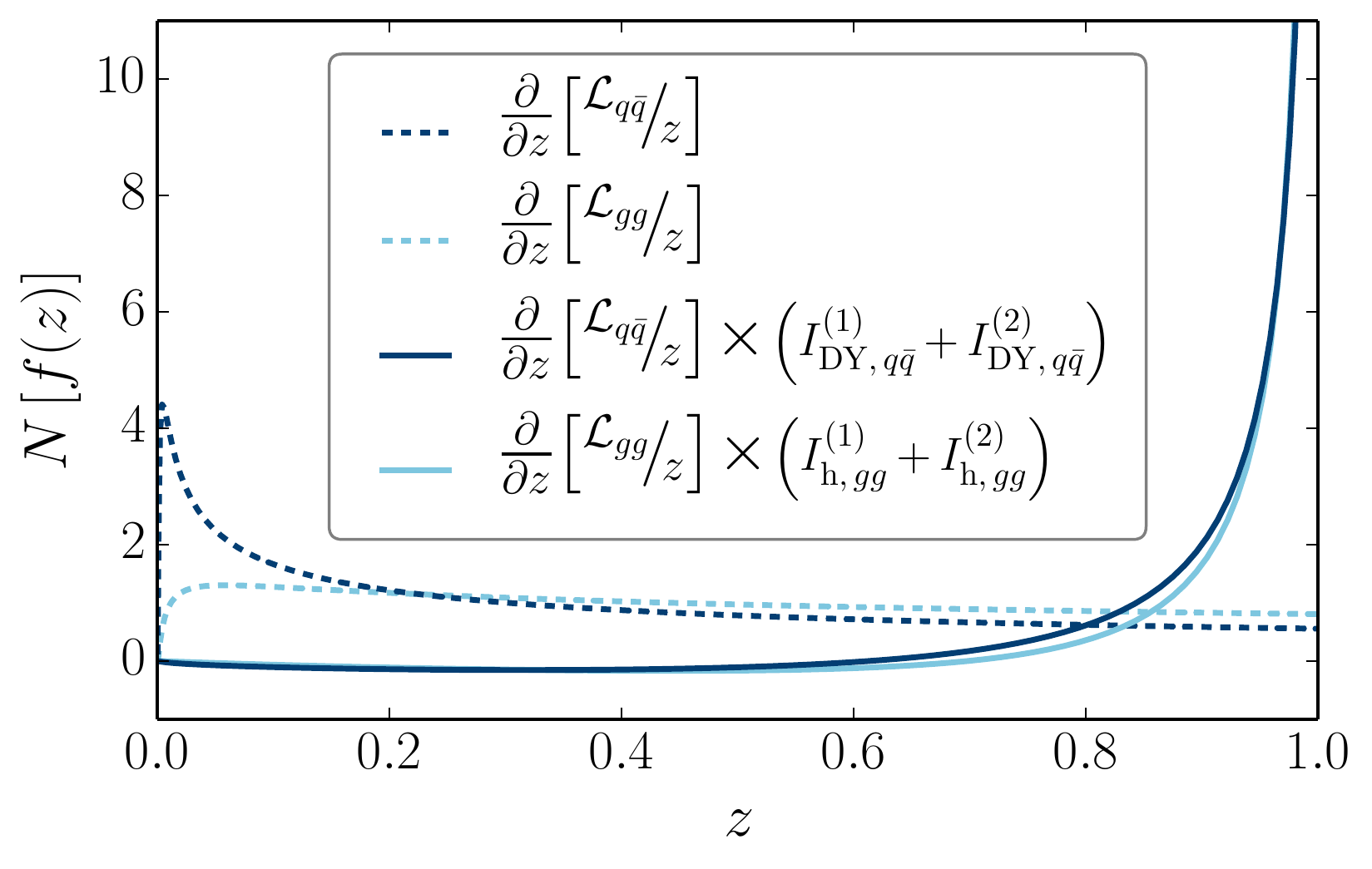}
        \caption{The derivative of the parton luminosity (dashed)  and its  point-by-point multiplication with the integrated LP partonic coefficient function up to NNLO (solid), for both DY (dark blue) and single Higgs (light blue) production at $Q=\mu=125$ GeV. All curves are normalized according to eq.~\eqref{norm_quantity}. }
        \label{fig:pLumder_times_pdistr}
    \end{figure}

\noindent In the previous subsection it became clear that the Higgs partonic coefficient functions have a more pronounced small-$z$ contribution than those of DY. In the hadronic cross sections, these coefficient functions are weighted by the parton flux as in eqs.~\eqref{eq:higgstotalxsec} and \eqref{eq:DYtotalxsec}, which possibly affects the quality of the threshold expansion for the hadronic cross section.

For non-singular terms in the partonic coefficient functions (those that do not contain plus-distributions), the convolutions in eq.~\eqref{eq:higgstotalxsec} and eq.~\eqref{eq:DYtotalxsec} correspond to a point-by-point multiplication with the parton luminosity function. The product, which is the integrand for the inclusive cross section at fixed $\tau$, is a function of $z$ only. This \emph{weight distribution} shows which parts of the coefficient functions are enhanced or suppressed by the parton flux, and thus gives much information about how well the threshold expansion can approximate the exact result. We emphasize that for understanding the quality of the threshold expansion, only the shape of the weight distributions for the non-singular contributions matters; the plus-distributions cannot be expanded any further in the threshold limit. 
For completeness, we nevertheless review the weight distributions for these LP terms briefly, as these notably involve the \emph{derivative} of the parton flux, as well as the separately integrated partonic coefficient function (see appendix~\ref{app:singular}). For brevity we denote the latter by
\begin{equation}
    I^{(n)}_{{\rm DY\!/h},\,ij}(z) \equiv -\int_0^z\! {\rm d}z^\prime\, \left.\Delta^{(n)}_{{\rm DY\!/h},\,ij}(z^\prime)\right|_{\rm LP}\,.
\end{equation}
In \fig{pLumder_times_pdistr} we plot the resulting weight distribution for DY and Higgs production, up to NNLO and for $Q=\mu=m_h$. To aid comparison we normalize the plotted functions (generically denoted by $f(z)$) as
\begin{equation}
    N\left[f(z)\right] = \frac{f(z)}{\left|\int_\tau^1 {\rm d}z\  f(z)\right|}\,, \label{norm_quantity}
\end{equation} 
 such that the absolute area under each curve equals unity. The $q\bar{q}$ parton luminosity is defined as a charge weighted sum of symmetrized parton flux contributions from the five lightest quark flavours:
 \begin{equation}
 \label{plumqqbar}
     \mathcal{L}_{q\bar{q}}\left(\frac{\tau}{z}\right) = \sum_{q\,\in\, \{u,d,c,s,b\}} e_{q}^2\,\int_{\tau/z}^1 \frac{{\rm d}x}{x}\left[f_q(x,\mu)\,f_{\bar{q}} \left(\frac{\tau/z}{x},\mu \right)+f_{\bar{q}} (x,\mu)\,f_q\left(\frac{\tau/z}{x},\mu\right)\right] \,,
 \end{equation} 
with $e_q$ the fractional charge of the quark, normalized to the electromagnetic charge $e$. The differentiated $q\bar{q}$-flux highlights the small-$z$ region more than the differentiated $gg$-flux, as shown by the dashed lines, but since the LP terms of the partonic coefficient functions for DY are small in that regime (see~\fig{qqbarNLODY} and \fig{qqbarNNLODY}), this affects their product with the coefficient functions (solid lines) only little. The product diverges (integrably so) for both channels for $z\rightarrow 1$, as one would expect for the LP terms. Overall, the LP behaviour of both processes is very similar. Around $z\sim 0.8$ a small difference appears, caused by the stronger decrease of the derivative of the $q\bar{q}$-flux. This results in a slightly more spread-out (i.e.~less threshold-centered) weight distribution for DY, but as we mentioned above, this cannot affect the quality of the threshold expansion. We point out that, somewhat counter-intuitively, the LP terms make up only a modest fraction of the total cross section, for both processes (we shall quantify this in \fig{HiggsexpansionHAD} and \fig{DYHAD}). 

 \begin{figure}[t]
    \mbox{
    \centering
    \begin{subfigure}{.465\textwidth}
    \centering
        \includegraphics[width=\textwidth]{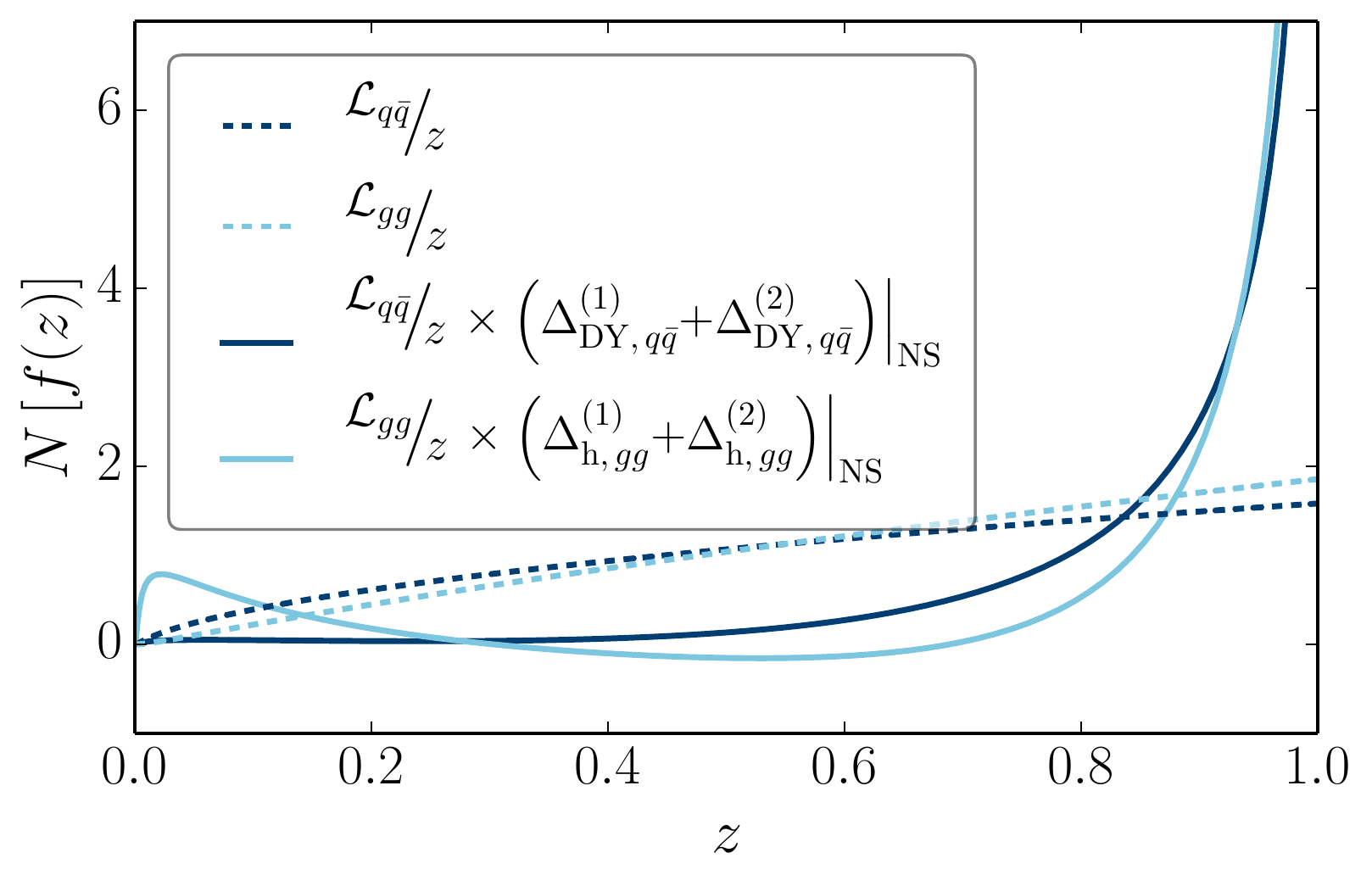}
          \caption{$Q = \mu = 125$ GeV}
        \label{fig:pLum_times_Kfactor125}
    \end{subfigure}
    \hspace{.3cm}
    \begin{subfigure}{.465\textwidth}
    \centering
        \includegraphics[width=\textwidth]{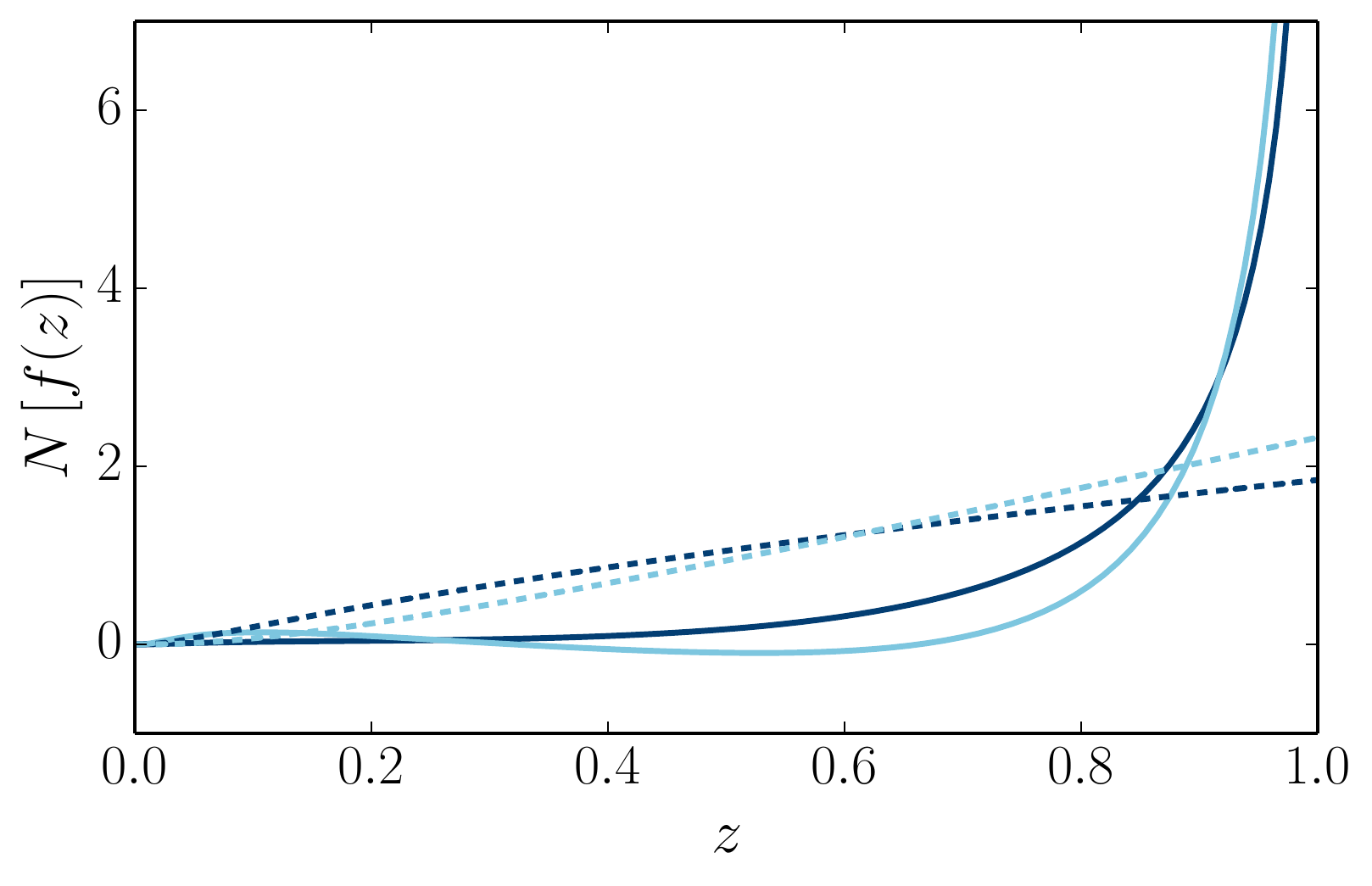}
       \caption{$Q = \mu = 500$ GeV}
        \label{fig:pLum_times_Kfactor500}
    \end{subfigure}
    }
    \vspace{0.2cm}
     \caption{The luminosity function (dashed) and its point-by-point multiplication with the non-singular part of the coefficient functions up to NNLO (solid), for the dominant channel in both DY ($q\bar{q}$, dark blue) and single Higgs ($gg$, light blue) production. All curves are normalized according to eq.~\eqref{norm_quantity}.}
     \label{fig:comparisonsNNLO}
\end{figure}
In \fig{comparisonsNNLO} we show the normalized non-singular part of the partonic coefficient functions up to NNLO, weighted by the respective parton luminosity (solid lines)~\footnote{Some contributions to the DY NNLO partonic coefficient in the $q\bar{q}$ channel ought to be summed over the quark flavours without charge weighing (see appendix~A of ref.~\cite{Hamberg:1990np}). We have explicitly verified that these terms contribute only from ${\rm N^6LP}$ onwards and that these contributions are numerically negligible. Therefore, the overall point-by-point multiplication with the parton luminosity function in \eq{plumqqbar} is deemed to be a valid approximation.}. We focus on the dominant production channel for each process and show the $z$-dependence of the parton fluxes themselves as well (dashed lines).  In addition to the natural comparison scale for on-shell Higgs production of $Q=\mu=125\ \rm GeV$ (\fig{pLum_times_Kfactor125}), we also consider one off-shell Higgs-production scenario with $Q=\mu=500\ \rm GeV$  (\fig{pLum_times_Kfactor500}). In both cases we see that the product of the parton luminosities and the coefficient functions are suppressed in the small $z$ regime, as a result of the small parton flux. This, in turn, is due to the momentum fractions $x_1$ and $x_2$ being large when $z\rightarrow \tau$ since $x_1 x_2 =\frac{\tau}{z}$. The individual sea quark and gluon PDFs are small in the large $x$ domain, suppressing the luminosity~\footnote{The valence quark contributions to the parton luminosity defined in \eq{plumqqbar} cause it to fall off less steeply with $x$ than for sea quarks (or gluons) alone. This is reflected in the stronger small-$z$ suppression of the parton luminosity for $gg$ compared to $q\bar{q}$.}. Larger values of $\mu$ in the flux strengthen this suppression. Therefore, the singular behaviour of the coefficient functions for single Higgs production near $z=0$  is more suppressed by the parton luminosity as $\mu$ increases. Thus at $Q=\mu\!=125$~GeV a notable peak is present in the small $z$ region (it is more pronounced at smaller $\mu$ values). It disappears due to the above-mentioned suppression for larger $\mu$ values, as seen in \fig{pLum_times_Kfactor500} for $Q=500$~GeV. However, at $Q=125$~GeV this feature should still affect the quality of the threshold expansion for the hadronic Higgs cross section. As the DY partonic coefficient function does not show singular behaviour in the small $z$-regime, the $\mu$-dependent suppression of the parton luminosity has less impact and we expect the quality of the threshold expansion to be mostly $Q$-independent. Based on these considerations we therefore expect that, especially for small-$Q$ values, the quality of the threshold expansion of the dominant channel for DY is better than that for Higgs production.  
\begin{figure}[t]
    \mbox{
    \begin{subfigure}{.495\textwidth}
    \centering
        \includegraphics[width=\textwidth]{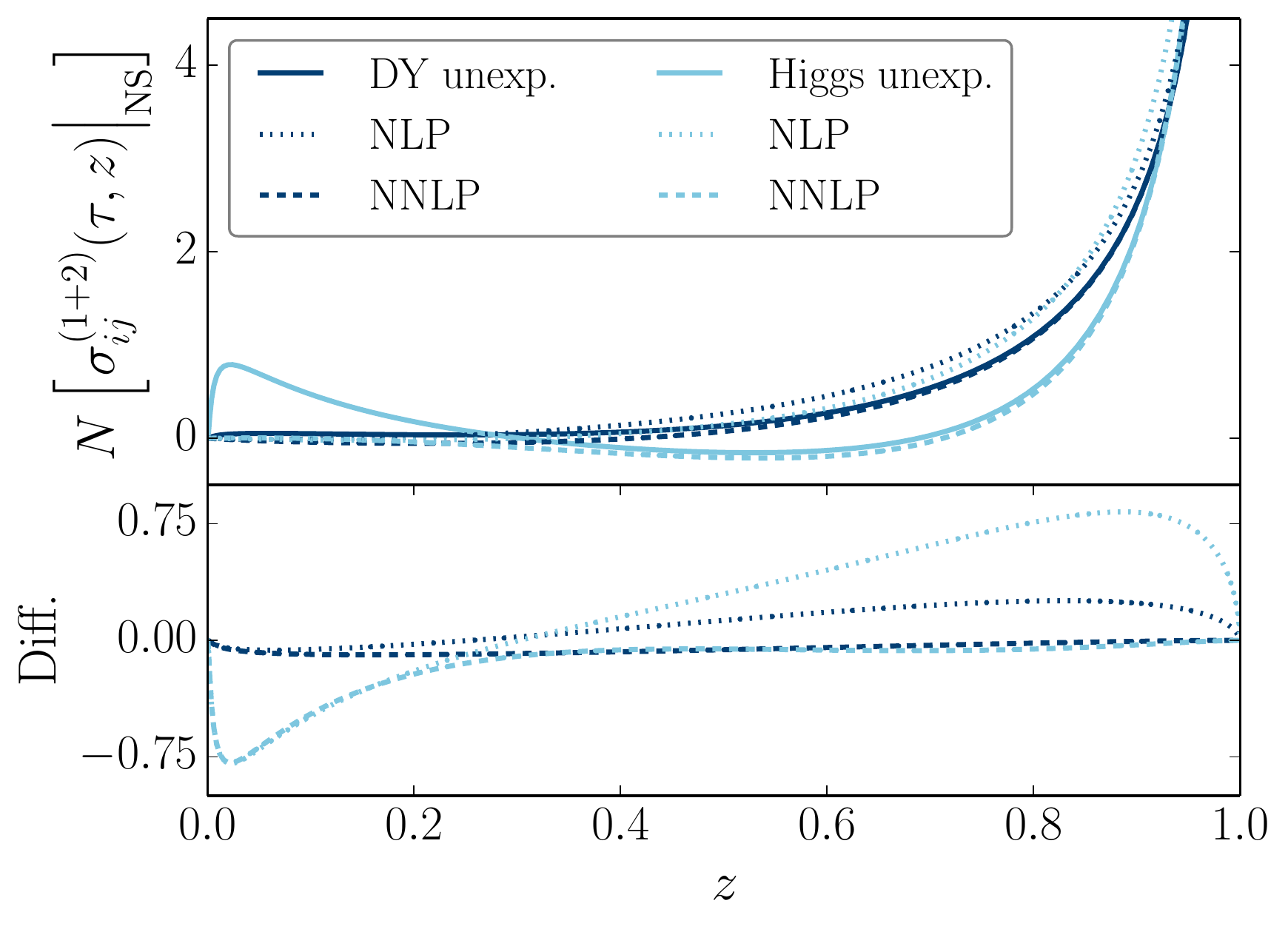}
          \caption{$Q = \mu =125$ GeV}
        \label{fig:z-weight_exp125}
    \end{subfigure}
    
    \begin{subfigure}{.495\textwidth}
    \centering
        \includegraphics[width=\textwidth]{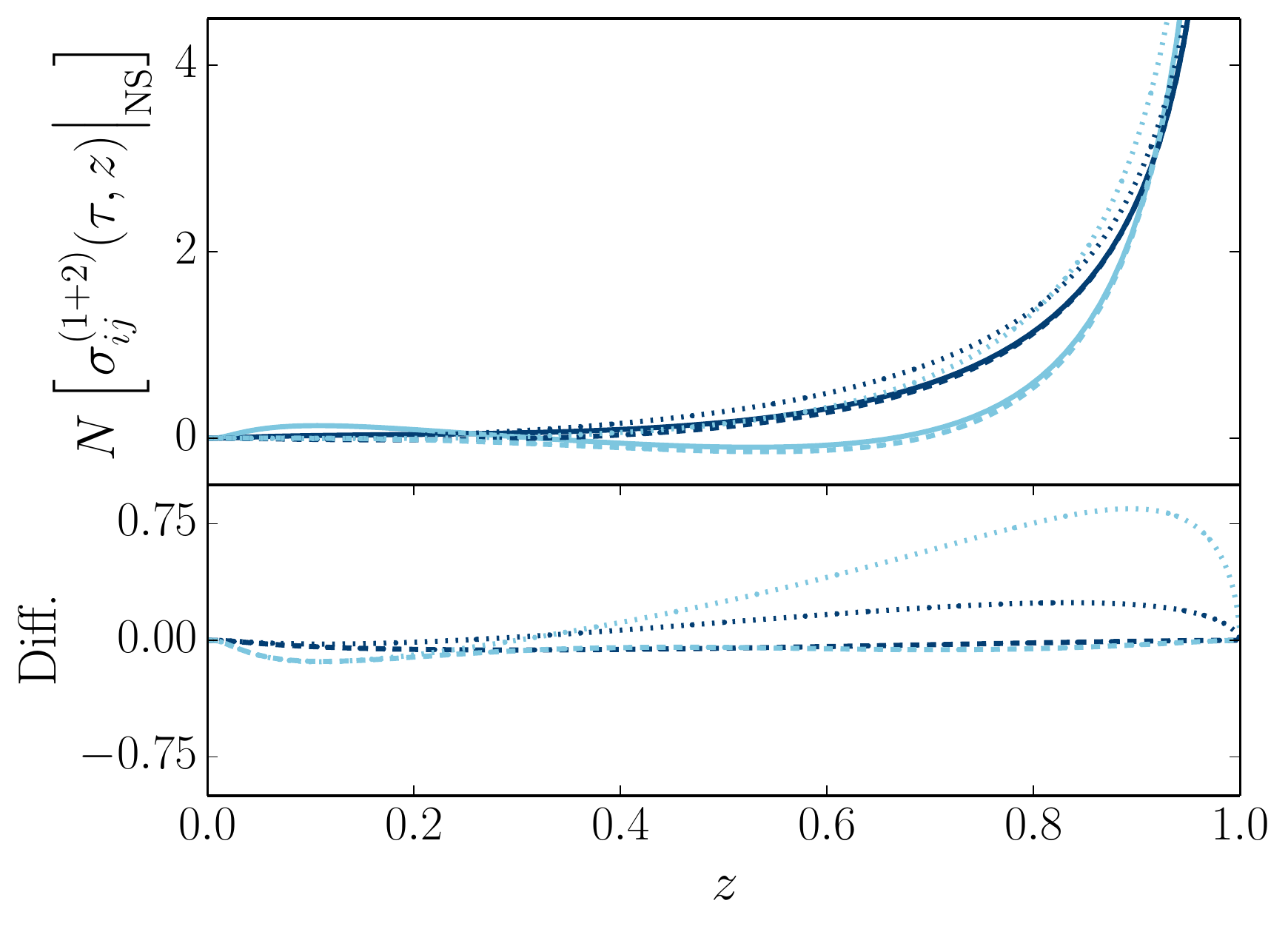}
       \caption{$Q = \mu =500$ GeV}
        \label{fig:z-weight_exp500}
    \end{subfigure}
    }
     \caption{Exact (solid lines) as well as truncated expressions at NLP (dotted) and NNLP (dashed) for the non-singular part of the coefficient functions up to NNLO, weighted with the parton luminosity, for the dominant channel of DY ($q\bar{q}$, dark blue) and single Higgs ($gg$, light blue) production. All curves are normalized according to eq.~\eqref{norm_quantity}.}
     \label{fig:zweight_exp}
\end{figure}

To test this expectation, we compare the parton-luminosity-weighted NLP and NNLP approximations of the partonic coefficient functions to the unexpanded NLO+NNLO result. This product, being the (approximated) integrand of the non-singular (NS) contribution to the hadronic cross section, is denoted by 
\begin{equation}
    \left.\sigma^{(1+2)}_{ij}(\tau,z)\right|_{\rm NS}^{\rm unexp./NLP/NNLP} = \frac{\mathcal{L}_{ij}(\tau/z)}{z}\left.\left(\Delta_{ij}^{(1)}(z)+\Delta_{ij}^{(2)}(z)\right)\right|_{\rm NS}^{\rm unexp./NLP/NNLP}\,.
\end{equation} 
 Fig.~\ref{fig:zweight_exp} shows normalized plots of these quantities. We see that the weight distribution for DY production is approximated well, for both $Q$ values, by the NLP truncation over the full range of $z$. At NNLP the agreement with the exact result is excellent. For Higgs production at $Q=125$~GeV, we again observe that the expansion does not capture the small-$z$ region well, neither at NLP nor at NNLP. The stronger parton luminosity suppression at $Q=500$~GeV does aid the convergence, with a NNLP truncation that is almost as good as for DY. Therefore, these plots confirm the expectation that the threshold expansions works better for the dominant production channel of DY than for Higgs production.
\begin{figure}[tbh]
    \centering
    \begin{subfigure}{.485\textwidth}
    \centering
        \includegraphics[width=\textwidth]{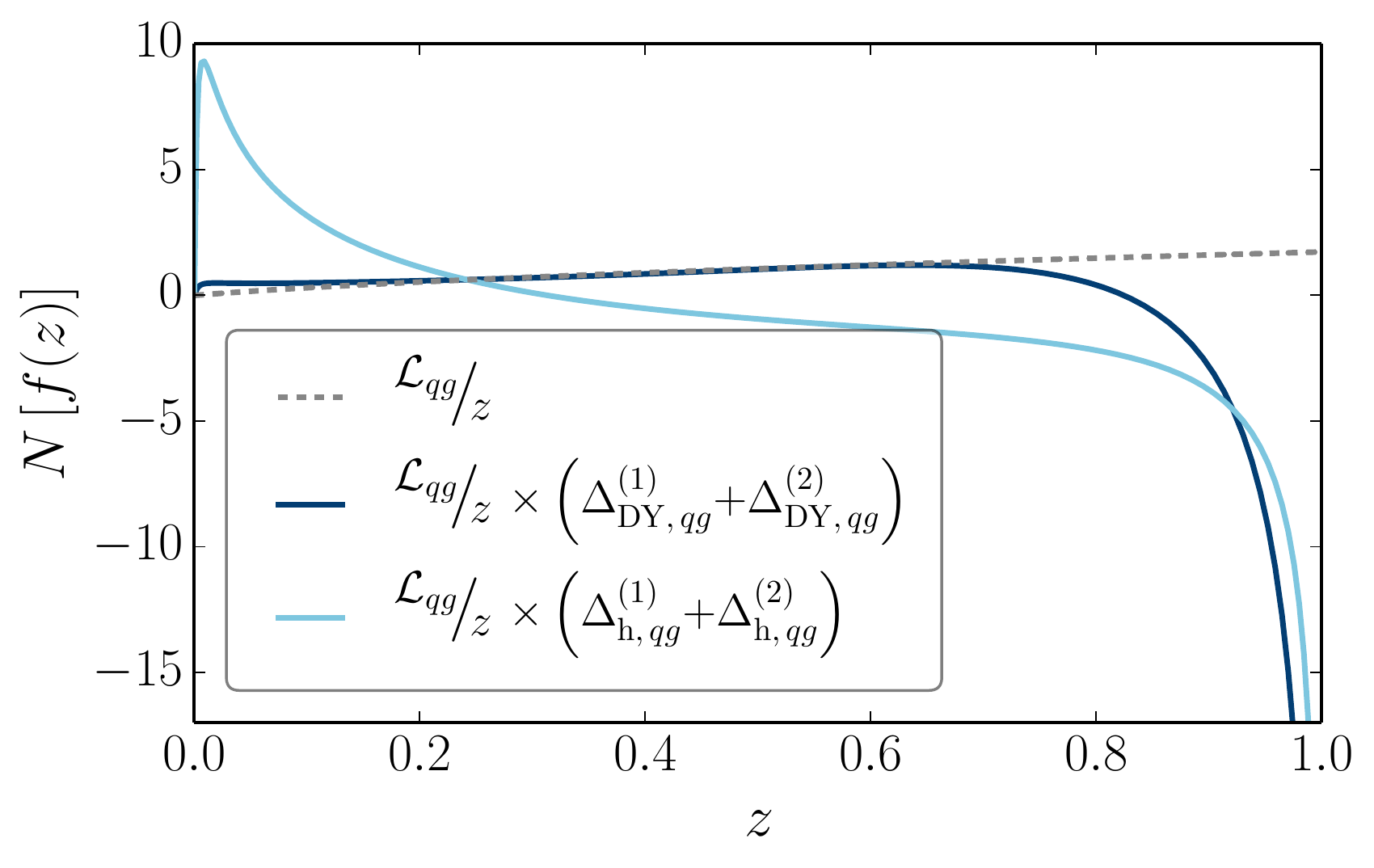}
          \caption{$Q = 125$ GeV}
        \label{fig:qg_flux_125}
    \end{subfigure}
    \hspace{0.2cm}
    \begin{subfigure}{.485\textwidth}
    \centering
        \includegraphics[width=\textwidth]{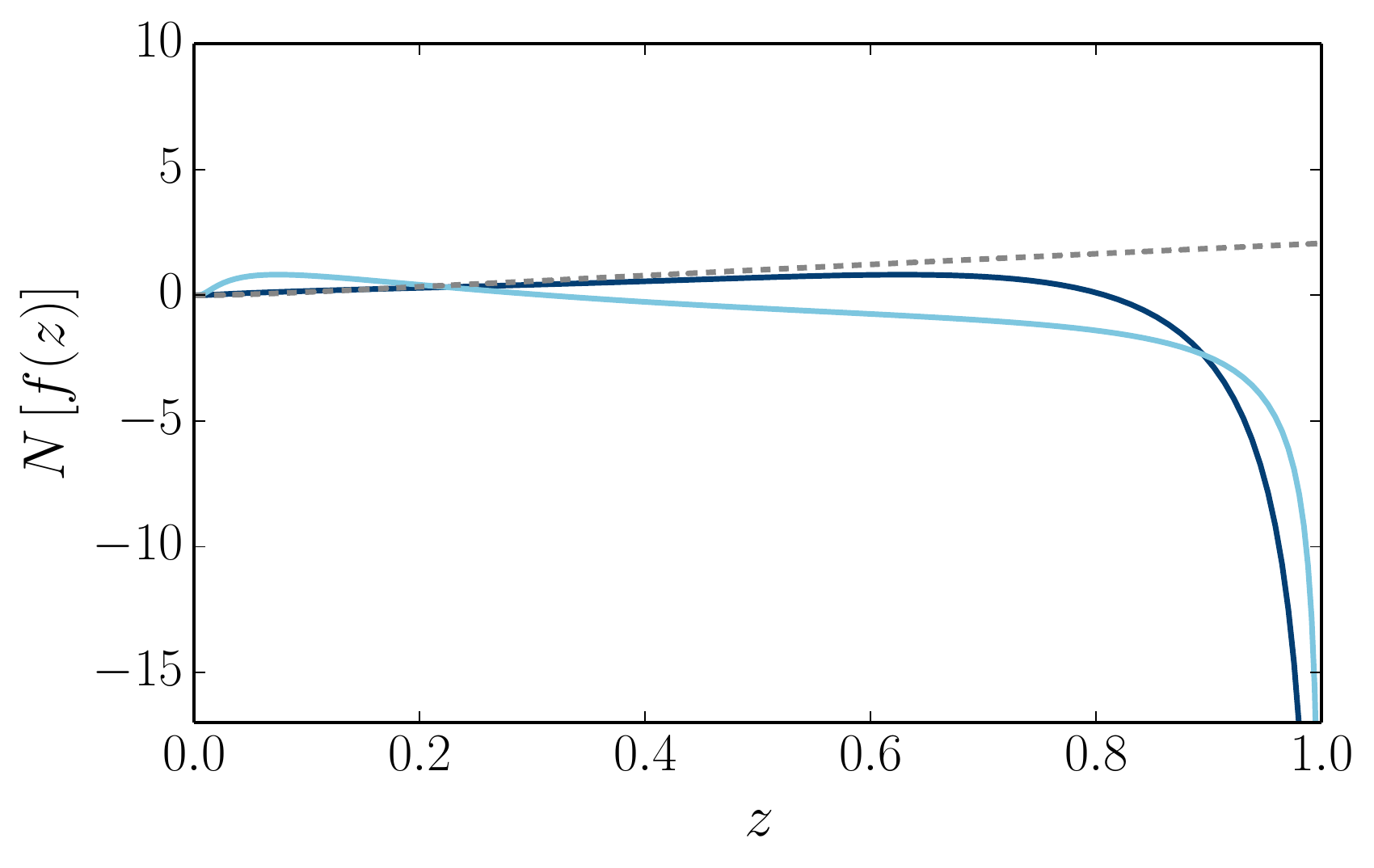}
       \caption{$Q = 500$ GeV}
        \label{fig:qg_flux_500}
    \end{subfigure}
    \caption{The luminosity function (grey dashed) and its point-by-point multiplication with the coefficient functions up to NNLO (solid), for the $qg$-channel of DY (dark blue) and single Higgs (light blue) production. All curves are normalized according to eq.~\eqref{norm_quantity}.}
    \label{fig:point_by_point_qg}
\end{figure}
\begin{figure}[h]
    \begin{subfigure}{.5\textwidth}
    \centering
        \includegraphics[width=\textwidth]{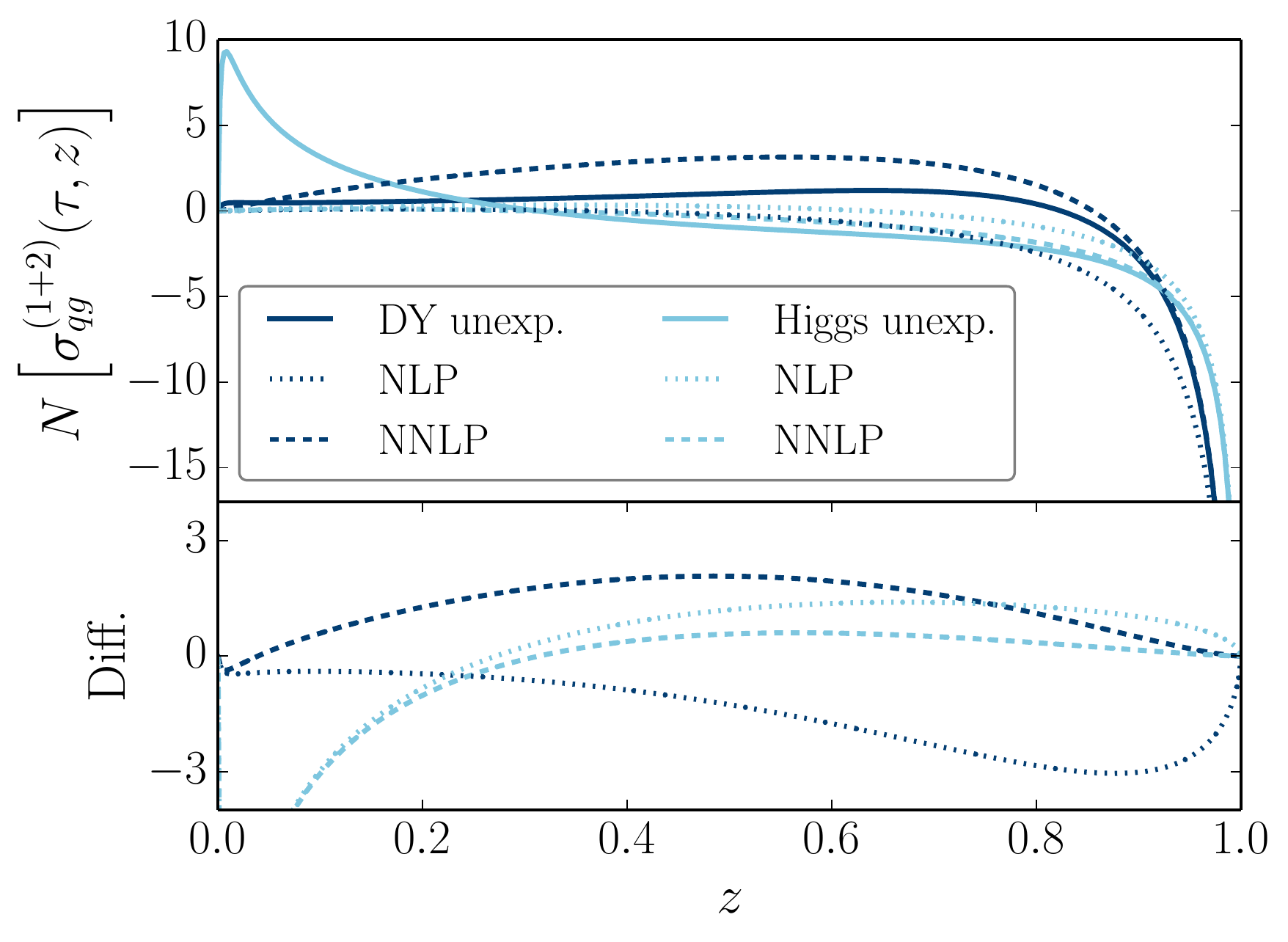}
          \caption{$Q = 125$ GeV}
        \label{fig:z-weight_qg_exp125}
    \end{subfigure}
    \begin{subfigure}{.5\textwidth}
    \centering
        \includegraphics[width=\textwidth]{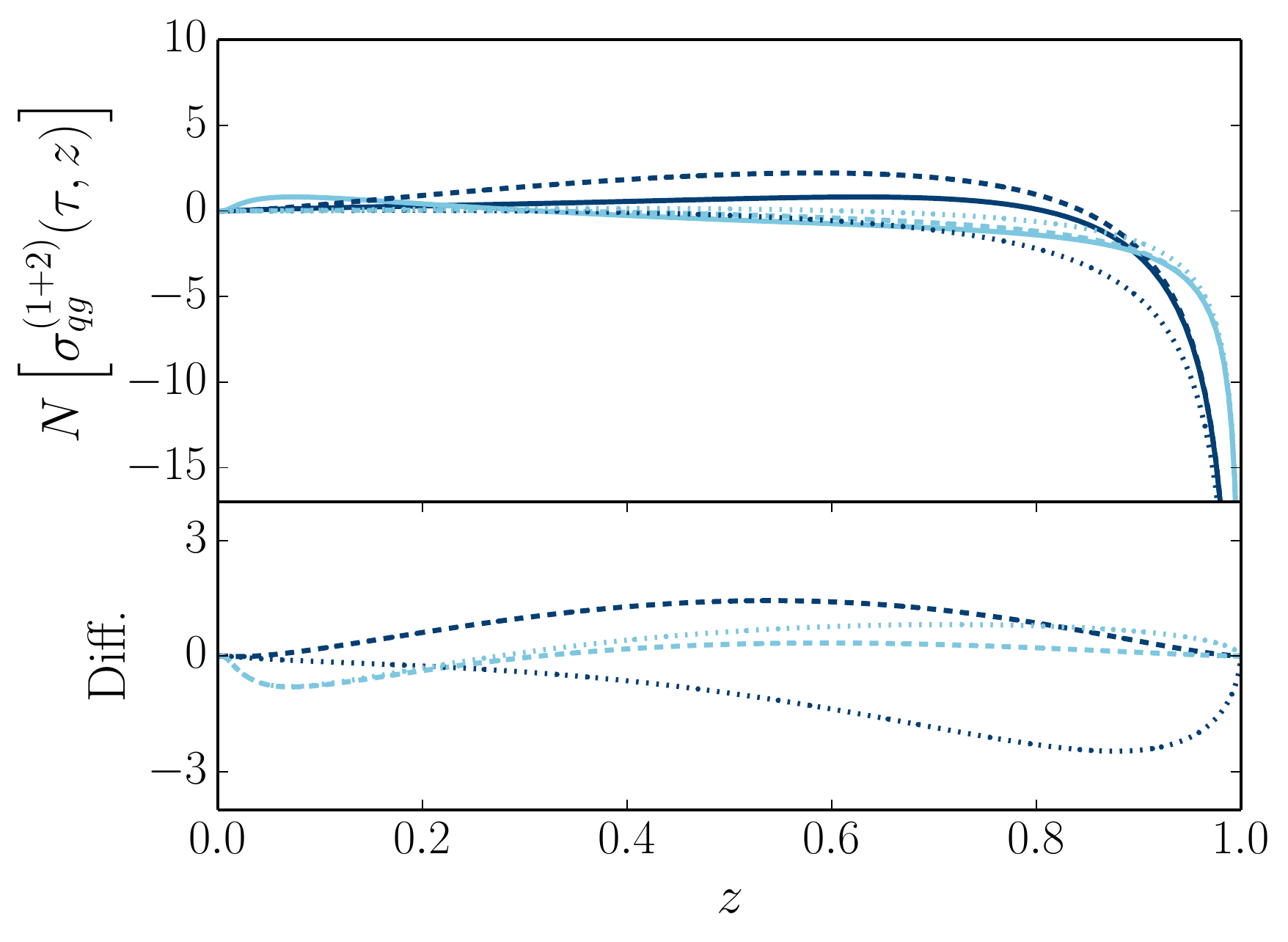}
       \caption{$Q = 500$ GeV}
    \label{fig:z-weight_qg_exp500}
    \end{subfigure}
    \caption{Exact and truncated expressions of the coefficient functions up to NNLO, weighted with the parton luminosity, for the $qg$-channel of DY and single Higgs production. Labeling is the same as for \fig{zweight_exp}. All curves are normalized according to eq.~\eqref{norm_quantity}.}
    \label{fig:zweight_qg_exp}
\end{figure}

In \fig{point_by_point_qg} and \fig{zweight_qg_exp} we show similar plots for the $qg$-channels of both processes. In particular, \fig{point_by_point_qg} shows that the $z$-dependence of the $qg$-flux is not qualitatively different from the $q\bar{q}$ and $gg$ flux~\footnote{We note that the parton luminosity shown is summed over the five lightest (anti)quark flavours. For DY this luminosity function is weighted with the quark charge as well, but since this does not alter the line shape significantly, we do not show it separately to improve readability.}. Just as for the $gg$-channel, the parton luminosity suppresses the small-$z$ domain, such that the enhancement in that region from the factor of $1/z$ in the Higgs partonic coefficient function is tempered. From \fig{zweight_qg_exp} we see that the power expansions approximate the full result less well for these channels than for the dominant production channels. Moreover, while the NLP and NNLP approximations for the Higgs $qg$-channel get significantly better for higher $Q$ values, those for the DY $qg$-channel do not, and the threshold expansion for Higgs outperforms the one for DY at $Q=500$~GeV. Again this is due to the parton luminosity suppression in the small $z$-region. For Higgs production the truncated expression deviates most from the exact (N)NLO coefficient function in this region, as seen in \fig{Higgsexpansion}, while for DY (\fig{DYexpansion}) this deviation is more spread out and therefore benefits less from going to large $Q$ values. 

We note that the results presented in this subsection contain a level of detail that is of course lost upon integration over $z$. As such, the integration over $z$ may lead to a seemingly contradictory result: cancellations between under- and overestimations of the exact result across the $z$ domain may cause crude approximations to look more favourable than expected. This is what we observe in the next subsection, where we consider the quality of the threshold expansion of the integrated hadronic cross section. For example, for $Q=125$~GeV the integrated NLP truncation approximates the exact NLO+NNLO result for the $qg$-channel in Higgs production better than the NNLP one, as will be seen in \fig{HiggsexpansionHAD}, since the more severe overshoot for $z\gtrsim 0.3$ provides a better cancellation for the undershoot at small $z$. 

\begin{figure}[t]
    \centering
    \mbox{
    \begin{subfigure}{.48\textwidth}
    \centering
        \includegraphics[width=\textwidth]{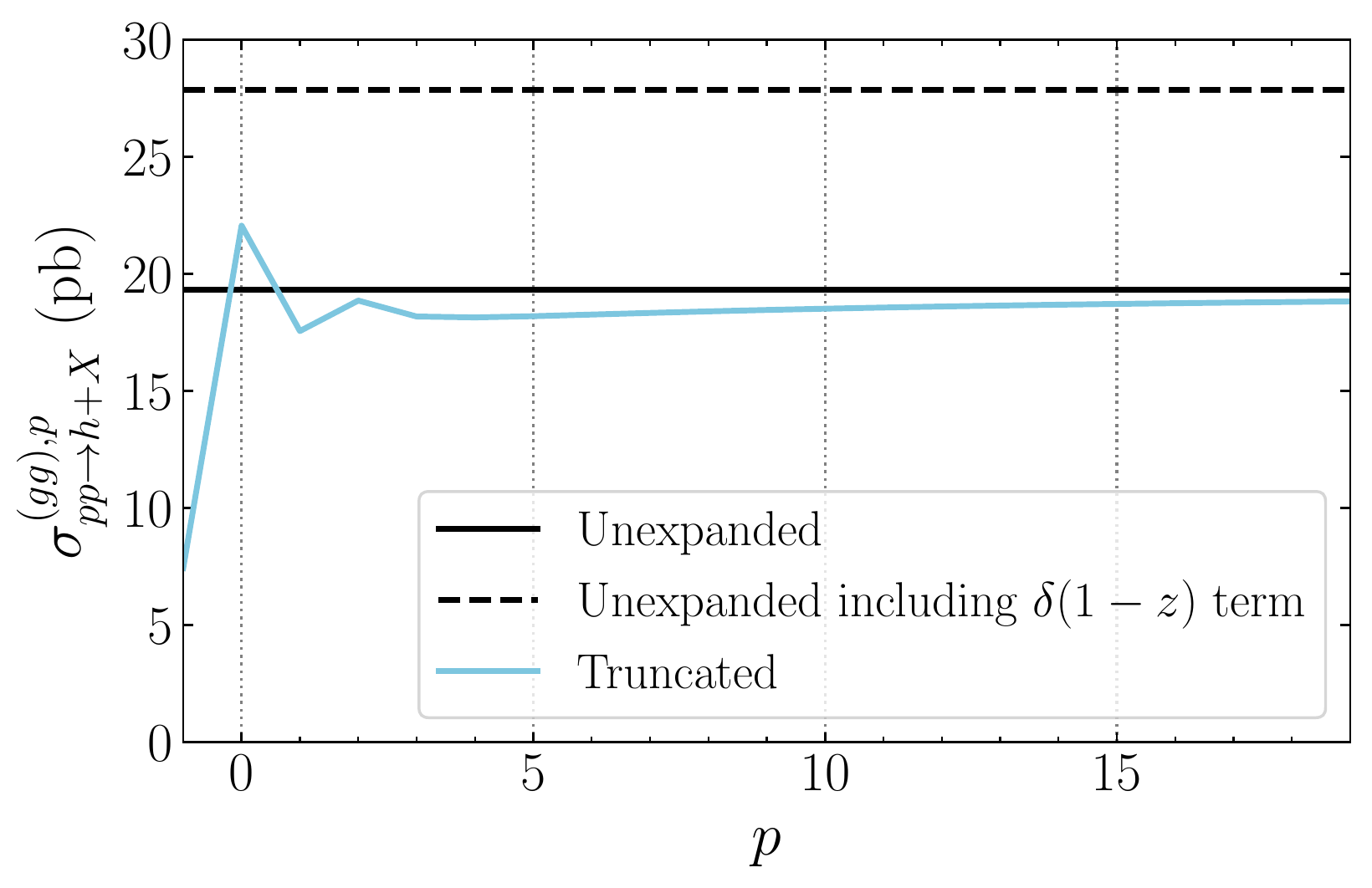}
    \end{subfigure}
    \hspace{0.1cm}
    \begin{subfigure}{.505\textwidth}
        \includegraphics[width=\textwidth]{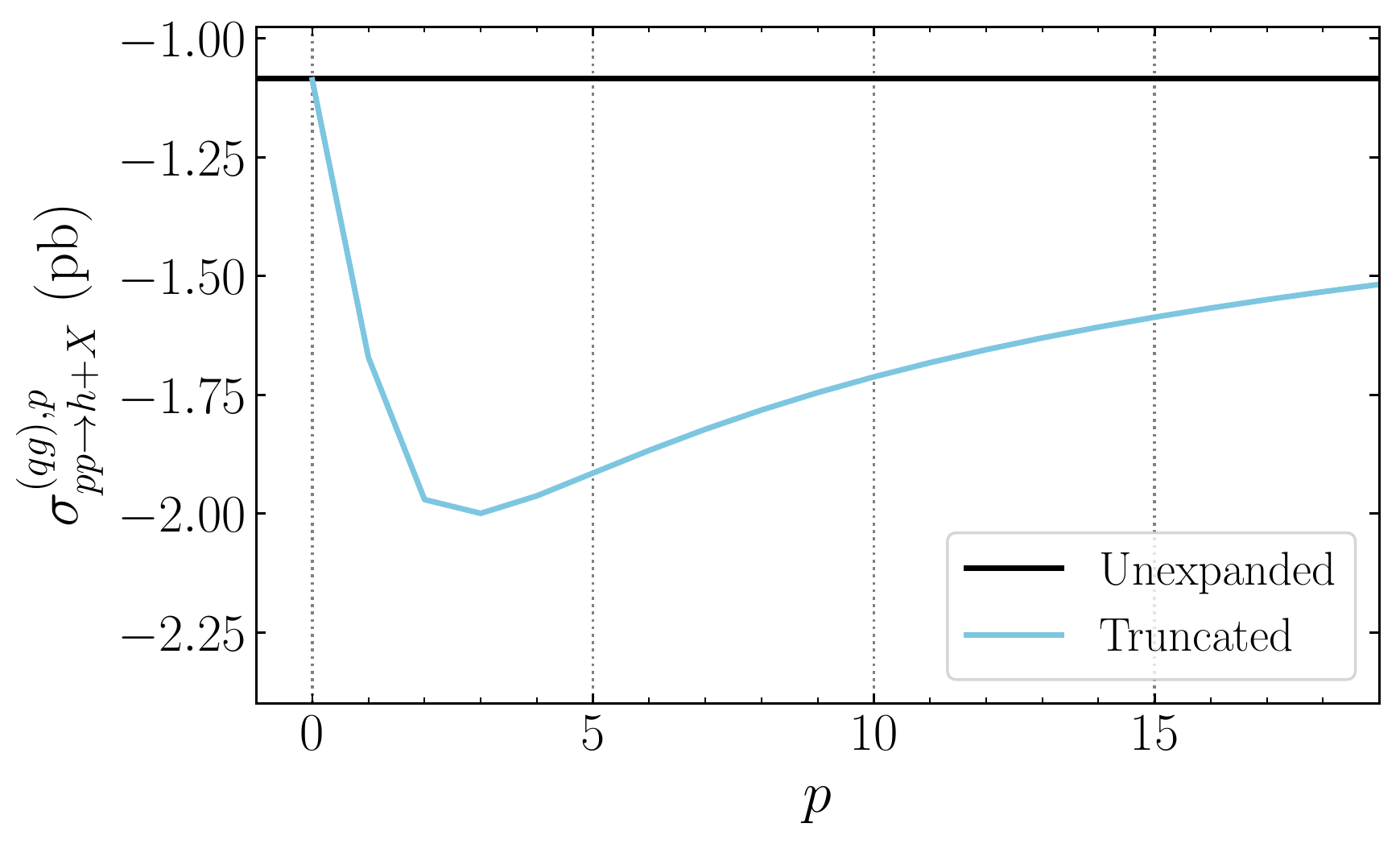}
    \end{subfigure}
    }
    \caption{Power expansions of the NLO+NNLO hadronic cross sections for single Higgs production for $Q = 125$~GeV. The results are shown in a cumulative way: in the expansion up to power $p$, each power $<p$ is included. The $gg$ ($qg$) contribution is shown on the left(right)-hand side. The $\delta(1-z)$ contribution in the $gg$-channels is shown separately, and the LO contribution is not included. }\label{fig:HiggsexpansionHAD}
\end{figure}

\subsection{Convergence of the threshold expansion in integrated hadronic cross sections} \label{int_had_exp}
We now turn our attention to the behaviour of the threshold expansions of the total hadronic cross section, starting with single Higgs production. We show the expansions up to N$^{p+1}$LP, which follows from
\begin{eqnarray}
\sigma^{(ij),\,p}_{pp\rightarrow h+X} = \sum_{n=1}^2\sigma_0^{\rm h}\left(\frac{\alpha_s}{\pi}\right)^n \,\int_{\tau}^1\frac{{\rm d}z}{z}\,\mathcal{L}_{ij}\,\left(\frac{\tau}{z}\right)~\left[\sum_{m=0}^{2n-1}\sum_{k=-1}^{p}(1-z)^k\,c_{nm}^{(ij),\,k}\,\ln^m (1-z)\right], 
\label{eq:expansionsigma}
\end{eqnarray}
with $c_{nm}^{(ij),\,-1}$ corresponding to $c_{nm}^{\rm LP}$ of eq.~\eqref{eq:defthreshold}, $c_{nm}^{(ij),\,0}$ to $c_{nm}^{\rm NLP}$, etc. As a possible $\delta$-contribution cannot be expanded around $z = 1$, we show this term separately if its coefficient is non-zero. As in the previous subsection, we add the NLO and NNLO contributions in eq.~\eqref{eq:expansionsigma}, and we restrict our discussion to those partonic channels that contribute at NLP.

The truncated cross sections can be seen in fig.~\ref{fig:HiggsexpansionHAD}, where the power expansions are shown up to N$^{20}$LP.  On the left-hand side of this figure, we show results for the $gg$-channel (up to NNLO) for $Q = 125$~GeV. The LP expansion severely underestimates the unexpanded part of the hadronic cross section (without the $\delta$-contribution), while the NLP expansion overestimates it by a much smaller amount. After including the NNLP term, the expansion stabilizes~\footnote{This is consistent with ref.~\cite{Herzog:2014wja}, eq.~(3) when we pick $g(z) =1$.}. We also observe a significant $\delta$-contribution, as is well known for this process.

For the $qg$-channel (right-hand side of fig.~\ref{fig:HiggsexpansionHAD}), we observe that the NLP truncation exactly produces the unexpanded cross section. This is however somewhat of a coincidence: the (N)NLO NLP truncation underestimates (overestimates) the magnitude of the negative unexpanded (N)NLO contribution by a similar amount. When added together, these under- and overestimates cancel. The NNLP truncation for the integrated cross section performs much worse, consistent with what we predicted based on \fig{z-weight_qg_exp125}. 
An overestimate of the negative contribution from this channel perseveres for higher-power truncations, and the expansion converges only very slowly to the full result. Again, this is due to the NLO and NNLO coefficients having a negative contribution at $z\rightarrow 1$, compensated by a large positive contribution for $z\rightarrow 0$, which is however only slowly reconstructed in a $1-z$ expansion.
\begin{figure}[t]
    \centering
    \mbox{\begin{subfigure}{.49\textwidth}
        \includegraphics[width=\textwidth]{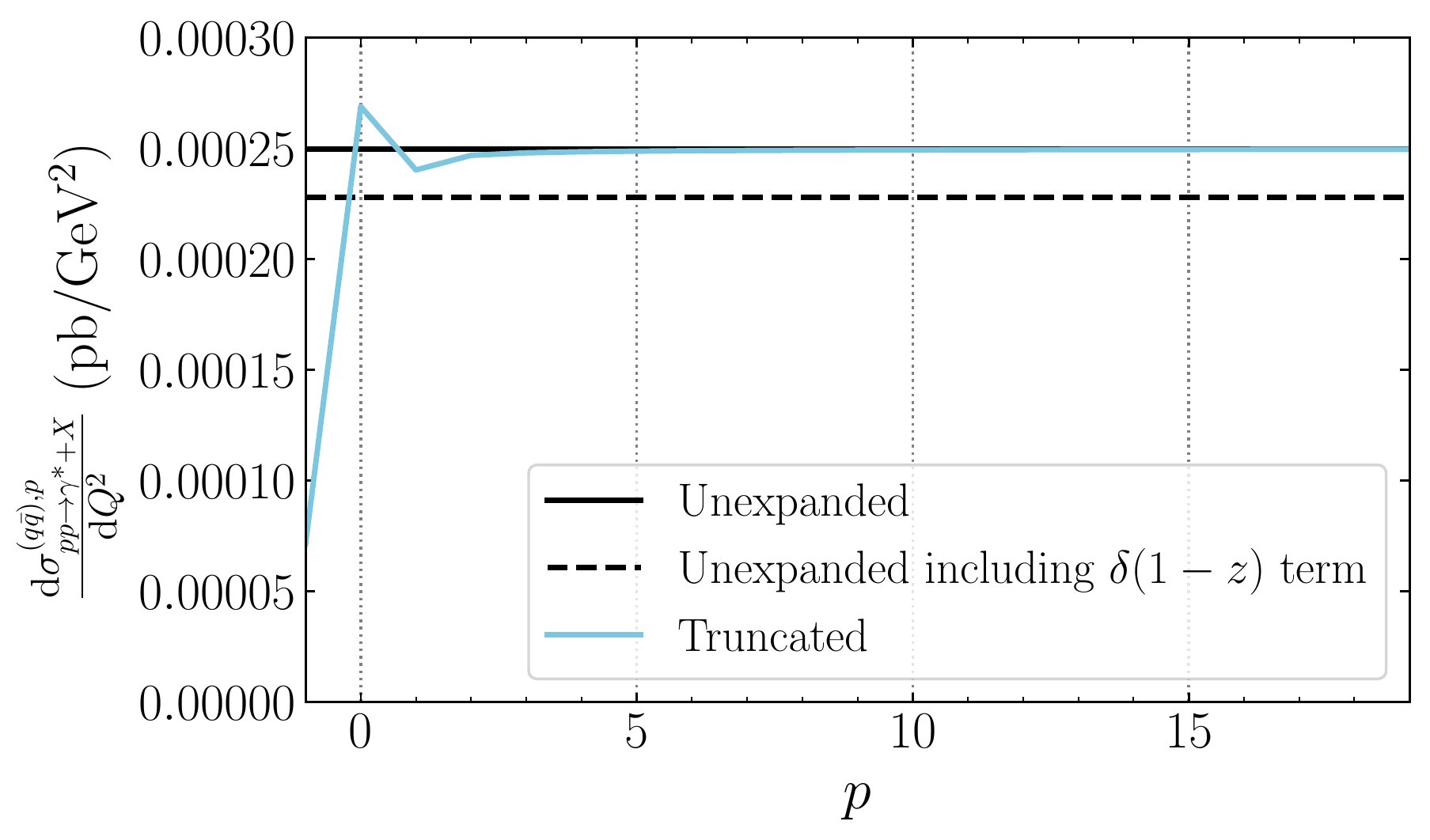}
    \end{subfigure}
    \hspace{0.1cm}
    \begin{subfigure}{.49\textwidth}
        \includegraphics[width=\textwidth]{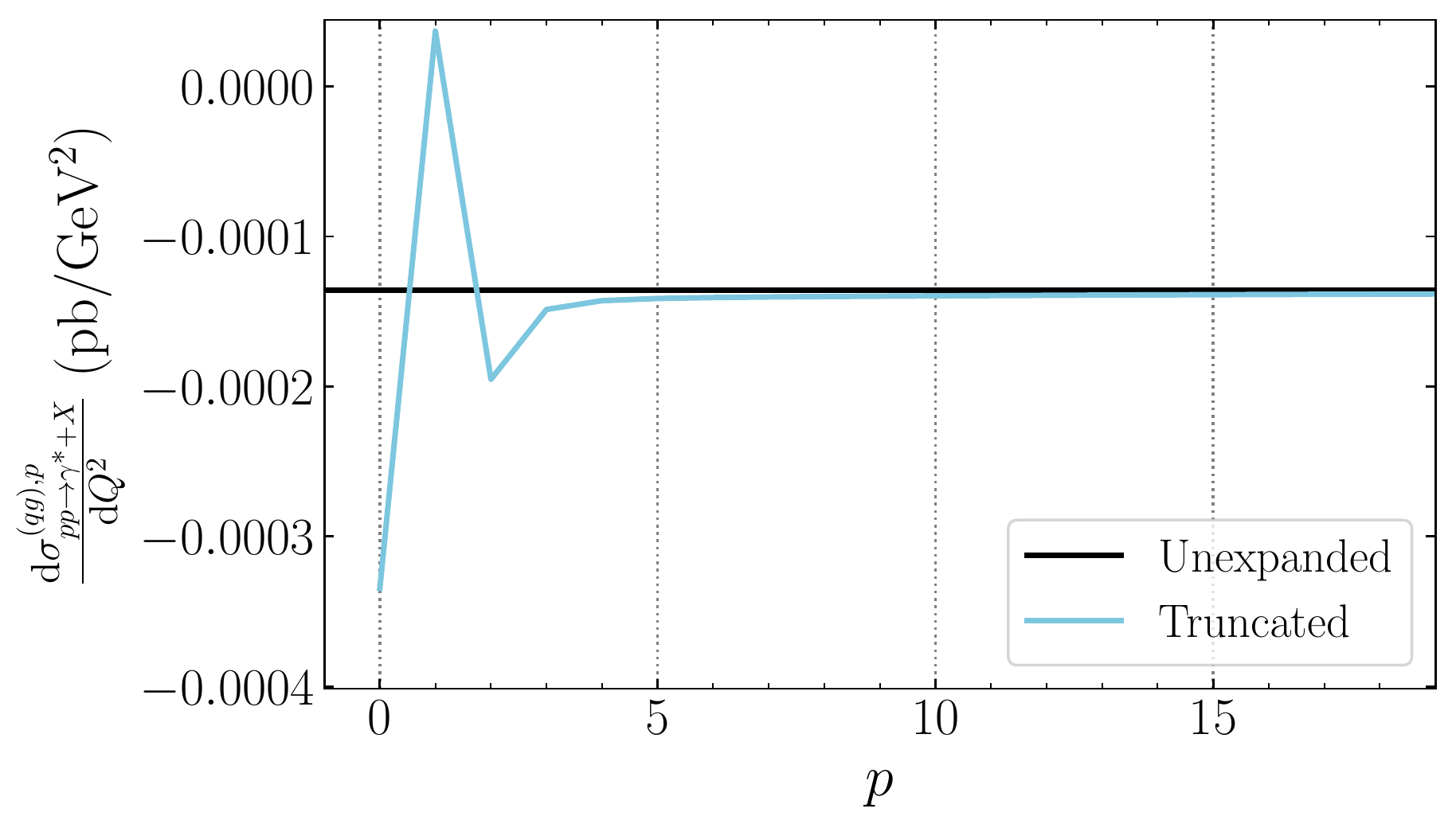}
    \end{subfigure}}
    \caption{Same as \fig{HiggsexpansionHAD}, but for the total cross section of the DY production of an off-shell photon with $Q=125$ GeV. The $q\bar{q}$ ($qg$) channel is shown on the left(right)-hand side.}\label{fig:DYHAD}
\end{figure}

 Also for DY production at $Q= 125~$GeV, the LP expansion of dominant channel is not a good approximation of the exact cross section, as seen in the left-hand side of fig.~\ref{fig:DYHAD}. The NLP expansion shows moderate overestimation, as seen in fig.~\ref{fig:z-weight_exp125} as well, but performs significantly better already. In general the power expansion converges quickly. For the $qg$-channel (right-hand side of fig.~\ref{fig:DYHAD}) we see that the NLP (NNLP) expansion overestimates (underestimates) the absolute value of the unexpanded NLO+NNLO coefficient, consistent with the deviations observed in fig.~\ref{fig:z-weight_qg_exp125}. Contrary to the Higgs $qg$-channel, the threshold expansion does converge after including the N$^4$LP contribution.

Before we conclude this section, we comment on the behaviour of the threshold expansions for other values of $Q$. For the $gg$-channel in single Higgs production, the convergence of the threshold expansion happens faster for higher values of $Q$, which is a direct consequence of the stronger $1/z$ suppression of the $gg$ luminosity function. Also for the Higgs $qg$-channel the threshold expansion converges more slowly for small $Q$ values than for larger ones, for the same reason.  On the other hand, the behaviour of the threshold expansion of the $q\bar{q}$ and $qg$ DY channels  marginally improves at higher $Q$ values. We conclude that the threshold expansion for NNLO cross sections is in general more reliable for DY than for single Higgs production for both the dominant ($q\bar{q}/gg$) and subleading $qg$ production channels.

%======================
%=======SECTION========
%======================
\section{NLP resummation in dQCD}
\label{sec:dQCDresum}
\noindent
Having exhibited the effect and quality of NLP approximations in fixed order cross sections, we now turn to consider NLP effects for resummed cross sections. 
This section discusses analytical aspects of NLP resummation in direct QCD (dQCD), whereas numerical results are shown in Section~\ref{sec:dQCD_results}. We concentrate on the dominant channels, and address LP and NLP resummation at the same time. 
\subsection{From LP to NLP resummation}
\label{dQCD_approach}
Resummation in dQCD is customarily performed in Mellin-moment ($N$) space, where the cross section is a product of $N$-space functions. Thus, for the gluon-fusion contribution to Higgs production in \eq{higgs_z} one obtains
\begin{equation} \label{higgs_gg_N}
\bm{\sigma}^{(gg)}_{pp\rightarrow h+X}(N) \equiv \int_0^1{\rm d}\tau \,\tau^{N-1}\sigma_{p p \rightarrow h + X}(\tau) = \sigma_0^{\rm h}\, \bm{f}_g(N,\mu)\bm{f}_g(N,\mu)\bm{\Delta}_{gg}(N,Q^2/\mu^2)\,,
\end{equation}
and similarly for the DY production (with $g \rightarrow q$ and $\sigma_0^{\rm h} \rightarrow \sigma_0^{\rm DY}$). We use a boldface notation for Mellin-transformed quantities. To perform the resummation in $N$-space, one uses the resummed perturbative coefficient $\bm{\Delta}_{aa}(N)$. The resummed hadronic cross section in momentum space is obtained after taking the inverse Mellin transform
\begin{equation}
\label{eq:inversemel}
\sigma^{(gg)}_{p p \rightarrow h + X}(\tau) = \frac{1}{2\pi i}\int_{c-i\infty}^{c+i\infty} {\rm d}N\, \tau^{-N} \bm{\sigma}^{(gg)}_{pp\rightarrow h+X}(N)\,,
\end{equation}
where we choose the Minimal Prescription~\cite{Catani:1996yz} for the $N$ integration contour. 
This corresponds to choosing $c=C_{\rm MP}$ to left of the large-$N$ branch-cut (starting for us at the branch-point $\bar{N}=\exp\left[1/(2\alpha_s b_0)\right]$) originating from the Landau pole, but to the right of all other singularities. We note that this approach works for both LP and NLP terms, as the branch cut is identical for both terms. To minimize numerical instabilities, the integration contour is usually bent towards the negative real axis. We include also $N$-space PDFs in our inverse Mellin transform, for which we use 
a fitted form of the PDFs (see the discussion around \eq{eq:fit}).

In ref.~\cite{Bahjat-Abbas:2019fqa} a resummed partonic coefficient for colour-singlet production processes was derived that also incorporates the LL NLP contributions for the dominant channel
\begin{align}
\bm{\Delta}_{aa}^{\rm dQCD,\,LP+NLP}(N,Q^2/\mu^2) &= g_{0}(\alpha_s)\,{\rm exp}\Bigg[\int_0^1{\rm d}z\,z^{N-1}\,\Bigg[ \frac{1}{1-z}D_{aa}\left(\alpha_s\left(\frac{(1-z)^2Q^2}{z}\right)\right) \nonumber \\
&\hspace{2.4cm} + 2 \int_{\mu^2}^{(1-z)^2Q^2/z}\frac{{\rm d}k_T^2}{k_T^2} P^{\rm LP+NLP}_{aa}\left(z,\alpha_s(k_T^2)\right)\Bigg]_+\,\Bigg] \nonumber \\
&\equiv g_{0}(\alpha_s)\,{\rm exp}\left[\bm{D}^{\rm LP + NLP}_{aa}+ 2 \bm{E}^{\rm LP + NLP}_a \right]\,.
\label{eq:resumgeneral}
\end{align}
The exponent $\bm{E}^{\rm LP + NLP}_a$ contains the diagonal DGLAP splitting function $P_{aa}$, expanded up to NLP in the threshold variable. The factor of $2$ multiplying this term reflects the initial state consisting of identical partons (either $gg$ or $q\bar{q}$). The soft wide-angle  contributions are collected in $D_{aa}(\alpha_s)$. Both terms are process-independent to the extent that they only depend on the colour structure of the underlying hard-scattering process. The overall $+$-subscript denotes that the plus-prescription needs to be applied to all LP contributions. After the Mellin transform has been carried out, the $\bm{D}^{\rm LP + NLP}_{aa}$ and $\bm{E}^{\rm LP + NLP}_a$ terms contain \emph{all} logarithmic $N$ dependence at LP and \emph{leading}-logarithmic $N$ dependence at NLP. It also contains some contributions that are beyond LL NLP (i.e.~the argument of $\alpha_s$ in the function $D_{aa}$ creates NLP contributions beyond LL, as we will see, and so does the $1/z$-dependence of the upper limit of the $k_T$ integral). The process-dependent function $g_0(\alpha_s)$ collects the $N$-independent contributions. For an N$^k$LL resummation, we need this function  only up to $\mathcal{O}(\alpha_s^{k-1})$, but if available we include the $\mathcal{O}(\alpha_s^k)$ terms as well, which upgrades N$^k$LL to N$^k$LL$^\prime$ resummation. The splitting function $P^{\rm LP+NLP}_{aa}$ can be written in the LP+NLP approximation as
\begin{align}
P^{\rm LP+NLP}_{aa}\left(z,\alpha_s(k_T^2)\right) &= \sum_{n=1}^{\infty}P^{(n)\rm LP+NLP}_{aa}(z)\left(\frac{\alpha_s(k_T^2)}{\pi}\right)^n \nonumber\\ &= A_a\left(\alpha_s(k_T^2)\right)\left[\left(\frac{1}{1-z}\right)_+-1\right],\label{eq:split}
\end{align}
where~\footnote{We include the usual factor of $2$ in eq.~\eqref{eq:resumgeneral}.}
\begin{equation}
A_a\left(\alpha_s(k_T^2)\right) \equiv \sum_{n=1}^{\infty} A_a^{(n)}\left(\frac{\alpha_s(k_T^2)}{\pi}\right)^n.
\label{eq:splitting}
\end{equation}
The coefficients $A^{(n)}_a$ are known up to fourth order \cite{Vogt:2018miu}, but for NNLL accurary we need them only up to  third order \cite{Moch:2004pa,Vogt:2004mw}. They are given explicitly in appendix~\ref{app:resummationcoeff}. The perturbative expansion for the function $D_{aa}(\alpha_s)$ reads
\begin{eqnarray}
D_{aa}(\alpha_s) = D_{aa}^{(1)}\frac{\alpha_s}{\pi} + D_{aa}^{(2)}\left(\frac{\alpha_s}{\pi}\right)^2 + \ldots\,,
\end{eqnarray}
which starts at the two-loop level since $D_{aa}^{(1)} = 0$. Only the coefficient $D_{aa}^{(2)}$  (see appendix~\ref{app:resummationcoeff}) is needed for a resummation at NNLL($^\prime$) accuracy. 

Upon substitution of \eq{eq:split} into eq.~\eqref{eq:resumgeneral}, we may compare it with its LP counterpart~\cite{Sterman:1987aj, Catani:1989ne}
\begin{align}
\label{eq:resumgeneralLP}
\bm{\Delta}_{aa}^{\rm dQCD,\,LP}(N,Q^2/\mu^2) = g_{0}(\alpha_s)\,{\rm exp}\Bigg[&\int_0^1{\rm d}z\, \frac{z^{N-1}-1}{1-z}D_{aa}\left(\alpha_s\left((1-z)^2Q^2\right)\right) \\
&+\, 2 \int_0^1{\rm d}z\,\frac{z^{N-1}-1}{1-z} \int_{\mu^2}^{(1-z)^2Q^2}\frac{{\rm d}k_T^2}{k_T^2} A_a\left(\alpha_s(k_T^2)\right)\Bigg]\,, \nonumber 
\end{align}
where the plus-prescription is already applied. We highlight two changes with respect to \eqref{eq:resumgeneral}. The first is that the splitting function is approximated to LP accuracy instead, removing the additional $-1$ term of eq.~\eqref{eq:split}. The second is that the upper limit of the $k_T$ integration on the second line, reflecting the exact phase-space constraint on the soft emission, has been replaced by its LP approximation. The same replacement has been made in the argument of $\alpha_s$ in $D_{aa}$ on the first line. How to calculate the integrals in eq.~\eqref{eq:resumgeneralLP} is outlined in ref.~\cite{Catani:1989ne,Forte:2002ni,Catani:2003zt}, and extended to accommodate NLP contributions to the splitting function in ref.~\cite{Laenen:2008ux}. However, the latter reference did not implement the exact phase-space constraint. If one does, a simpler formula can be derived that obtains the NLP contribution directly from the LP exponent using a derivative with respect to the Mellin moment $N$, which we show in the next subsection. 

\subsection{The NLP exponent through differentiation}
Let us calculate $\bm{E}^{\rm LP + NLP}_{a}$ at LL accuracy, since we are interested in the NLP LL contribution for which we only need $A_a^{(1)}$
\begin{equation}
    \bm{E}^{\rm LP + NLP,\, LL}_{a} =  \frac{A_a^{(1)}}{\pi} \int_0^1{\rm d}z\,z^{N-1}\Bigg[\left(\frac{1}{1-z}-1\right) \int_{\mu^2}^{(1-z)^2Q^2/z}\frac{{\rm d}k_T^2}{k_T^2}\alpha_s(k_T^2)\Bigg]_+. \label{ELL}
\end{equation}
The integral over $k_T$ may be evaluated using the QCD $\beta$-function, for which we only need the one-loop coefficient $b_0$ (see appendix~\ref{app:resummationcoeff}),  and write
\begin{align}
    \int_{\mu^2}^{(1-z)^2 Q^2/z}\frac{{\rm d}k^2_T}{k_T^2} &= -\frac{1}{b_0}\int_{\alpha_s}^{\alpha_s((1-z)^2Q^2/z)}\frac{{\rm d}\alpha_s}{\alpha_s} \nonumber \\ &= \frac{1}{b_0}\ln\left(\frac{\alpha_s}{\alpha_s((1-z)^2 Q^2/z)}\right),
\end{align}
where we use $\alpha_s \equiv \alpha_s(\mu^2)$. This may be written as
\begin{align}
    \int_{\mu^2}^{(1-z)^2 Q^2/z}\frac{{\rm d}k^2_T}{k_T^2} &= \frac{1}{b_0}\ln\left(1+\alpha_s\,b_0\,\ln\left(\frac{(1-z)^2 Q^2/z}{\mu^2}\right)\right) \nonumber \\ &= -\frac{1}{b_0}\sum_{k=1}^\infty \frac{(-\alpha_s\,b_0)^k}{k}\left[\ln\left(\frac{(1-z)^2}{z}\right)+\ln{\left(\frac{Q^2}{\mu^2}\right)}\right]^k,
\end{align}
where we have separated the logarithms with scale-dependence from those with $z$-dependence. 
Since we wish to resum the highest power of the threshold logarithms at each order in $\alpha_s$, we select the pure $\ln^k((1-z)^2/z)$ term from the above expression, discarding the explicit scale logarithms that only contribute at NLL accuracy and beyond (we would have obtained the same result by making the replacement $\mu^2\rightarrow Q^2$ in the lower integration boundary of \eq{ELL}). Expanding
\begin{equation}
    \ln^k\left(\frac{(1-z)^2}{z}\right) = 2^k\ln^k(1-z) + k\,(1-z)\, 2^{k-1}\ln^{k-1}(1-z) +\mathcal{O}\left((1-z)^2\right), \label{eq:expzdep}
\end{equation}
and retaining the first two terms, Eq.~\eqref{ELL} becomes
\begin{align}
    \bm{E}^{\rm LP + NLP,\, LL}_{a} &=  -\frac{A_a^{(1)}}{\pi\,b_0} \int_0^1{\rm d}z\,\sum_{k=1}^\infty \frac{(-2\,\alpha_s\,b_0)^k}{k}\Bigg( \frac{z^{N-1}-1}{1-z}\,\ln^k(1-z)-z^{N-1}\,\ln^k(1-z) \nonumber \\ &\hspace{180pt}+ \frac{k}{2}\, z^{N-1}\,\ln^{k-1}(1-z)\Bigg) \nonumber \\ &\equiv  -\frac{A_a^{(1)}}{\pi\,b_0} \,\sum_{k=1}^\infty \frac{(-2\,\alpha_s\,b_0)^k}{k}\Bigg( \mathcal{D}_k-\mathcal{J}_k + \frac{1}{2}\mathcal{J^\prime}_k\Bigg).\label{ELL2}
\end{align} In the second line we introduced shorthand notation for the Mellin integrals, which are evaluated through their generating functions ${\cal G_F}$. We define 
\begin{equation}
    \mathcal{F}_k (N) = \left.\frac{{\rm d}^k}{{\rm d}\eta^k}G_\mathcal{F} (N,\eta)\right|_{\eta=0}, \label{Gdef}
\end{equation}
for ${\cal F = \{D,J,J^\prime\}}$ and 
\begin{subequations}
\label{genfunc}
\begin{align}
    G_{\cal D}(N,\eta) &= \int_0^1\! {\rm d}z\, \left(z^{N-1}-1\right)(1-z)^{\eta-1}  = \frac{\Gamma(N)\Gamma(\eta)}{\Gamma(N+\eta)} - \frac{1}{\eta}\,, \\
    G_{\cal J}(N,\eta) &= \int_0^1\! {\rm d}z\, z^{N-1}(1-z)^{\eta} = \frac{\Gamma(N)\Gamma(1+\eta)}{\Gamma(1+N+\eta)}\,, \\
    G_{\cal J^\prime}(N,\eta) &= \eta\, G_{\cal J}(N,\eta)\,. \label{eq:Jprime}
\end{align}
\end{subequations}
This method was proposed in~\cite{Forte:2002ni} for the LP contributions $\mathcal{D}_k$ and generalized to the NLP integrals $\mathcal{J}_k$ in~\cite{Laenen:2008ux}. The terms labeled by $\mathcal{J^\prime}_k$ were not included before. Although of sub-leading logarithmic accuracy, they are important for our final result. Expanding \eq{genfunc} around the limit $N\rightarrow \infty$ yields, up to $\mathcal{O}(1/N^2)$ corrections
\begin{subequations}
\label{genfuncexp}
\begin{align}
     \label{GD} G_{\cal D}(N,\eta) &= \frac{1}{\eta}\left[\frac{\Gamma(1+\eta)}{N^\eta}\left(1+\frac{\eta\,(1-\eta)}{2N}\right)-1\right], \\ \label{GJ} 
    G_{\cal J}(N,\eta) &= \frac{\Gamma(1+\eta)}{N^{1+\eta}}\, ,\\
    G_{\cal J^\prime}(N,\eta) &= \frac{\eta\, \Gamma(1+\eta)}{N^{1+\eta}}\,.
\end{align}
\end{subequations}
We see that pure LP contributions in $z$-space, as contained in $\mathcal{D}_k$, give NLP contributions in Mellin space (i.e.~terms proportional to $1/N$). In dQCD, the resummation is done in Mellin space, and in what follows, when we refer to `NLP' in dQCD, we mean terms that are proportional to $1/N$. We now observe that all $\mathcal{O}(1/N)$ terms can be generated from the LP contributions in $N$-space by means of a derivative with respect to $N$
\begin{align}
    G_{\cal D}(N,\eta) - G_{\cal J}(N,\eta) + \frac{1}{2}G_{{\cal J}^\prime}(N,\eta) &=  \frac{1}{\eta}\left[\Gamma(1+\eta)\left(1-\frac{\eta}{2N}\right)N^{-\eta}-1\right] \nonumber \\ &=\left(1+\frac{1}{2}\frac{\partial}{\partial N}\right)\frac{1}{\eta}\left[\Gamma(1+\eta)\exp\left[-\eta\,\ln N\right]-1\right]. \label{derID}
\end{align}
Note that this only holds when including ${\cal J^\prime}$, as it cancels the $\eta^2$ term in the square brackets of \eq{GD}. Taylor expansion of the LP contribution (i.e.~the term in eq.~\eqref{derID} without the derivative) around $\eta=0$ yields
\begin{align}
    \frac{1}{\eta}\left[\Gamma(1+\eta)N^{-\eta}-1\right] &= \frac{1}{\eta}\left(\sum_{m=0}^\infty\sum_{n=0}^m\frac{1}{n!\,(m-n)!}\left[\frac{\partial^n}{\partial\eta^n}\Gamma(1+\eta)\,\frac{\partial^{m-n}}{\partial\eta^{m-n}}\exp\left[-\eta\, \ln N\right]\right]_{\eta=0}\eta^m-1\right) \nonumber \\ 
    &= \sum_{m=1}^\infty\sum_{n=0}^m\frac{(-1)^{m-n}}{n!\,(m-n)!}\Gamma^{(n)}(1)\ln^{m-n}N\, \eta^{m-1}\,,
    \label{eq:idid}
\end{align} 
with $\Gamma^{(n)}$ denoting the $n$-th derivative of the gamma function, i.e.
\begin{eqnarray}
\Gamma^{(n)}(1) = \frac{{\rm d}^n}{{\rm d} \eta^n}\Gamma(1+\eta)\big|_{\eta\rightarrow 0}\,.
\end{eqnarray}
In this way, we have isolated the $\eta$ behaviour in one simple factor, such that the derivative of \eq{Gdef} amounts to the relation
\begin{eqnarray}
\frac{{\rm d}^k\eta^{m-1}}{{\rm d}\eta^k}\Big|_{\eta = 0} = \delta_{k,m-1}\,.
\end{eqnarray}
Rewriting \eq{derID} using \eq{eq:idid} we obtain, via \eq{Gdef}, for \eq{ELL2}
\begin{align}
    \bm{E}^{\rm LP + NLP,\, LL}_{a} &= \frac{A_a^{(1)}}{\pi\,b_0}\left(1+\frac{1}{2}\frac{\partial}{\partial N}\right)\sum_{k=1}^\infty \frac{(2\,\alpha_s\,b_0)^k}{k\,(k+1)}\sum_{n=0}^{k+1}\binom{k+1}{n}(-1)^{n}\Gamma^{(n)}(1)\ln^{k+1-n}N\,, \label{ELL3}
\end{align}
Terms in the sum with $n>0$ are sub-leading logarithmic terms in $N$, but some of those are easily included by redefining $N$ to $\bar{N}=\exp\left[\gamma_E\right]N$. To illustrate this, we note that
\begin{equation}
    (-1)^n\,\Gamma^{(n)}(1) = \gamma_E^n + \frac{\zeta(2)}{2} n\,(n-1)\, \gamma_E^{n-2} + \mathcal{O}\left(\gamma_E^{n-3}\right)\,, \label{approxGammaID}
\end{equation}
where all corrections contain constants of higher transcendental weight ($\zeta(n)$ for $n\ge 3$). Using \eq{approxGammaID} and shifting the summation index $n\rightarrow n^\prime+2$ for the second term in \eq{approxGammaID} (where $n=0$ and $n=1$ do not contribute) yields
\begin{align}
     \bm{E}^{\rm LP + NLP,\, LL}_{a} =&\, \frac{A_a^{(1)}}{\pi\,b_0}\left(1+\frac{1}{2}\frac{\partial}{\partial N}\right)\sum_{k=1}^\infty (2\,\alpha_s\,b_0)^k \nonumber \\&\times \left[\frac{1}{k\,(k+1)}\sum_{n=0}^{k+1}\binom{k+1}{n}\gamma_E^n\,\ln^{k+1-n}N+\frac{\zeta(2)}{2}\sum_{n^\prime=0}^{k-1}\binom{k-1}{n^\prime}\gamma_E^{n^\prime}\,\ln^{k-1-n^\prime} N \right]\nonumber \\ =&\, \frac{A_a^{(1)}}{\pi\,b_0}\left(1+\frac{1}{2}\frac{\partial}{\partial N}\right)\sum_{k=1}^\infty (2\,\alpha_s\,b_0)^k\left[\frac{\ln^{k+1}(\bar{N})}{k\,(k+1)}+\frac{\zeta(2)}{2}\ln^{k-1}(\bar{N})\right],
\end{align}
having recognized the binomial series for $(\ln N + \gamma_E)^{k\pm1} \equiv \ln^{k\pm1}\bar{N}$ in the last line. We see that the $\zeta(2)$ term in \eq{approxGammaID} contributes only at NNLL accuracy. Defining $\lambda = b_0\,\alpha_s\ln \bar{N}$ and performing the summation over $k$, this factor results in 
\begin{align}
     \bm{E}^{\rm LP + NLP,\, NNLL}_{a} =&\, \alpha_s \frac{A_a^{(1)}}{\pi}\left(1+\frac{1}{2}\frac{\partial}{\partial N}\right)\frac{\zeta(2)}{1-2\lambda}\, \nonumber \\
     =&\, \alpha_s \zeta(2) \frac{A_a^{(1)}}{\pi}\left(1+\frac{1}{2}\frac{\partial}{\partial N}\right)\left[\frac{2\lambda}{1-2\lambda}+1\right]. 
\label{eq:NNLLcontribution}
\end{align}
The $+1$ contribution in the last line is included in the $\mathcal{O}(\alpha_s)$ contribution of $g_0$. The first term in the square brackets is instead included in the NNLL contribution to the resummed exponent (see eq.~\eqref{eq:g3} of appendix~\ref{app:resummationcoeff}). At LL accuracy we have
\begin{align}
     \bm{E}^{\rm LP + NLP,\, LL}_{a} &= \frac{A_a^{(1)}}{2\pi\,b_0^2\,\alpha_s}\left(1+\frac{1}{2}\frac{\partial}{\partial N}\right)\left[2\lambda+(1-2\lambda)\ln(1-2\lambda)\right]\nonumber \\ &=\frac{A_a^{(1)}}{2\pi\,b_0^2\,\alpha_s}\left[2\lambda+(1-2\lambda)\ln(1-2\lambda)\right]-\frac{A_a^{(1)}}{2\pi\,b_0}\frac{\ln(1-2\lambda)}{N} \nonumber \\ &\equiv \frac{1}{\alpha_s}g_a^{(1)}(\lambda)+h_a^{(1)}(\lambda,N) = \frac{1}{\alpha_s}\left(1+\frac{1}{2}\frac{\partial}{\partial N}\right)g_a^{(1)}(\lambda) \,,
     \label{eq:derivativetrick}
\end{align} where the LL NLP resummation function $h_a^{(1)}$ is obtained from the LL LP function $g_a^{(1)}$ by taking the derivative towards $N$. Let us stress that the logarithmic accuracy of the generated NLP terms is limited by that of the LP function on which it acts. Therefore, this result is strictly of LL accuracy at both LP and NLP. We will use the resulting NLP function for the numerical studies in section~\ref{sec:dQCD_results}. 

By using the full $A(\alpha_s)$ function for $\bm{E}^{\rm LP+NLP}_a$ the result is straightforwardly extended to higher logarithmic accuracy at LP, and including \emph{partial} N$^{i}$LL NLP terms through the derivative in $N$ of the relevant LP resummation functions
\begin{eqnarray}
\label{eq:Elpnlpll6}
\bm{E}^{\rm LP+NLP}_a = \left(1+\frac{1}{2}\frac{\partial}{\partial N}\right)\int_0^1{\rm d}z\,\frac{z^{N-1}-1}{1-z}\int_{\alpha_s}^{\alpha_s((1-z)^2Q^2)}\frac{{\rm d}\alpha_s}{\beta(\alpha_s)}A_a(\alpha_s)\,.
\end{eqnarray}
A somewhat different form of this LP + NLP result proves to be useful in section~\ref{sec:comparison}, which is obtained by acting with the derivative on the result of ref.~\cite{Catani:2003zt} where the Mellin transform
has been expressed differently
\begin{eqnarray}
\label{eq:Elpnlpll7}
\bm{E}^{\rm LP+NLP}_a = \left(1+\frac{1}{2}\frac{\partial}{\partial N}\right){\rm e}^{\gamma_E \frac{\partial}{\partial\ln N}}\Gamma\left(1-\frac{\partial}{\partial \ln N}\right)\int_{1/\bar{N}}^{1} \frac{{\rm d}y}{y} \int_{\alpha_s}^{\alpha_s(y^2Q^2)}\frac{{\rm d}\alpha_s}{\beta(\alpha_s)}A_a(\alpha_s)\,.
\end{eqnarray}
This is valid to arbitrary LP logarithmic accuracy, and to NLP LL accuracy, as we do not have control over subleading logarithmic NLP contributions that originate from wide-angle soft gluon emissions. 

For the wide-angle contribution in eq.~\eqref{eq:resumgeneral}, we can use the same method to arrive at a derivative formula, but where the derivative term changes sign
\begin{eqnarray}
\label{eq:wideref}
\bm{D}^{\rm LP + NLP}_{aa} &=& \int_0^1{\rm d}z\, \frac{z^{N-1}-1}{1-z}D_{aa}\left(\alpha_s\left(\frac{(1-z)^2Q^2}{z}\right)\right) \\
&=& \frac{1}{2}\left(1-\frac{1}{2}\frac{\partial}{\partial N}\right){\rm e}^{\gamma_E \frac{\partial}{\partial\ln N}}\Gamma\left(1-\frac{\partial}{\partial \ln N}\right)\int_{\alpha_s(Q^2/\bar{N}^2)}^{\alpha_s(Q^2)} \frac{{\rm d}\alpha_s}{\beta(\alpha_s)}D_{aa}\left(\alpha_s\right)\,. \nonumber
\end{eqnarray}
For the first non-zero term in the perturbative expansion of $D_{aa}$ the result is
\begin{eqnarray}
\bm{D}^{\rm LP + NLP}_{aa} &=& -\alpha_s\frac{D_{aa}^{(2)}}{b_0\pi^2} \left(1-\frac{1}{2}\frac{\partial}{\partial N}\right)\frac{\lambda}{1-2\lambda}\,. \label{eq:wideangleD}
\end{eqnarray}
This is an LP NNLL contribution that does not give rise to LL contributions at NLP, hence we may drop the derivative term.

Our final form for the dQCD resummation exponent accurate to LP NNLL$^\prime$ and NLP LL order reads
\begin{eqnarray}
\label{eq:resumfunctions}
\bm{\Delta}^{\rm dQCD}_{aa}(N,Q^2/\mu^2)& =& g_0(\alpha_s)\,{\rm exp}\Big[\frac{2}{\alpha_s}g_a^{(1)}(\lambda) + 2g_a^{(2)}\left(\lambda, Q^2/\mu^2\right) \\
&& \hspace{5cm} + 2\alpha_s g_a^{(3)}\left(\lambda, Q^2/\mu^2\right) + 2h_a^{(1)}(\lambda, N)\Big]. \nonumber 
\end{eqnarray}
The wide-angle contribution of eq.~\eqref{eq:wideangleD} is contained in $g_a^{(3)}$. Explicit expressions for these resummation exponents are collected in Appendix~\ref{app:resummationcoeff}. We will use eq.~\eqref{eq:resumfunctions} for our numerical studies in section~\ref{sec:dQCD_results}. 

%======================
%=======SECTION========
%======================
\section{Numerical studies of LP and NLP resummation cross sections in dQCD}
\label{sec:dQCD_results}
Having set-up the formalism for dQCD resummation at NNLL$^\prime$ at LP and NLP LL, we turn to the numerical study of this formalism for various processes, in the context
of LHC collisions at $\sqrt{S} = 13$~TeV. We use fitted PDFs that allow for an analytical evaluation of the Mellin transform, as explained in appendix \ref{app:PDFs}. For DY and Higgs production we will show resummed observables that are matched to the fixed-order result at NNLO. The matching is defined by
\begin{eqnarray}
\sigma^{(\text{matched})} = \sigma^{{\rm dQCD,LP+NLP}} - \sigma^{{\rm dQCD,LP+NLP}}|_{\text{(fixed order)}} + \sigma^{\text{(fixed order)}}.
\end{eqnarray}
Note that we include the full fixed-order result in $\sigma^{\text{(fixed order)}}$, i.e.~also the sub-dominant production channels.
The second term on the right is the expanded resummed observable.
We calculate this term by Taylor-expanding the resummed coefficient function in $N$-space 
\begin{eqnarray}
\label{eq:expansionDeltaaa}
\mathbf{\Delta}_{aa}^{\rm dQCD,LP+NLP}(N,Q^2/\mu^2)\Big|_{\text{(fixed order)}} = \sum_{j = 0}^{n}\frac{\alpha_s^j}{j!}\left[\frac{\partial^j}{\partial \alpha_s^j}\mathbf{\Delta}_{aa}^{\rm dQCD}(N,Q^2/\mu^2)\right]_{\alpha_s = 0},
\end{eqnarray}
where we only need the terms up to $\mathcal{O}(\alpha_s^n)$ for matching with an N$^n$LO fixed-order calculation. The result is substituted into eq.~\eqref{higgs_gg_N}, after which the Mellin-space inversion is handled via eq.~\eqref{eq:inversemel}. Terms that are of higher power than $\mathcal{O}(1/N)$ are created in the expansion \eqref{eq:expansionDeltaaa}. They are kept in the matching and thus subtracted from the resummed result as they are also contained in the complete fixed-order expression. We perform the LP resummation at NNLL$^{\prime}$ accuracy, where NNLL$^{\prime}$ resummation offers an improvement over NNLL by the inclusion of the exact $N$-independent terms at NNLO (i.e.~we include $g_0$ up to $\mathcal{O}(\alpha_s^2)$). This is not strictly necessary for NNLL resummation, but it is relevant in case of large virtual corrections at the two-loop level, such as  for single Higgs production. We stress again that we only resum the channels that contribute at LP. To examine the impact of NLP resummation on colour-singlet processes other than Drell-Yan or single Higgs production, we also show at the end of this section results for di-boson and di-Higgs production, which we include to LP NLL accuracy and do not match to fixed higher-order results.   

\subsection{Single Higgs production}
To obtain the total resummed Higgs cross section in $N$-space, we start from eq.~\eqref{higgs_gg_N}
\begin{eqnarray}
\bm{\sigma}_{pp\rightarrow h+X}(N) = \sigma_0^{\rm h}\bm{f}_g(N,\mu)\bm{f}_{g}(N,\mu)\bm{\Delta}_{gg}^{\rm dQCD}(N,Q^2/\mu^2)\,,
\end{eqnarray} 
with $\sigma_0^{\rm h}$ given in eq.~\eqref{eq:higgsLO} and the resummed contribution $\bm{\Delta}_{gg}^{\rm dQCD}(N,Q^2/\mu^2)$ of eq.~\eqref{eq:resumfunctions} with $a = g$. As mentioned above we include $g_0$ up to $\mathcal{O}(\alpha_s^2)$. We vary $Q$($=m_h$) with the aim of exploring the resummation effects more widely than only for the physical scale $Q = 125$~GeV. 

\begin{figure}[t]
\centering
\mbox{\begin{subfigure}[h]{0.485\textwidth}
         \centering
         \includegraphics[width=\textwidth]{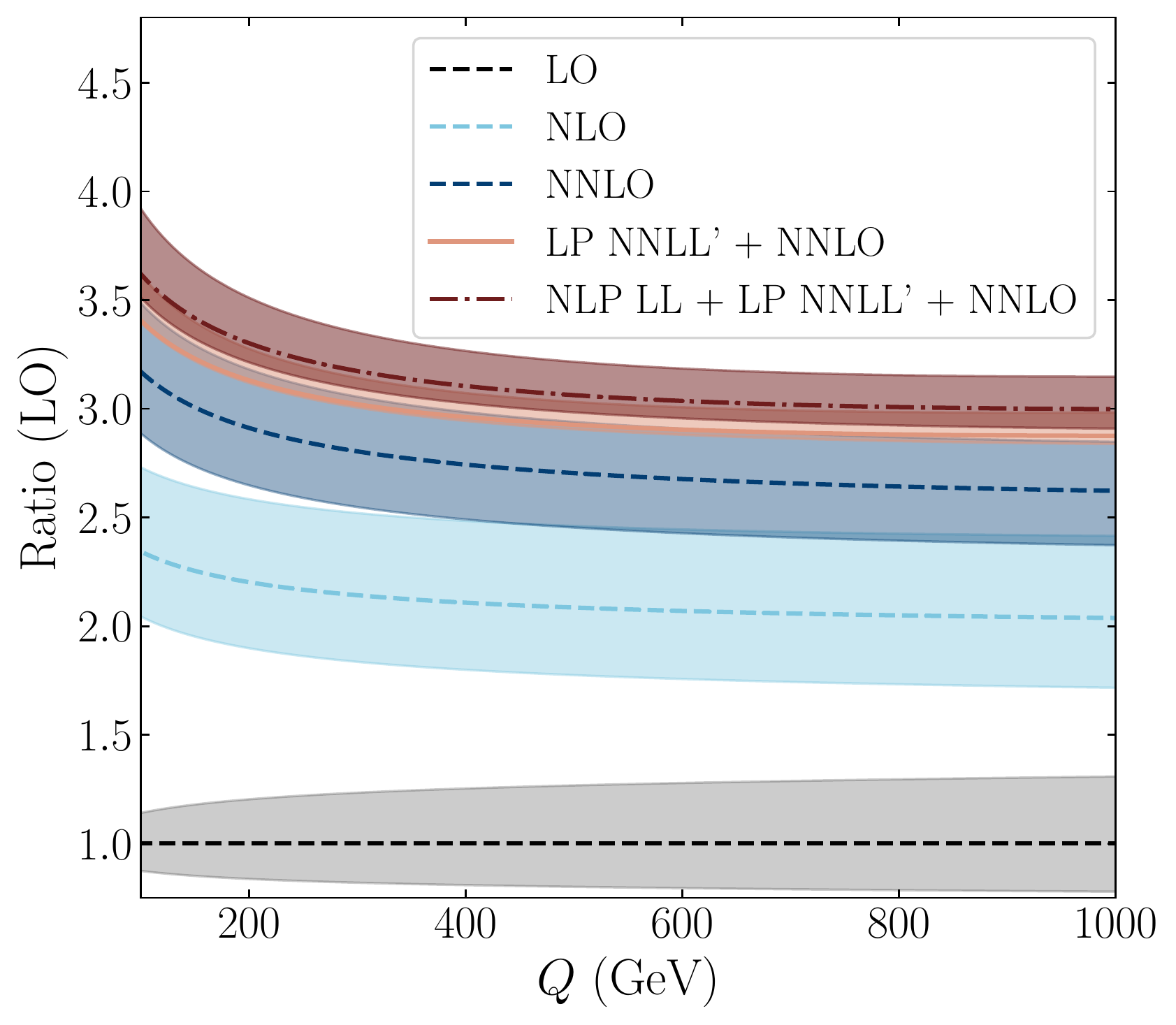}
         \caption{Ratio with respect to LO}
         \label{fig:higgswrtLO}
     \end{subfigure}
     \hfill
     \begin{subfigure}[h]{0.505\textwidth}
         \centering
         \includegraphics[width=\textwidth]{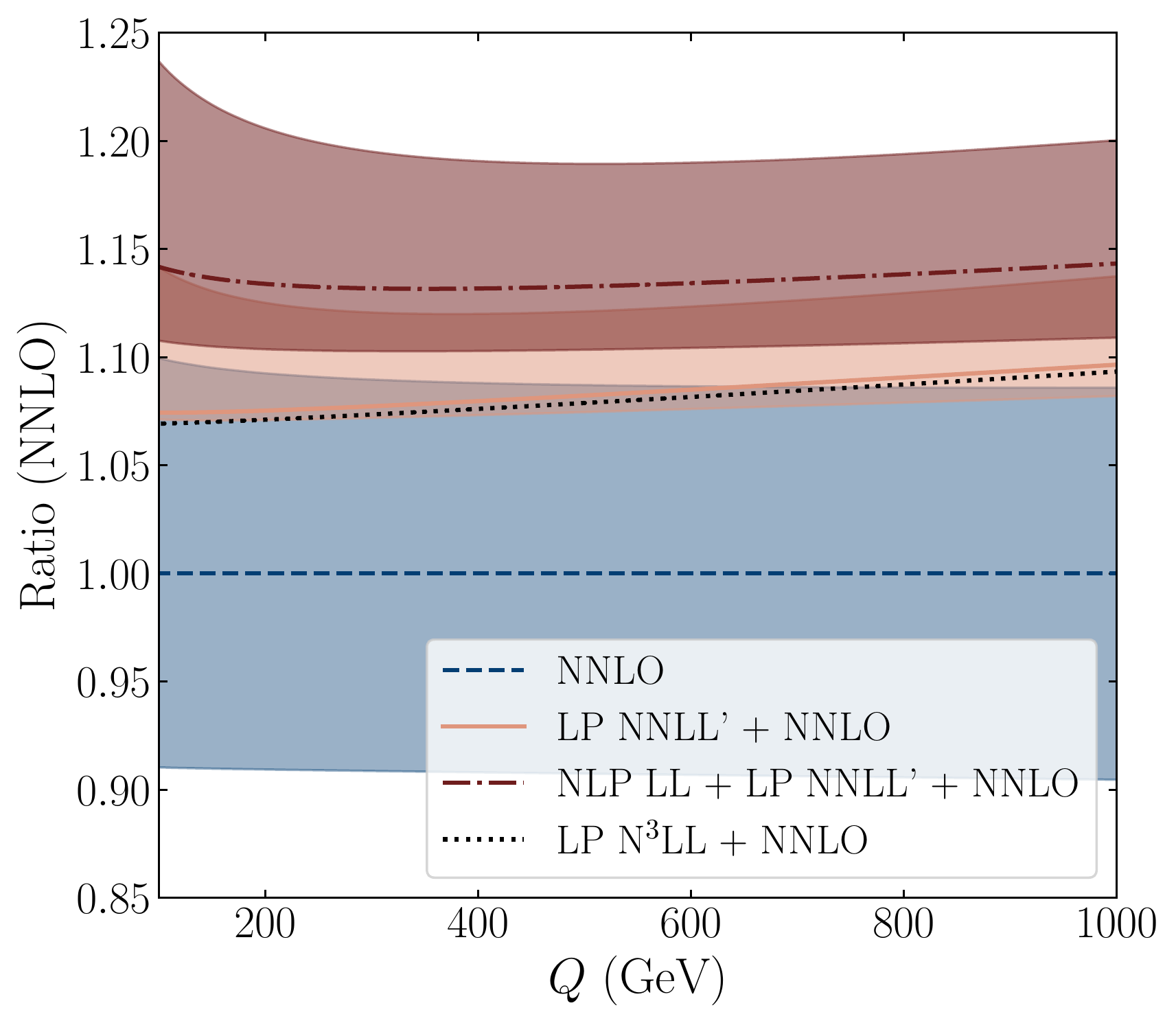}
         \caption{Ratio with respect to NNLO}
         \label{fig:higgswrtNNLO}
     \end{subfigure}}
    \caption{Left: ratio plot for the Higgs total cross section, normalized to the LO result. The fixed-order results at LO, NLO and NNLO are shown in black, light-blue and dark-blue dashed lines respectively. The LP NNLL$^{\prime}$ (+NLP LL) resummed result matched to NNLO is indicated by the orange solid (red dash-dotted) lines. The scale variation, obtained by varying $\mu$ between $\mu = Q/2$ and $\mu = 2Q$, is indicated by the coloured bands. Right: ratio for the same distribution normalized to the NNLO distribution. We also show the N$^3$LL resummed result, indicated by the black dotted line. The scale-uncertainty band for the N$^3$LL distribution is not shown, as it overlaps with the one obtained for the NNLL$^{\prime}$ distribution.}\label{fig:higgsdQCD}
\end{figure}

\begin{figure}[t]
\centering
\mbox{\begin{subfigure}[h]{0.47\textwidth}
         \centering
         \includegraphics[width=\textwidth]{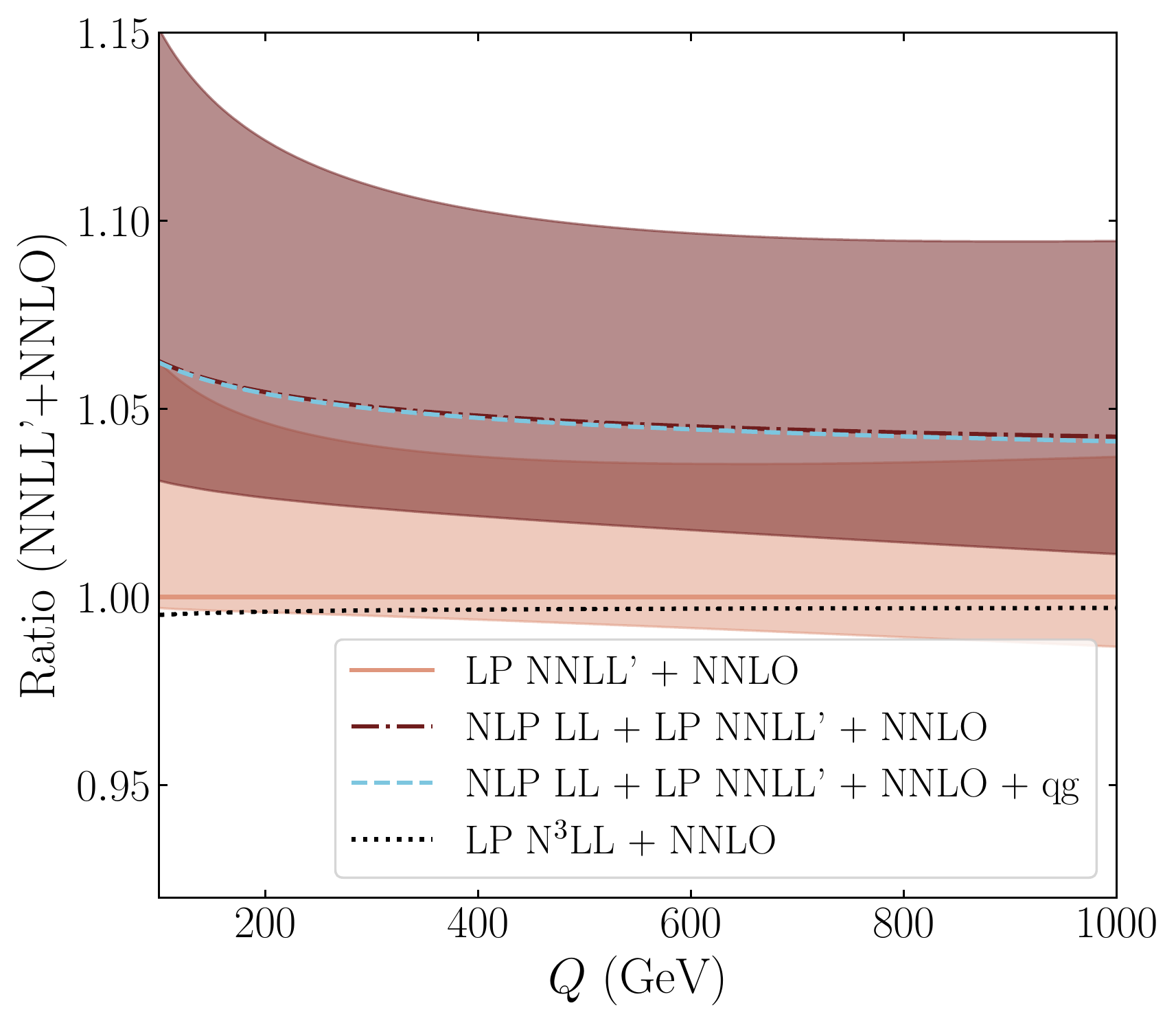}
         \caption{Ratio of the Higgs total cross section with respect to NNLL$^{\prime}$ + NNLO.}
         \label{fig:higgswrtNNLL}
     \end{subfigure}
     \hspace{0.5cm}
     \begin{subfigure}[h]{0.47\textwidth}
         \centering
         \includegraphics[width=\textwidth]{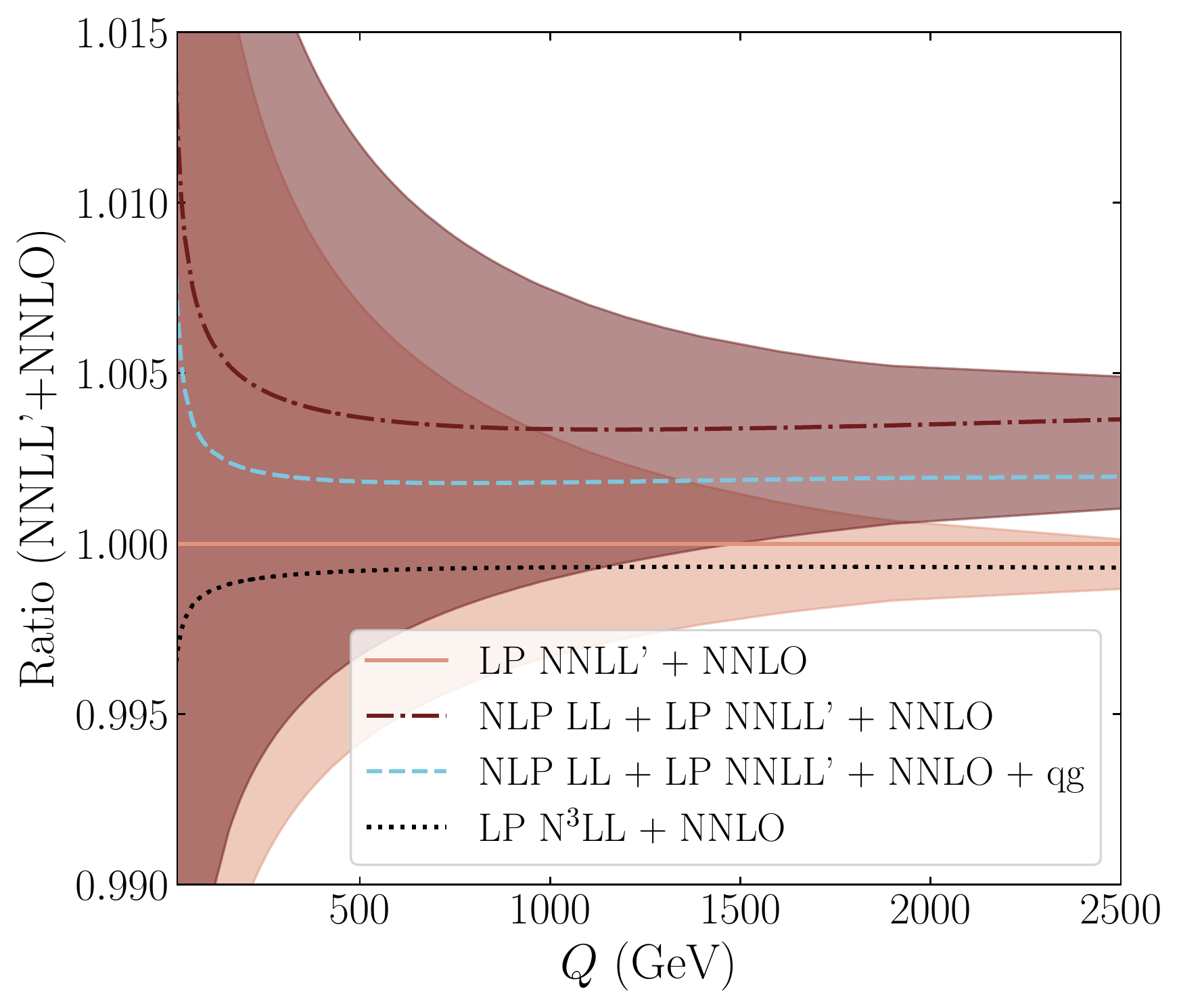}
         \caption{Ratio of the DY invariant mass distribution with respect to NNLL$^{\prime}$ + NNLO.}
         \label{fig:DYwrtNNLL}
     \end{subfigure}}
    \caption{Ratio plot for the Higgs total cross section (left) and DY invariant mass distribution (right), normalized to the  NNLL$^{\prime}$ + NNLO result. The NNLL$^{\prime}$ (+NLP LL) result is again shown by the orange solid (red dash-dotted) lines. The NNLL$^{\prime}$ NLP LL result obtained by adding the N$^3$LO NLP LL $qg$ result is shown by the dotted light-blue line.}\label{fig:NNLLratios}
\end{figure}

The results are shown in fig.~\ref{fig:higgsdQCD}, where all lines have been normalized either to LO~(\subref{fig:higgswrtLO}) or NNLO~(\subref{fig:higgswrtNNLO}). All resummed results are matched to the NNLO fixed-order result. The large enhancement due to the NNLL$^{\prime}$ matched resummation with respect to the LO result is caused by the sizable $\delta(1-z)$ contribution to both the NLO and NNLO fixed-order cross sections, which enter the resummed distribution via $g_0(\alpha_s)$. The enhancement of the LP NNLL$^{\prime}$+ NLP LL resummed result, with respect to the LO distribution (using $\mu = Q$), is roughly $300-360 \%$ in the considered $Q$ range, the strongest increase occurring for smaller $Q$ values. When compared to the matched LP NNLL$^{\prime}$ resummed cross section, we find an NLP enhancement between $4.3-6.3\%$, with larger effects for smaller $Q$ values. This increase can be contrasted with the effect of increasing the LP accuracy to N$3$LL. To obtain this we need to include only the $g^{(4)}_a(\lambda, Q^2/\mu^2)$ function in the resummation exponent~\footnote{We extract this function from the publicly available TROLL code~\cite{Bonvini:2014joa}.}, given that $g_0(\alpha_s)$ is already included up to $\mathcal{O}(\alpha_s^2)$ for NNLL$^{\prime}$ accuracy. Only a modest (negative) N$^3$LL correction to the NNLL$^{\prime}$ resummed cross section is found: between $-(0.3-0.5)\%$. We have verified that the NLP corrections are competitive with the numerical increase from NLL to NNLL, and we show explicitly that they dominate over the increase from NNLL$^{\prime}$ to N$^3$LL~\footnote{One could also compare to the N$^3$LL$^{\prime}$ results, where one includes the $\mathcal{O}(\alpha_s^3)$ contribution to $g_0(\alpha_s)$. Here we choose not to do so, in order to directly compare the numerical contribution of terms that are of $\mathcal{O}(\alpha_s^k\ln^{k-2}(\bar{N}))$ with those that are of $\mathcal{O}(\alpha_s^k\ln(\bar{N})^k/N)$ (where $k$ runs from $1$ to $\infty$). If we would compare at LP N$^3$LL$^{\prime}$ instead, the $\mathcal{O}(\alpha_s^k\ln^{k-2}(\bar{N}))$ terms in the exponent would be multiplied by a different hard function than the LP NNLL$^{\prime}$ + NLP LL result, which pollutes the comparison with a possibly sizeable constant contribution. }. The scale uncertainty increases somewhat after including the NLP resummation. For the NNLL$^{\prime}$ resummed result, the scale uncertainty is between $-0.03\%$ and $+6.2\%$, whereas the NLP resummed result shows a scale uncertainty between $-2.9\%$ and $+8.3\%$ (we expect that the inclusion  NLP NLL terms would reduce this scale uncertainty).  Below, we report explicitly on the Higgs cross sections at $Q = 125$~GeV, corresponding with the physical Higgs mass
\begin{eqnarray*}
\sigma^{\text{(NNLO)}}_{pp\rightarrow h+X}(m_{\rm h}) &=& 39.80^{+9.7\%}_{-9.0\%} \text{ pb}, \\
\sigma^{\text{(LP NNLL$^{\prime}$)}}_{pp\rightarrow h+X}(m_{\rm h}) &=& 42.76^{+5.6\%}_{-0.3\%}\text{ pb},\\
\sigma^{\text{(LP N}^3\text{LL)}}_{pp\rightarrow h+X}(m_{\rm h}) &=& 42.64^{+5.8\%}_{-0.1\%}\text{ pb},\\
\sigma^{\text{(LP NNLL$^{\prime}$ + NLP LL)}}_{pp\rightarrow h+X}(m_{\rm h}) &=&  45.32^{+7.5\%}_{-2.8\%}\text{ pb}.
\end{eqnarray*}
We thus find a notable NLP LL contribution in the diagonal channel. Of course, the resummation of the $qg$-channel is missing, which would also result in NLP LL enhanced terms (see fig.~\ref{fig:HiggsexpansionHAD}), and may potentially alter the observed NLP effect. To estimate the size of such contributions we include the NLP LL contribution from the $qg$-channel at N$^3$LO, the order at which we expect the largest contribution from a potential resummation in that channel. The full N$^3$LO results for the Higgs total cross section are available in the infinite-top-mass limit~\cite{Mistlberger:2018etf}, and can be inferred from the iHixs2 code~\cite{Dulat:2018rbf}. We find that the $qg$ induced NLP LL contribution reads, in Mellin space,
\begin{equation}
\bm{\Delta}_{{\rm h},qg}|_{\alpha_s^3}(N)
= -\left(\frac{\alpha_s}{\pi}\right)^3\frac{C_F(115C_A^2+50C_AC_F+27C_F^2)}{96}\frac{\ln^5{\bar{N}}}{N}, \label{eq:qg_N3LO_NLP_LL}
\end{equation}
which coincides with the prediction of ref.~\cite{Presti:2014lqa}. This $qg$-contribution results in a correction of the NLP LL resummation effect of $-0.5\%$ ($-3\%$) for small (large) $Q$ values (see fig.~\ref{fig:higgswrtNNLL}). The $\mathcal{O}(\alpha_s^3)$ contribution of the $qg$-channel thus gives a negligible contribution to the NLP LLs. Adding the terms of \eq{eq:qg_N3LO_NLP_LL} does not lead to a noticeable reduction of the scale uncertainty (not shown in the figure). Note that the smallness of the $qg$-initiated result is not caused by the partonic flux: the $qg$ flux exceeds the $gg$ flux at $\mu = 125$~GeV. Instead, this contribution is relatively small because, in contrast to the NLP LL contribution from the $gg$-channel, the $qg$-contribution is not proportional to the sizable higher-order constant terms of the LP $gg$-channel contained in the $\mathcal{O}(\alpha_s)$ and $\mathcal{O}(\alpha_s^2)$ contributions to the hard function $g_0(\alpha_s)$. A similar hard function is not included for the $qg$-channel, as cannot use an exponentiated form for the $qg$ contribution, nor know what the hard function in that case would be.
We expect that the $qg$-channel will play a larger role in the DY process, where the $g_0^1$ and $g_0^2$ coefficients are comparatively small. 

\subsection{DY invariant mass distribution}
\begin{figure}[t]
\centering
\mbox{\begin{subfigure}[h]{0.484\textwidth}
         \centering
         \includegraphics[width=\textwidth]{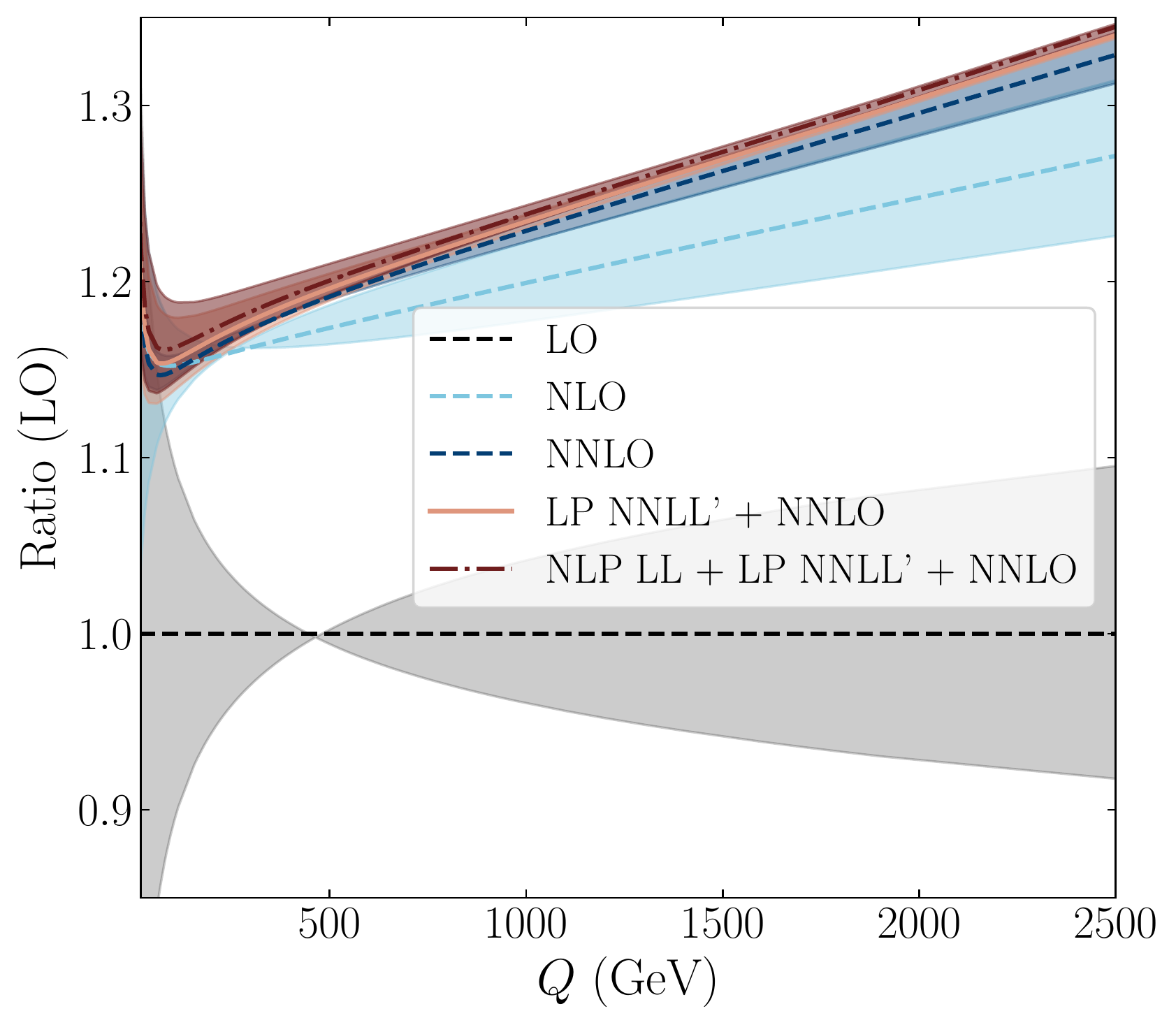}
         \caption{Ratio with respect to LO}
         \label{fig:DYwrtLO}
     \end{subfigure}
     \hfill
     \begin{subfigure}[h]{0.493\textwidth}
         \centering
         \vspace*{-2.2pt}
         \includegraphics[width=\textwidth]{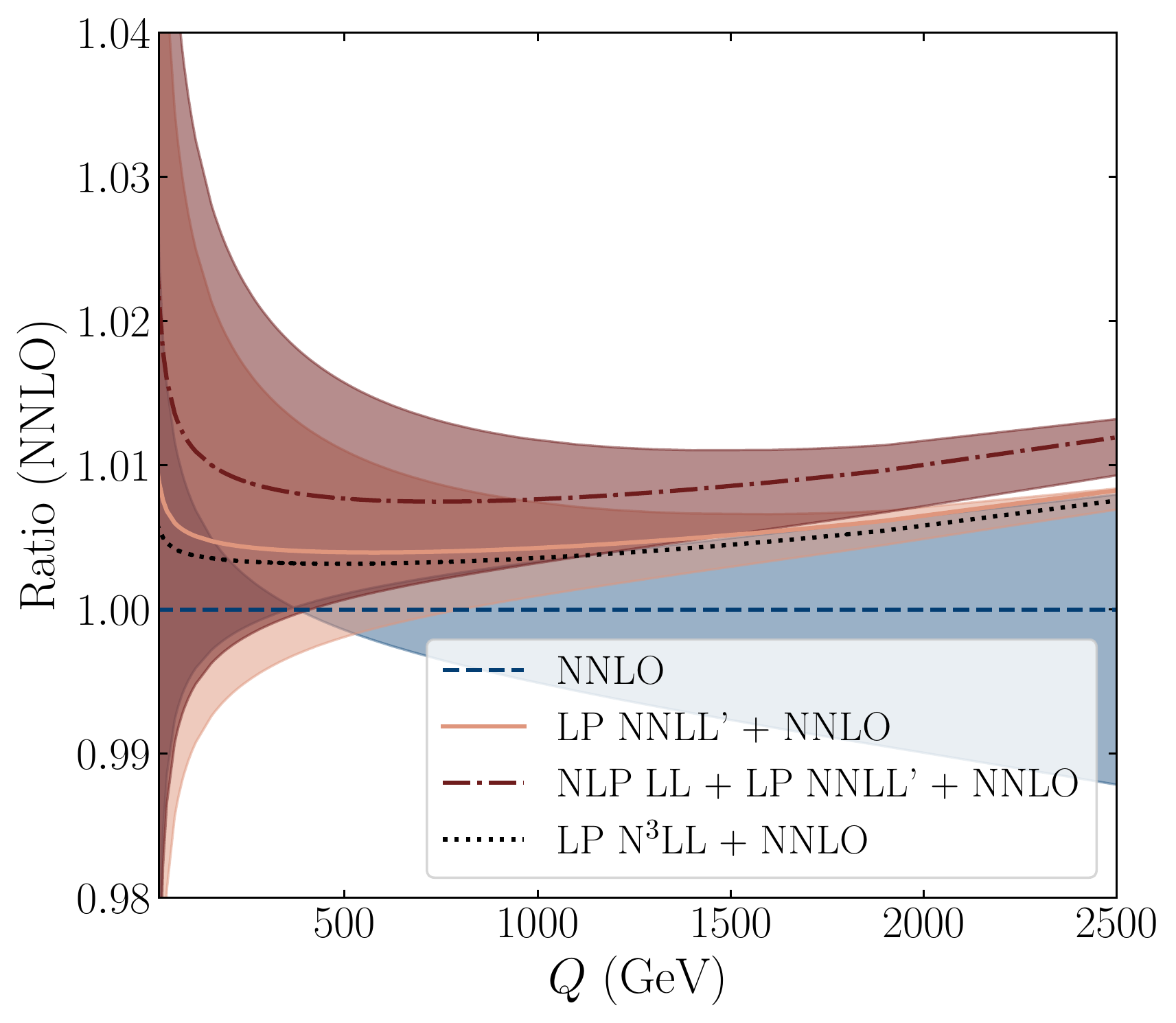}
         \caption{Ratio with respect to NNLO}
         \label{fig:DYwrtNNLO}
     \end{subfigure}}
    \caption{Ratio plot for the DY invariant mass distribution, normalized to the LO one (left) or NNLO one (right). The colour coding is the same as in fig.~\ref{fig:higgsdQCD}. The scale variation is obtained by setting either to $\mu = Q/2$ or $\mu = 2Q$. }\label{fig:DYinvdQCD}
\end{figure}

For the computation of the $N$-space resummed DY invariant mass distribution we use
\begin{equation}
\label{eq:qg_offdiag_higgs}
\frac{{\rm d} \bm{\sigma}_{pp\rightarrow \gamma^*+X}}{{\rm d} Q^2} = \sigma_0^{\rm DY}\,\sum_{q} e_q^2\,\bm{f}_q(N,\mu)\,\bm{f}_{\bar{q}}(N,\mu)\,\bm{\Delta}_{q\bar{q}}^{\rm dQCD}(N,Q^2/\mu^2),
\end{equation}
with $\sigma_0^{\rm DY}$ given in eq.~\eqref{eq:DYLOcoeff}, $\bm{\Delta}_{q\bar{q}}^{\rm dQCD}(N,Q^2/\mu^2)$ in eq.~\eqref{eq:resumfunctions} and the sum runs over quarks  treated as massless, $q\,\in\, \{u,d,c,s,b\}$. We show the resulting ratio plots in fig.~\ref{fig:DYinvdQCD}, where as in the Higgs case, all resummed results are matched to the NNLO result.

The LP NNLL$^{\prime}$ + NLP LL resummation enhances the LO distribution (using $\mu = Q$) by $15-34\%$, with increasing effect for larger $Q$ values. The NLP contribution provides an increase with respect to the LP NNLL$^{\prime}$ resummed (and matched) distribution of only $0.35-1.15\%$. Larger NLP enhancements are found for very small values of $Q$, where one moves further away from threshold. As for the Higgs production case, the NLP increase dominates over the N$^3$LL effect, which is $-(0.1-0.3)\%$ as shown in fig.~\ref{fig:DYwrtNNLO}, where the larger deviation is only found for small values of $Q$ (as best seen in \fig{DYwrtNNLL}). The larger size of the NLP LL with respect to the N3LL is not as pronounced as in the Higgs production case. 

The scale uncertainty of the NNLL$^{\prime}$ resummed result lies between $-4.4\%$ and $+5.4\%$ for the range of $Q$ values shown in fig.~\ref{fig:DYinvdQCD}, while that of the NLP resummed result is between $-4.8\%$ and $+5.8\%$. Therefore, by including the NLP LL contribution, we find a modest increase in the scale uncertainty of the resummed result, which would expect to decrease if NLL resummation at NLP were available. Note that for large $Q$ values ($Q>1$~TeV), the central value of the NLP correction lies outside the LP resummed uncertainty band.

As in the Higgs production case, the NLP LL resummation is not complete, since the $qg$-channel is missing (see fig.~\ref{fig:DYHAD}). Using the results of ref.~\cite{Presti:2014lqa}, we may obtain the N$^3$LO NLP LL contribution stemming from the $qg$-channel, resulting in
\begin{equation}
\label{eq:qg_offdiag_DY}
\bm{\Delta}_{{\rm DY},qg}|_{\alpha_s^3}(N) =- \left(\frac{\alpha_s}{\pi}\right)^3 T_F\frac{27 C_A^2+50C_AC_F+115 C_F^2}{96}\frac{\ln^5{\bar{N}}}{N}.
\end{equation}
As already anticipated, in contrast to what was observed for single Higgs production, we see that the addition of this result to the resummed $q\bar{q}$-channel significantly alters the NLP effect (fig.~\ref{fig:DYwrtNNLL}). The $qg$-contribution gives a $-54\%$ correction to the NLP effect of the dominant channel for small $Q$ values, while for larger $Q$ values it gives a $-44\%$ correction. As for Higgs production, we find that the scale uncertainties do not decrease after adding the $qg$ NLP LL contribution at $\mathcal{O}(\alpha_s^3)$. This is perhaps not surprising: for both the Higgs and DY processes we introduce additional scale-dependence via  $\alpha_s$ in either the NLP LL function $h^{(1)}_a$ for the leading channel, or via eq.~\eqref{eq:qg_offdiag_higgs}/\eqref{eq:qg_offdiag_DY} for the off-diagonal channel. This scale dependence is not balanced at LL, but could (for both channels) be balanced at NLP NLL via the introduction a scale term proportional to $\ln(Q^2/\mu^2)$. However, since we have no control over any other contribution that might appear at NLP NLL, we refrain from adding this contribution.  

\subsection{Other colour-singlet production processes}
To see how general our results for the impact of NLP effects are we  extend the analysis to the di-Higgs and di-boson production processes. As it is not our aim to provide a precise phenomenological prediction, but rather analyse the numerical effects of NLP LL resummation, we consider the (unmatched) resummation of these processes up to LP NLL + NLP LL, and include only the leading-order contribution to $g_0(\alpha_s)$. 

We start by considering the di-Higgs production process, which is also dominated by gluon fusion. Threshold resummation has been achieved first up to NLL in the SCET framework in the heavy-top-mass limit, including top-quark-mass dependent form factors to partially correct for this~\cite{Shao:2013bz} approximation. This study has been extended to NNLL in ref.~\cite{deFlorian:2015moa}, and the inclusion of the full top-mass effects was studied up to NLL+NLO in ref.~\cite{deFlorian:2018tah}. Using full top-mass dependence, which we study here, the lowest-order expression for the hadronic differential distribution $\frac{{\rm d}\sigma_{pp\rightarrow hh}}{{\rm d}Q}$ is given by~\cite{Plehn:1996wb}
\begin{eqnarray}
\label{eq:sigma0dihiggs}
\frac{{\rm d}\sigma_{pp\rightarrow hh}}{{\rm d}Q} = \frac{2Q}{S} \int_{\tau}^1 \frac{{\rm d}x}{x}f_g\left(x,\mu\right)f_g\left(\frac{\tau}{x},\mu\right)\sigma_{gg\rightarrow hh}(Q^2),
\end{eqnarray}
with $\tau = Q^2/S$ and 
\begin{eqnarray}
\sigma_{gg\rightarrow hh}(Q^2) = \frac{G_F^2\alpha_s^2}{256(2\pi)^3}\int_{t_-}^{t^+}{\rm d}t \left(\left|C_{\Delta}F_{\Delta}+C_{\square}F_{\square}\right|^2+\left|C_{\square}G_{\square}\right|^2\right).
\end{eqnarray}
where the integration variable $t$ is given by
\begin{eqnarray}
t = -\frac{1}{2}\left(Q^2 - 2m_{h}^2 - Q^2 \sqrt{1-\frac{4m_{h}^2}{Q^2}}\cos\theta\right),
\end{eqnarray}
while the integration limits are obtained by setting $\cos\theta = \pm 1$. The exact expressions of ref.~\cite{Plehn:1996wb} are used for $C_{\Delta}$, $C_{\square}, F_{\Delta}$, $F_{\square}$ and $G_{\square}$, where no approximation on the mass ratio between the top-quark and the Higgs boson has been applied. \\
For the resummation of this process, it suffices to follow the same procedure as before. That is, we Mellin transform eq.~\eqref{eq:sigma0dihiggs} with respect to the hadronic threshold variable $\tau$. Then we replace the partonic LO coefficient $\sigma_{gg\rightarrow hh}(Q^2)$ by its resummed version, $\sigma_{gg\rightarrow hh}(Q^2)\bm{\Delta}_{gg}^{\rm dQCD}\left(N,Q^2/\mu^2\right)$. This replacement is valid at LP (see~ref.~\cite{Catani:2014uta}). We exploit the universal structure of next-to-soft gluon emissions~\cite{DelDuca:2017twk} to justify the same substitution at NLP. The resummed expression then reads
\begin{eqnarray}
\frac{{\rm d}\sigma_{pp\rightarrow hh+X}}{{\rm d}Q} = \sigma_{gg\rightarrow hh}(Q^2)\,\frac{2Q}{S}\int \frac{{\rm d}N}{2\pi i} \tau^{-N} \bm{f}_g(N,\mu)\bm{f}_g(N,\mu)\bm{\Delta}^{\rm dQCD}_{gg}\left(N,Q^2/\mu^2\right).
\end{eqnarray}
Since we work at NLL accuracy, we use the expression of \eq{eq:resumfunctions} for  $\bm{\Delta}^{\rm dQCD}_{gg}\left(N,Q^2/\mu^2\right)$ with $g_0(\alpha_s) = 1$ and neglect the $g^{(3)}$ contribution in the exponent. 

\begin{figure}[t]
   \centering
        \includegraphics[width=0.5\textwidth]{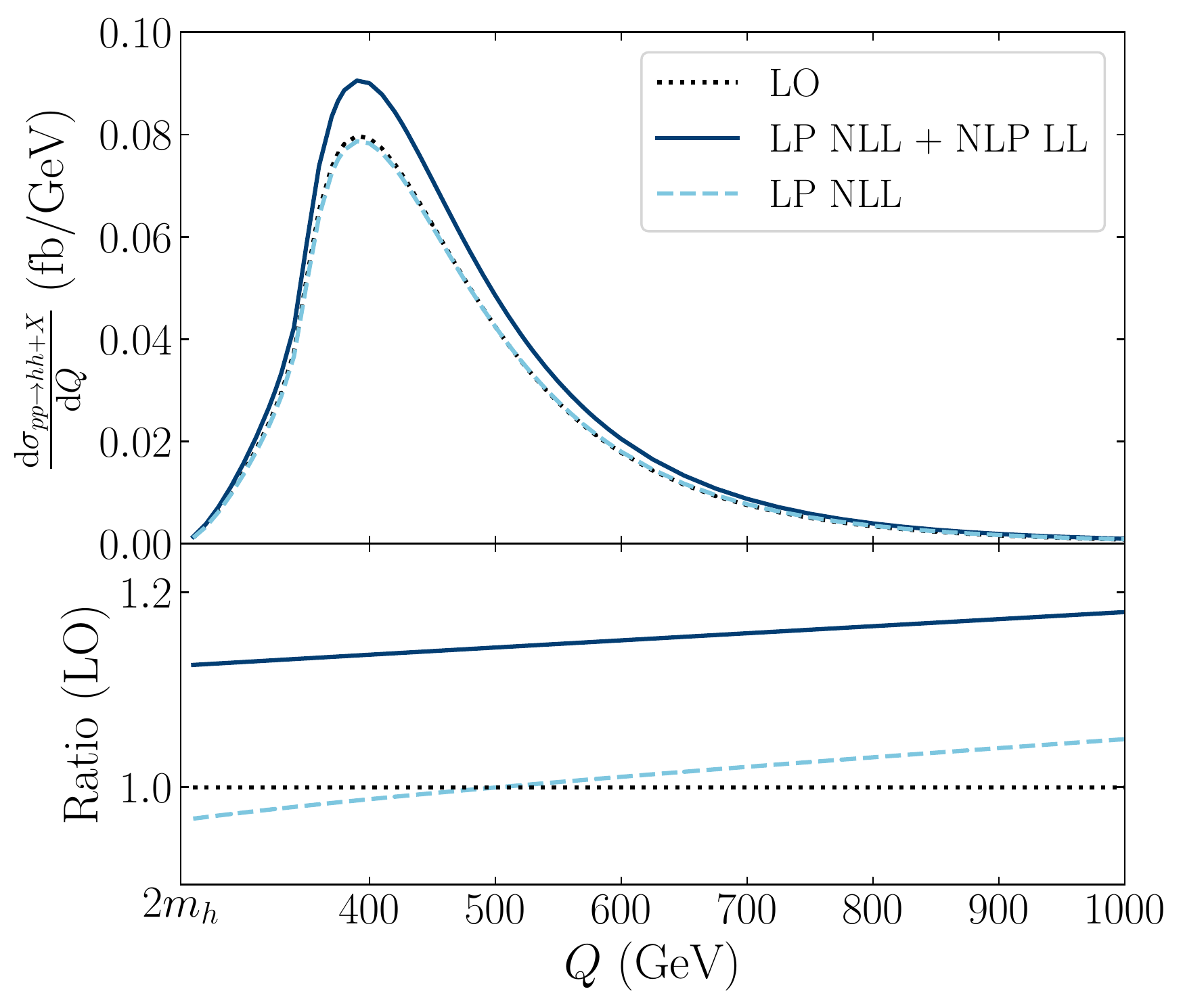}
    \caption{Differential cross section for $hh$ production showing the LO (black dotted), LP NLL (dashed light-blue), and NLP LL + LP NLL (solid dark-blue) results. The scale $\mu$ is set to $Q/2$. The bottom pannel shows the ratio with respect to the LO distribution. }\label{fig:dihiggsressumed}
\end{figure}

The results are shown in fig.~\ref{fig:dihiggsressumed}, where we used a factorisation/renormalisation scale $\mu=Q/2$, which is the scale for which higher-order corrections are smallest~\cite{DiMicco:2019ngk}. We see that the LP NLL result gives a correction to the LO distribution of $-3.2\%$ to $4.7\%$, while the NLP LL terms cancel the partially negative LP correction, leading to a substantial enhancement of the LO distribution of $12.6-17.8\%$. Larger corrections are found for higher values of $Q$. An increase of the NLP LL + LP NLL result with respect to the LP NLL result is found to be between $12.5\%$ and $16.3\%$. As for the DY and single Higgs processes, we thus again find that the NLP LL effect is sizeable. For comparison, the increase of going from LP LL to LP NLL (not shown here) is of the same order~\footnote{Note that the numerical size of the LP NLL contribution primarily comes from the term proportional to $\ln(Q^2/\mu^2)$ in $g^{(2)}_g$.}. 

\begin{figure}[t]
\centering
   \mbox{\begin{subfigure}{0.5\textwidth}\centering
        \includegraphics[width=\textwidth]{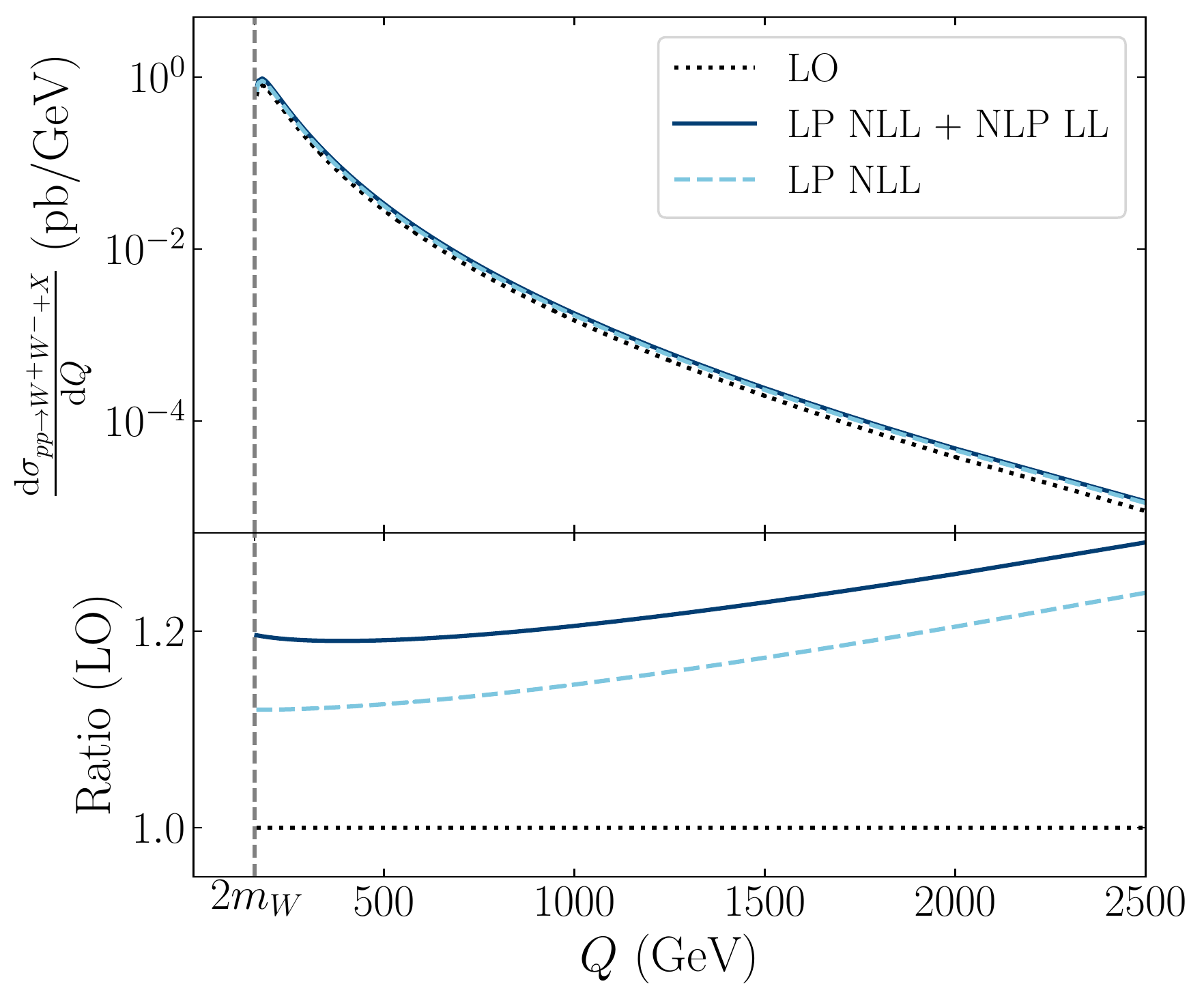}
    \end{subfigure}
    \begin{subfigure}{0.5\textwidth}\centering
        \includegraphics[width=\textwidth]{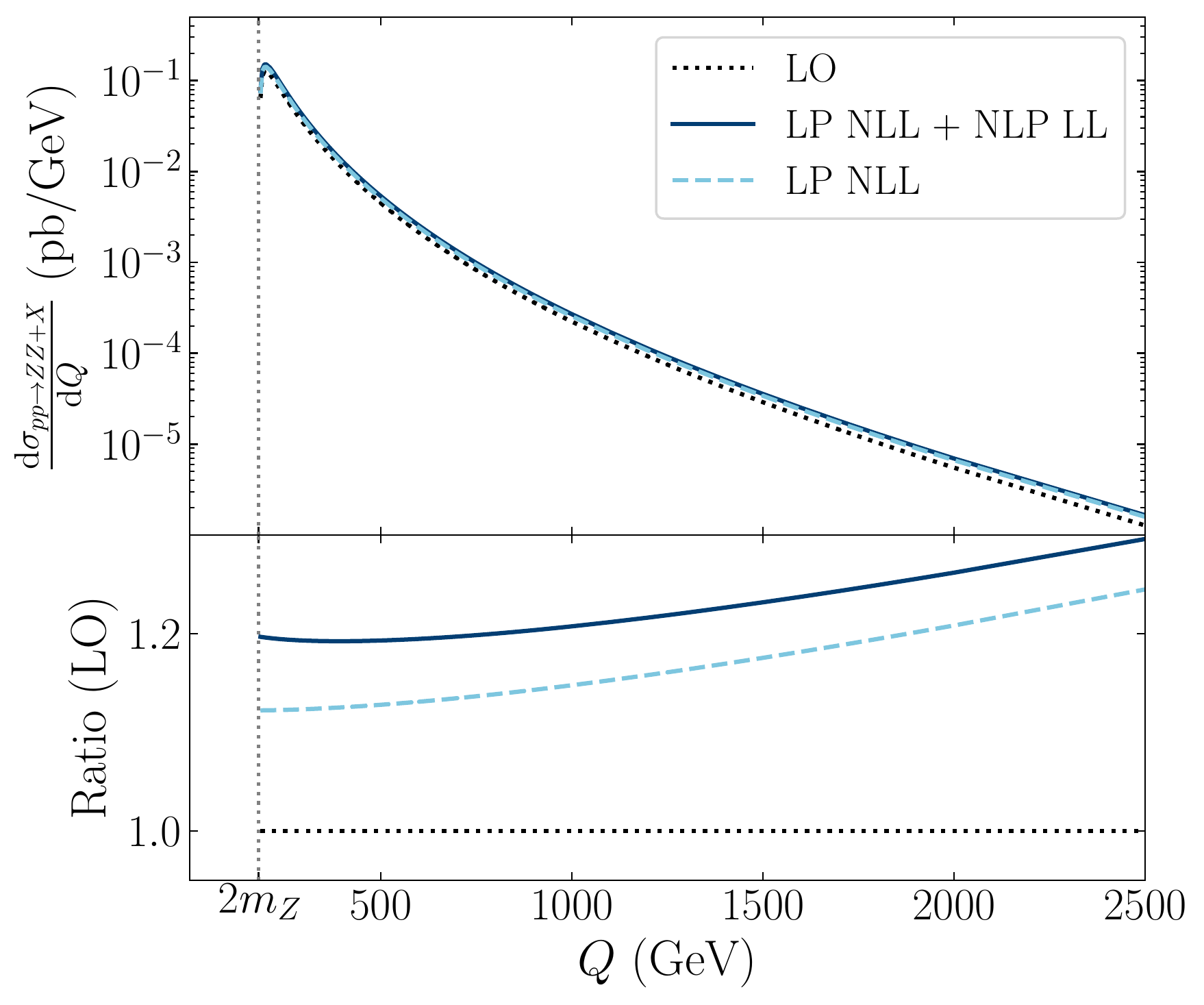}
    \end{subfigure}}
     \caption{Differential cross section for $W^+W^-$ production (left) and $ZZ$ production (right) showing the LO (black dotted) and  resummed LP NLL (dashed light-blue), and NLP LL + LP NLL (solid dark-blue) results. The scale $\mu$ is set to $Q$. }\label{fig:dibosonressumed}
\end{figure}
We now move on to $W^+W^-$ and $ZZ$ production. Threshold resummation up to NNLL was considered in the direct-QCD framework in ref.~\cite{Dawson:2013lya}, and for the SCET framework in ref.~\cite{Wang:2014mqt,Wang:2015mvz}. Transverse-momentum resummation for both processes has also been performed in ref.~\cite{Grazzini:2015wpa}. We use the LO expressions from ref.~\cite{Frixione:1993yp, Brown:1978mq}~\footnote{Note that in ref.~\cite{Frixione:1993yp}, there is a misprint in the LO integrated coefficients (eq.~(3.12)). The second terms in the expressions for $\mathcal{F}_i(s)$ and $\mathcal{J}_i(s)$ need to be multiplied by a factor of $16$.}. We may write, similarly to \eq{eq:sigma0dihiggs}
\begin{eqnarray}
\label{eq:dibosonLO}
\frac{{\rm d}\sigma_{p p \rightarrow VV}}{{\rm d}Q} &=& \frac{2Q}{S}\sum_{i = q,\bar{q}}\int_{\tau}^1 \frac{{\rm d} x}{x}f_i\left(x,\mu\right)f_{\bar{i}}\left(\frac{\tau}{x},\mu\right)\sigma_{q_i\bar{q}_i \rightarrow VV}(Q^2),
\end{eqnarray}
with the LO partonic cross sections given by
\begin{eqnarray}
\sigma_{q_i\bar{q}_i \rightarrow W^+W^-}(Q^2) &=& \frac{1}{64\pi C_A Q^4}\left(c_i^{tt} \mathcal{F}_i(Q^2) - c_i^{st}(Q^2)\mathcal{J}_i(Q^2)+c_i^{ss}(Q^2) \mathcal{K}_i(Q^2)\right),\\
\sigma_{q_i\bar{q}_i \rightarrow ZZ}(Q^2) &=& \frac{g_{V,i}^4 + g_{A,i}^4 + 6 \left(g_{V,i}g_{A,i}\right)^2}{4\pi C_A Q^2} \left(\frac{1+\frac{4m_Z^4}{Q^4}}{1-\frac{2m_Z^2}{Q^2}}\ln\left(\frac{1+\beta_Z}{1-\beta_Z}\right)-\beta_Z\right),
\end{eqnarray}
where
\begin{eqnarray}
\beta_Z = \sqrt{1-\frac{4m_Z^2}{Q^2}}.
\end{eqnarray}
The expressions for the coefficients $c_i^{tt}$, $c_i^{st}$, and $c_i^{ss}$, and the functions $\mathcal{F}_i$, $\mathcal{J}_i$ and $\mathcal{K}_i$  may be found in ref.~\cite{Frixione:1993yp}. Note that the partonic coefficient functions $\sigma_{q_i\bar{q}_i\rightarrow VV}$ depend on the left- and right-handed couplings of the quarks with the $Z$-boson, which are of course different for up-type quarks and down-type quarks. Following the same procedure as for the di-Higgs results, we obtain a resummed expression for both cases that reads
\begin{eqnarray}
\frac{{\rm d}\sigma_{pp\rightarrow VV+X}}{{\rm d}Q} = \sigma_{q\bar{q}\rightarrow VV}(Q^2)\,\frac{2Q}{S} \int \frac{{\rm d}N}{2\pi i} \tau^{-N} \bm{f}_q(N,\mu)\bm{f}_{\bar{q}}(N,\mu)\bm{\Delta}_{q\bar{q}}\left(N,Q^2/\mu^2\right).
\end{eqnarray}
The results are shown in fig.~\ref{fig:dibosonressumed}. The two processes are identical from a resummation point-of-view, therefore, we find an LP NLL increase of the LO distribution between $12.2-20.8\%$ for both processes, whereas the NLP LL + LP NLL increases the LO distribution by $19.2-26.2\%$. As for the other processes, larger enhancements are found for higher values of $Q$. 

The increase induced by the NLP LL contributions with respect to the LP NLL result is between $4.5-6.8\%$, which is smaller than the correction that was found for the di-Higgs process. This is not surprising: it is well known that gluon-initiated processes receive larger LL threshold corrections due to $C_A > C_F$. The difference between LP LL and LP NLL resummation for the $VV$ processes is around $1\%$, which is not dissimilar from the increase found for the inclusion of the NLP terms. We therefore again see the numerical importance of the NLP LL terms. 

What we have observed for all processes is that the NLP LL contribution is not negligible, and in many cases exceeds the size of higher-logarithmic LP contributions. To understand this, note that the saddle-point approximation for the inverse Mellin transform integral leads to values of $\bar{N}$ around $1.5-3$~\cite{Bonvini:2010tp,Bonvini:2012an,Bonvini:2014qga, vanBeekveld:2020cat}. With $\alpha_s \simeq 0.12$, we have 
$ \alpha_s\ln(\bar{N}) \simeq \alpha_s\ln(\bar{N})/N$, while the NNLL term $\alpha_s^2\ln(\bar{N})$ is suppressed by $\alpha_s$. Naturally, we do not suggest to cease pursuing effects of higher-logarithmic LP contributions. Firstly, one can include the $\mathcal{O}(\alpha_s)$ contribution in the hard function
$g_0(\alpha_s)$ (i.e.~leading to LP NLL$^\prime$) accuracy), resulting in a large correction if the constant contribution of the NLO correction is sizeable, as we have seen for single Higgs production, and also seems to be the case for di-Higgs production (see e.g.~\cite{deFlorian:2018tah,Borowka:2016ehy,Nandan:2016ohb}). Secondly, although the central result might change minimally by including higher-logarithmic contributions at LP, such contributions stabilize
the scale dependence of the prediction.
\subsection{Some final remarks on NLP LL resummation in dQCD}
\label{sec:someremarks}
In this section we have explored the numerical behaviour of the NLP $h^{(1)}_a$ correction (eq.~\eqref{eq:resumfunctions}) on a few selected colour-singlet production processes. For all processes we reach a similar conclusion: the inclusion of NLP LL resummation leads to a numerically noticeable result. We briefly comment on the relation between the work on NLP resummation presented here and in refs.~\cite{Bonvini:2013td,Bonvini:2014joa} for single Higgs production. The methods employed in this paper may best be compared to their $\psi-{\rm soft}_2$ prescription without exponentiation of the constant contributions~\footnote{Another prescription, $A-{\rm soft}_2$ in ref.~\cite{Bonvini:2014joa}  uses the Borel Prescription~\cite{Abbate:2007qv,Forte:2006mi} to handle the asymptotic summation of the perturbative expansion. Numerical differences between $A-{\rm soft}_2$ and  $\psi-{\rm soft}_2$ are shown to be small there.}. The $\psi-{\rm soft}_2$ prescription consists of replacing $\lambda$ in the resummation exponents $g^{(i)}(\lambda)$ by the combination of polygamma functions $\lambda = \alpha_s b_0 (2\psi_0(N)-3\psi_0(N+1)+2\psi_0(N+2))$. 
Denoting their resummation exponent at LP NLL with $\mathbf{\Delta}^{\psi-{\rm soft}_2}_{aa}$, the difference with our LP NLL + NLP LL resummation exponent (eq.~\eqref{eq:resumfunctions}  without $g^{(3)}_a$ and with $g_0(\alpha_s) = 1$) may be evaluated at $\mathcal{O}(\alpha_s)$, where we obtain
\begin{eqnarray}
\mathbf{\Delta}^{\psi-{\rm soft}_2}_{aa} - \mathbf{\Delta}^{\rm dQCD}_{aa}  &=& \frac{\alpha_s}{\pi} \Bigg[ -A^{(1)}_a\left(2\gamma_E^2 - 2\gamma_E\ln\frac{Q^2}{\mu^2}\right)
- \frac{A^{(1)}_a}{N}\ln\frac{Q^2}{\mu^2}  \\
&& \hspace{1cm}
+ \frac{A^{(1)}_a}{N^2}\left(\frac{1}{2} - \frac{25}{3}\left(\ln(N)+\gamma_E\right)+\frac{25}{6}\ln\frac{Q^2}{\mu^2}\right) +\mathcal{O}\left(\frac{1}{N^3}\right)\Bigg] + \mathcal{O}(\alpha_s^2)\,.\nonumber
\end{eqnarray}
The difference at LP is of NNLL accuracy, and originates from using $N$ exponentation in the $\psi-{\rm soft}_2$ prescription rather then $\bar{N}$ exponentiation, which is what we have used in our work. At NLP, the difference is of NLL accuracy, whereas the discrepancy at NNLP starts at LL with the term proportional to $\ln(N)/N^2$. They observe that the correction on the fixed-order single Higgs production cross section is increased with respect to standard resummation by making the $\psi-{\rm soft}_2$ replacement, consistent with our observations on the effect of NLP LL resummation. No soft-quark correction was considered in their work. To the best of our knowledge, a similar analysis at the hadronic level for DY has not been performed. 

Returning to our results, we found in this section that for single Higgs-production the central value of the LP NNLL$^{\prime}$ + NLP LL resummed coefficient lies outside the uncertainty band for the LP resummed (and matched) result. For DY production processes this is observed for $Q>1$~TeV. Both processes show that the numerical effect of resumming the NLP LL terms exceeds that of improving LP resummation to N$^3$LL accuracy. Scale uncertainties seem to slightly increase after the NLP LL result is included. We expect that the scale uncertainty can be reduced only after the inclusion of the (unavailable) NLP NLL contributions. For di-boson production the numerical enhancement of upgrading a LP LL calculation to NLP LL rivals (in the case of di-Higgs) or exceeds (for $ W^+W^-$/$ ZZ$) that of going from LP LL to LP NLL, for central scale choices. In general, NLP LL effects originating from next-to-soft-gluon emissions are found to be larger for $gg$-induced processes than for $q\bar{q}$-induced processes. On the other hand, we find that the off-diagonal $qg$ NLP LL contribution is more important for the $q\bar{q}$-induced DY process than for the $gg$-induced Higgs process. 

%======================
%=======SECTION========
%======================
\section{Comparison of NLP resummation in SCET and dQCD}
\label{sec:comparison}
NLP corrections have been calculated and resummed not only in direct QCD, as described in section~\ref{dQCD_approach}, but also within soft-collinear effective field theory (SCET). While the two approaches should formally give the same result, numerical differences have turned out to be sizable for LP resummation, and originate from power-suppressed contributions that are included differently in the two methods~\cite{Ahrens:2008nc,Bonvini:2012az,Bonvini:2013td,Sterman:2013nya,Almeida:2014uva,Bonvini:2014qga}. Now that the exact LL NLP contribution (in $z$-space) has been determined in both approaches, it is important to perform such a comparison again, and we do so in this section. We shall see that the inclusion of the LL NLP contribution reduces the numerical differences between the two approaches. For the comparison we consider Drell-Yan and Higgs production, at LP NNLL (i.e.~we use eq.~\eqref{eq:resumgeneralLP} with $g_0$ up to $\mathcal{O}(\alpha_s)$ for dQCD) and NLP LL accuracy. We refrain from matching to fixed-order calculations as this has no bearing on the comparison of resummed calculations.

Below we first establish notation for SCET resummation, and briefly discuss the results of \cite{Beneke:2018gvs,Beneke:2019mua,Beneke:2019oqx}, 
revisiting in particular the so-called kinematic correction. We obtain a slightly modified SCET resummation formula, which will ease our task of relating the SCET formalism to dQCD. 
We then proceed to the comparison of the NLP LL contributions within the two theories. Note that while presenting the analytical comparisons, we use the $\beta$-function coefficients $b_n$ and $\beta_n = (4\pi)^{n+1}b_n$ (eq.~\eqref{eq:beta1} and \eqref{eq:beta2}) interchangeably, as the $b_n$ are often used in dQCD, whereas $\beta_n$ are usually preferred in SCET. The section is concluded by discussing numerical results for both methods. 

\subsection{NLP Resummation in SCET}
\label{SCET_approach}
Using the results of ref.~\cite{Beneke:2019oqx, Beneke:2019mua, Beneke:2018gvs}, we write the analogue of eq.~\eqref{eq:resumgeneralLP} in SCET up to NLP in $z$-space as
\begin{eqnarray}
\label{eq:scetuptoNLP}
\Delta^{\rm SCET}(z) = 
\Delta^{\rm SCET,LP}(z) + 
\Delta^{\rm SCET,NLP}(z),
\end{eqnarray}
where $\Delta^{\rm SCET,LP/NLP}(z)$ resums large logarithms at LP and NLP, respectively, and the scale dependence of these factors is suppressed. The NLP term $\Delta^{\rm SCET,NLP}(z)$  consists of two contributions: a kinematic factor, which originates from taking the partonic cross section to LP in the SCET operator product expansion, but retaining the exact kinematics up to NLP, and a dynamical correction that arises by considering sub-leading-power operators
\begin{eqnarray}
\label{NLPSplitKinDyn}
\Delta^{\rm SCET,NLP}(z) = 
\Delta^{\rm SCET,NLP}_{\rm kin}(z) 
+ \Delta^{\rm SCET,NLP}_{\rm dyn}(z).
\end{eqnarray}
We first focus on the kinematic correction, which together with its LP counterpart, is expressed in the factorised form  \cite{Beneke:2018gvs,Beneke:2019oqx}
\begin{eqnarray}
\label{LPplusNLPkinDY}
\Delta_{\rm kin}^{\rm DY, SCET,LP+NLP}(z) = H_{\rm DY}(s)\frac{Q}{z}\left(1+\frac{1}{Q}\left(\frac{\partial}{\partial \vec{x}}\right)^2\frac{\partial}{\partial \Omega_*}\right)S^{\rm DY}_0\left(\Omega_*, \vec{x}\right)\Big|_{\vec{x} = 0},
\end{eqnarray}
for the DY partonic coefficient function of the $q\bar{q}$-channel, and 
\begin{eqnarray}
\label{LPplusNLPkinH}
\Delta_{\rm kin}^{\rm h, SCET,LP+NLP}(z) =H_{\rm h}(s)\frac{Q}{z^2}\left(1+\frac{1}{Q}\left(\frac{\partial}{\partial \vec{x}}\right)^2\frac{\partial}{\partial \Omega_*}\right)S^{\rm h}_0\left(\Omega_*, \vec{x}\right)\Big|_{\vec{x} = 0},
\end{eqnarray}
for the $gg$-channel contribution to Higgs production, \cite{Beneke:2019mua}. The hard function $H$ depends on short-distance physics, and is different for DY and Higgs. The function $S_0$ denotes the Fourier transform of the position-space soft function
\begin{equation}
S_0\left(\Omega_*, \vec{x}\right) 
= \int \frac{dx^0}{4\pi} \, e^{i \frac{\Omega_*}{2} x^0} \widetilde S(x^0,\vec x),
\end{equation}
where the latter is defined as a vacuum-expectation value of time-ordered products of Wilson lines associated to the directions of the external partons~\cite{Li:2011zp,Beneke:2018gvs}
\begin{equation}
\widetilde S(x) = \frac{1}{{\cal C}} 
\, {\rm Tr} \langle 0 | 
{\bf \bar T}\big[Y^{\dag}_+(x)Y_-(x)\big]\,
{\bf T}\big[Y^{\dag}_-(0)Y_+(0)\big] | 0 \rangle.
\end{equation}
Here ${\cal C} = N_c$ for Drell Yan, and 
${\cal C} = N_c^2-1$ for Higgs production. Furthermore, the soft Wilson lines are defined in terms of soft gluon fields in SCET
\begin{equation}
Y_{\pm}(x) = {\cal P}  \exp 
\bigg[ i g_s \int_{-\infty}^0 ds\, n_{\mp}A_s(x+s n_{\mp}) \bigg], 
\end{equation}
where the light-like vectors $n_{\pm}$ are oriented along the directions of the incoming partons. At NLO the Drell-Yan and Higgs soft functions are related by the simple replacement $C_F \to C_A$. The normalizations for the Higgs and DY case are chosen, as explained in section \ref{sec:fo_threshold} and further discussed in appendix \ref{AppNormalization}, such that we have a universal NLP LL correction for both processes, resulting in a $1/z$ difference between the Higgs and DY factorisation formulae, as it is evident by comparing \eq{LPplusNLPkinDY} with \eq{LPplusNLPkinH}. The variable $\Omega_*$ is the reciprocal energy, and is given by $\Omega_* = 2\sqrt{s}(1-\sqrt{z})$. 

Eqs.~(\ref{LPplusNLPkinDY}) and~(\ref{LPplusNLPkinH})
are expanded in powers of $1-z$, which implies in particular that $\Omega_{*}$ is expanded around $\Omega \equiv Q(1-z)$. Then, the LP contribution reads 
\begin{eqnarray}
\label{eq:choice1}
{\Delta}^{\rm DY /h, SCET, LP}(z) = H(Q^2)Q S_0\left(\Omega, \vec{x}\right)\Big|_{\vec{x} = 0}\,,
\end{eqnarray}
where we have suppressed the ${\rm DY/h}$ labels of $H$ and $S_0$. At fixed-order, both $H$ and $S_0$ depend on the factorisation scale. The hard function contains factors of $\ln(Q^2/\mu^2)$, and the soft function contains $\ln(Q^2(1-z)^2/\mu^2)$. Therefore, depending on the choice of $\mu$, either $H$ or $S_0$ will result in a large logarithmic correction. However, the hard and soft functions obey renormalisation-group equations (RGEs). According to the effective-field theory approach, it is then possible to choose a natural scale, $\mu_h \simeq Q$ for the hard function, and $\mu_s \simeq Q(1-z)$ for the soft function, such that fixed-order logarithms are small. The original large logarithms are then resummed using the RGEs to evolve the hard and soft function from their natural scale to the scale at which the cross section is evaluated. We will recall the resummed partonic cross section one obtains starting from \eq{eq:choice1} in what follows, but let us first revisit the NLP terms originating from the expansion eqs.~(\ref{LPplusNLPkinDY}) and~(\ref{LPplusNLPkinH}), i.e. the kinematic correction. 

To obtain \eq{eq:choice1}, we made the choice to expand \eq{LPplusNLPkinDY} and~\eq{LPplusNLPkinH} around $\Omega_* = \Omega$ and set $s = Q^2$. At NLP, this choice leads to the kinematic NLP corrections $K_1,\ldots,K_4$ 
\begin{eqnarray} \label{kinematicSoftDef}
\Delta^{\rm DY/h, SCET,NLP}_{\rm kin}(z) = H (Q^2)S_K (\Omega)\,, \quad \text{with}\quad S_K(\Omega) = \sum_{i=1}^4 S_{K_i}(\Omega)\,,
\end{eqnarray}
where
\begin{subequations}
\begin{eqnarray}
S_{K_1}(\Omega) &=& \frac{\partial}{\partial \Omega}\left(\frac{\partial}{\partial \vec{x}}\right)^2 S_0\left(\Omega,\vec{x}\right)\Big|_{\vec{x} = 0}, \\
S_{K_2}(\Omega) &=& \frac{3}{4}\Omega^2 \frac{\partial}{\partial \Omega} S_0\left(\Omega,\vec{x}\right)\Big|_{\vec{x} = 0}, \\
S^{\rm DY}_{K_3}(\Omega) &=& \Omega S_0\left(\Omega,\vec{x}\right)\Big|_{\vec{x} = 0}, 
\qquad\qquad S^{\rm h}_{K_3}(\Omega) = 2\Omega S_0\left(\Omega,\vec{x}\right)\Big|_{\vec{x} = 0}, \\
S_{K_4}(\Omega) &=& \frac{\partial}{\partial \ln(\mu^2)}\ln(H(\mu^2))\Big|_{\mu^2 = Q^2} \Omega  S_0\left(\Omega,\vec{x}\right)\Big|_{\vec{x} = 0}.
\end{eqnarray}
\end{subequations}
The $K_1$ term comes from identifying $\frac{\partial}{\partial \Omega_*} = \frac{\partial}{\partial \Omega} +\mathcal{O}(1-z)$, and $K_2$ originates from the Taylor expansion of the soft function around $\Omega_* = \Omega$.  The $K_3$ term is the first order of the expansion of $Q/z$ in the case of DY, and $Q/z^2$ in the case Higgs, hence the resulting contribution $S_{K_3}$ is different for these processes. Finally, $K_4$ contains the Taylor expansion of the hard function around $s = Q^2/z$.
Evaluating these terms up to NLP yields for their sum~\footnote{We quote here the renormalised kinematic soft functions.}~\cite{Beneke:2018gvs,Beneke:2019mua,Beneke:2019oqx}
\begin{subequations}
\label{KinSoft:choice1}
\begin{eqnarray}
S^{\rm DY}_K(\Omega) &=& 2 \frac{\alpha_s\, C_F}{\pi} \theta(\Omega)\,, \qquad\qquad\qquad\qquad\, \\
S^{\rm h}_K(\Omega) &=& 2 \frac{\alpha_s\, C_A}{\pi}\left[ -2\ln\frac{\mu}{\Omega}+1\right]\theta(\Omega)\,.
\end{eqnarray}
\end{subequations}
With these definitions, the LP partonic cross section with full kinematic dependence is expanded as
\begin{eqnarray}
\Delta^{\rm DY/h, SCET,LP+NLP}_{\rm kin}(z) = \Delta^{\rm DY/h, SCET,LP}(z)
+ \Delta^{\rm DY/h, SCET,NLP}_{\rm kin}(z)\,,
\end{eqnarray}
where $\Delta^{\rm LP}(z)$ is strictly LP, 
as defined in eq.~(\ref{eq:choice1}), and the corresponding NLP kinematic correction is given eq.~(\ref{kinematicSoftDef}). While this choice is consistent with a systematic power expansion, it is also possible to consider other expansion schemes, for which parts of the kinematic correction are not expanded, but instead kept within the LP term. Here we discuss one such case, motivated by the comparison with dQCD resummation. In this expansion scheme we keep factors of $Q/\sqrt{z}$ exact, i.e.~we expand $\Omega^*$ around $\hat{\Omega} = \frac{Q}{\sqrt{z}}(1-z)$. Comparing with \eq{ELL}, we immediately see that this choice consistently keeps the same kinematic factor in the exponent, both in the dQCD and in the SCET approach. The LP factorisation then reads
\begin{eqnarray}
\label{eq:choice2}
\widehat \Delta^{\rm DY/h, SCET,LP}(z) = H(Q^2)\frac{Q}{\sqrt{z}} S_0\left(\hat{\Omega}, \vec{x}\right)\Big|_{\vec{x} = 0}\,.
\end{eqnarray}
This choice leads for the $K_1$ and $K_4$ contributions of the kinematic correction (eq.~\eqref{kinematicSoftDef}) to a replacement of $\Omega\rightarrow \hat{\Omega}$. The kinematic corrections $K_2$ and $K_3$ change and become
\begin{subequations}
\begin{eqnarray}
\hat S_{K_2}(\hat{\Omega}) &=& \frac{1}{4}\hat{\Omega}^2 \frac{\partial}{\partial \hat{\Omega}} S_0\left(\hat{\Omega},\vec{x}\right)\Big|_{\vec{x} = 0}, \\
\hat S_{K_3}^{\rm DY}(\hat{\Omega}) &=& \frac{\hat{\Omega}}{2} S_0\left(\hat{\Omega},\vec{x}\right)\Big|_{\vec{x} = 0}, \qquad\qquad \hat S_{K_3}^{\rm h}(\hat{\Omega}) = \frac{3\hat{\Omega}}{2} S_0\left(\hat{\Omega},\vec{x}\right)\Big|_{\vec{x} = 0}.
\end{eqnarray}
\end{subequations}
This expansion scheme presents several advantages. First of all, the constant (NLP NLL) contribution, which was originally present in $\Delta_{\rm kin}^{\rm DY/h, SCET,NLP}(\Omega)$, is now absent in the new kinematic correction $\widehat \Delta_{\rm kin}^{\rm DY/h, SCET,NLP}(\Omega)$. At the same time, the NLP
LL contribution, which is absent for the DY case, and equal to the scale logarithm in the Higgs case, is unchanged. Explicitly, we have
\begin{subequations}
\label{KinSoft:choice2}
\begin{eqnarray}
\hat S^{\rm DY}_K(\Omega) &=& 0\,, \\
\hat S^{\rm h}_K(\Omega) &=& -4\frac{\alpha_s\, C_A}{\pi} \ln\frac{\mu}{\hat{\Omega}}\,\theta(\hat \Omega)\,.
\end{eqnarray}
\end{subequations}
The NLP NLL contribution that was present in eq.~\eqref{KinSoft:choice1} is now removed by the new expansion scheme. This contribution represents a phase space correction, which is universal. Instead of residing in the kinematic correction, it is now contained in the modified LP term of eq.~(\ref{eq:choice2}); in this way it too can be readily resummed.

With this result in mind, we move to the resummed SCET LP expression, obtained by evolving the hard and soft function to an equal scale. If we use the LP factorisation as in \eq{eq:choice1} one obtains the resummed partonic coefficient function
\begin{equation}
\label{eq:scetlp}
{\Delta}^{{\rm SCET,LP}}(z,Q,\mu_h,\mu_s,\mu) = H(Q,\mu_h)\,U(Q,\mu_h,\mu_s,\mu)\,\tilde{s}_a\Big(
\ln\tfrac{Q^2}{\mu_s^2}+\tfrac{\partial}{\partial \eta_a},\mu_s\Big)\frac{1}{(1-z)^{1-2\eta_a}}\frac{e^{-2\gamma_E\eta_a}}{\Gamma(2\eta_a)}\,,
\end{equation}
while using the modified scheme of \eq{eq:choice2}
we obtain
\begin{equation}
\label{eq:scetlpwoBN}
\widehat \Delta^{\rm SCET,LP}(z,Q,\mu_h,\mu_s,\mu) = H(Q,\mu_h)\,U(Q,\mu_h,\mu_s,\mu)\,\tilde{s}_a\Big(
\ln\tfrac{Q^2}{\mu_s^2}+\tfrac{\partial}{\partial \eta_a},\mu_s\Big)\frac{z^{-\eta_a}}{(1-z)^{1-2\eta_a}}\frac{e^{-2\gamma_E\eta_a}}{\Gamma(2\eta_a)}\,.
\end{equation}
In these equations the function $\eta_a$ is fixed to
\begin{eqnarray}
\label{eq:etaa}
\eta_a(\mu_s^2,\mu^2) = -\int_{\mu_s^2}^{\mu^2}\frac{{\rm d}\mu^2}{\mu^2}\Gamma_{{\rm cusp}, a}\left(\alpha_s(\mu^2)\right), 
\end{eqnarray}
after the derivative with respect to $\eta_a$ has been taken. We provide the process-dependent hard function $H(Q,\mu_h)$ and the Laplace transform of the soft function, $\tilde{s}_a\left(
\ln\frac{Q^2}{\mu_s^2}+\frac{\partial}{\partial \eta_a},\mu_s\right)$, for Drell-Yan and Higgs production in appendix~\ref{app:resummationcoeff}, up to NLO, which is sufficient to achieve resummation at NNLL accuracy. The corresponding expression up to NNLO, necessary to achieve resummation at N$^3$LL can be found in \cite{Becher:2007ty} and \cite{Ahrens:2008nc}, respectively~\footnote{Concerning Higgs production, note that we directly integrate out the top quark, and do not include a separate coefficient $C_t(m_t)$ with its corresponding evolution factor, as for instance in \cite{Ahrens:2008nc}. The reason is that here we are interested in comparing the SCET result with dQCD, where the factor $C_t(m_t)$ is included in the 
function $g_0$. Thus we include the fixed order $C_t(m_t)$ into $H_h(Q,\mu_h)$, where at NLO it contributes a factor $11/6$.}. In eqs.~(\ref{eq:scetlp}) and~(\ref{eq:scetlpwoBN}) the actual resummation is organized in the exponent $U$, given by~\cite{Becher:2007ty}
\begin{eqnarray}
\label{eq:Uscet}
U_{aa}(Q,\mu_h,\mu_s,\mu) = {\rm exp}\left[S_a(\mu_h^2,\mu_s^2)-a_{\gamma_ {V/S}}(\mu_h^2,\mu_s^2)+2a_{\gamma_a}\left(\mu_s^2,\mu^2\right)-a_{\Gamma_{{\rm cusp}, a}}(\mu_h^2,\mu_s^2)\ln\frac{Q^2}{\mu_h^2}\right],
\end{eqnarray}
with $U_{\rm DY} \equiv U_{qq}(Q,\mu_h,\mu_s,\mu)$ and $\gamma_{V/S} = \gamma_V$ (see appendix~\ref{app:resummationcoeff}) for DY production . For the Higgs production case we have $\gamma_{V/S} = \gamma_S$ and include an additional contribution~\cite{Ahrens:2008nc}
\begin{eqnarray}
\label{eq:UscetHiggs}
U_{{\rm h}}(Q,\mu_h,\mu_s,\mu) \equiv \frac{\alpha_s^2(\mu_s^2)}{\alpha_s^2(\mu^2)}\left[\frac{\beta(\alpha_s(\mu_s^2))/\alpha_s(\mu_s^2)}{\beta(\alpha_s(\mu_h^2))/\alpha_s(\mu_h^2)}\right]^2 U_{gg}(Q,\mu_h,\mu_s,\mu)\,,
\end{eqnarray}
whose role will become clear later. The resummation factor $U_{aa}(Q,\mu_h,\mu_s,\mu)$ is expressed in terms of the Sudakov exponent $S_a$ and the anomalous dimension exponents $a_{\gamma}$, defined as
\begin{eqnarray}
\label{eq:softfuncSCET}
S_a(Q^2,\mu_s^2)&=& -\int_{\alpha_s(Q^2)}^{\alpha_s(\mu_s^2)}\frac{{\rm d}\alpha_s}{\beta\left(\alpha_s\right)}\Gamma_{{\rm cusp}\,, a}(\alpha_s)\int_{\alpha_s(Q^2)}^{\alpha}\frac{{\rm d}\alpha_s'}{\beta(\alpha_s')}, \\
a_{\gamma}(\mu^2,\nu^2) &=& -\int_{\alpha_s(\mu^2)}^{\alpha_s(\nu^2)}\frac{{\rm d}\alpha_s}{\beta\left(\alpha_s\right)}\gamma(\alpha_s)\,.
\end{eqnarray}
In appendix~\ref{app:resummationcoeff} we provide the explicit expansion in $\as$ of both $S_a(Q^2,\mu_s^2)$ and $a_{\gamma_i}(\mu^2,\nu^2)$, as well as the anomalous dimensions they depend upon. Note that the functions $A_a(\alpha_s)$ (from dQCD) and $\Gamma_{{\rm cusp}, a}(\alpha_s)$ coincide up to $\mathcal{O}(\alpha_s^3)$
\begin{eqnarray}
\Gamma_{{\rm cusp}, a}(\alpha_s) = A_a(\alpha_s),
\end{eqnarray}
as can be seen by comparing the equations in appendix~\ref{app:resummationcoeff}. 

Comparing \eq{eq:scetlp} with \eq{eq:scetlpwoBN}
we see that the difference between the two schemes 
amounts to a factor of $z^{-\eta_a}$, which resums NLP NLL contribution of kinematic origin. The resummation formula defined in \eq{eq:scetlp} coincides with the one found in \cite{Becher:2007ty,Ahrens:2008nc}. What is new from our discussion above, is the fact that the NLP LL resummed contribution can be added consistently 
to both eq.~(\ref{eq:scetlp}), as was done in 
\cite{Beneke:2018gvs,Beneke:2019mua}, or to eq.~(\ref{eq:scetlpwoBN}), as we are doing here. The latter is particularly useful for the comparison with dQCD. The ultimate reason that allows one to proceed in this way is the fact that the NLP LL contribution to the kinematic soft function is unaltered by the different convention that leads from \eq{KinSoft:choice1} to  \eq{KinSoft:choice2}, which affects only NLLs at NLP. Thus, the LL contribution from the NLP kinematic soft function is added unaltered to the LL contribution from the NLP dynamical soft function, according to \eq{NLPSplitKinDyn}, preserving the universality of the total NLP LL contribution, whose resummed expression takes the form~\cite{Beneke:2019mua,Beneke:2018gvs}
\begin{equation}
\label{eq:NLPSCET}
\Delta^{\rm SCET,NLP}(z,Q,\mu_s,\mu) = -\frac{2C_a}{\pi b_0}\ln\frac{\alpha_s(\mu^2)}{\alpha_s(\mu_s^2)}{\rm exp}\left[S_{a,{\rm LL}}(Q^2,\mu^2)-S_{a,{\rm LL}}(\mu_s^2,\mu^2)\right],
\end{equation}
where the function $S_{\rm LL}(\mu^2,\nu^2)$ reads
\begin{eqnarray}
\label{eq:sll}
S_{a,{\rm LL}}(\mu^2,\nu^2)  = \frac{A^{(1)}_a}{b_0^2\pi}\left[\frac{1}{\alpha_s(\mu^2)}-\frac{1}{\alpha_s(\nu^2)} -\frac{1}{\alpha_s(\mu^2)}\ln\left(\frac{\alpha_s(\nu^2)}{\alpha_s(\mu^2)}\right) \right].
\end{eqnarray}
Interestingly, we can directly obtain the NLP LL SCET contribution from the LP LL one. The LP LL SCET contribution is obtained after setting $z^{-\eta_a} = 1$, $\tilde{s}_a = 1$ and $H(Q,\mu_h) = 1$. Furthermore $a_{\gamma} = 0$ at NLL, such that the evolution exponent \eq{eq:Uscet} simply becomes
\begin{eqnarray}
U_{aa,{\rm LL}}(Q,\mu_s) = {\rm exp}\left[S_{a,{\rm LL}}(Q^2,\mu_s^2)\right] = {\rm exp}\left[S_{a,{\rm LL}}(Q^2,\mu^2) - S_{a,{\rm LL}}(\mu_s^2,\mu^2)\right].
\end{eqnarray}
The NLP contribution can then be obtained from this evolution exponent via a derivative 
\begin{eqnarray}
\label{eq:derivSCET}
\Delta^{\rm SCET,NLP}(z,Q,\mu_s,\mu) &=&  -\beta(\alpha_s(\mu_s^2))\frac{\partial}{\partial\alpha_s(\mu_s^2)}U_{aa,{\rm LL}}(Q,\mu_s)\,. \\
 &=&  \frac{\beta\left(\alpha_s(\mu_s^2)\right)}{\alpha_s^2b_0^2}\frac{A^{(1)}_a}{\pi}\ln\frac{\alpha_s(\mu^2)}{\alpha_s(\mu_s^2)}{\rm exp}\left[S_{a,{\rm LL}}(Q^2,\mu^2) - S_{a,{\rm LL}}(\mu_s^2,\mu^2)\right].\nonumber
\end{eqnarray}
By now evaluating $\beta\left(\alpha_s(\mu_s^2)\right)$ to its lowest order and recognizing that $A_a^{(1)} = C_a$, we readily find that this result is equal to eq.~\eqref{eq:NLPSCET}. It is remarkable that we could perform a similar derivative trick in dQCD (eq.~\eqref{eq:Elpnlpll6}) to obtain the LL NLP results. There the derivative is taken with respect to $N$, while here it is taken with respect to the coupling evaluated at the soft scale. We will see in what follows that these forms are actually equal. 

\subsection{Analytical comparison of dQCD and SCET at NLP}

We are now in a position to relate resummation in dQCD vs SCET, where we closely follow ref.~\cite{Bonvini:2012az,Bonvini:2013td,Bonvini:2014qga}, where this was performed at LP, up to NNLL. We keep the same LP logarithmic accuracy, but we extend their analysis to NLP LL (in $N$-space). To compare resummation in dQCD vs SCET, it is convenient to convert the SCET resummation formula to Mellin space. We set the hard and factorisation scale ($\mu_h$ and $\mu$) equal to $Q$, and to obtain a function whose Mellin transform may be performed analytically, we assume that the soft scale $\mu_s$ is independent of $z$~\cite{Sterman:2013nya}. Furthermore, we will work with the expansion scheme in eq.~\eqref{eq:scetlpwoBN}, and comment on the role of $z^{-\eta_a}$ at the end of this section. Then we can write~\footnote{Note that this is only formally true when $\eta_a > 0$, otherwise we have to regulate eq.~\eqref{eq:scetlpwoBN} by using a $+$-description for the term involving $(1-z)^{-1+2\eta_a}$. The results remain unchanged in this procedure~\cite{Becher:2006nr}.}
\begin{eqnarray}
\label{eq:LPalmostSCET}
\bm{\widehat \Delta}^{\rm SCET,LP}(N,Q,\mu_s) &=& H(Q,Q)\,U(Q,\mu_s)\,\tilde{s}_a\left(\partial_{\eta_a}+\ln\frac{Q^2}{\mu_s^2},\mu_s\right)\frac{\Gamma(N-\eta_a)e^{-2\gamma_E\eta_a}}{\Gamma(N+\eta_a)}\,.
\end{eqnarray}
Up to NNLL accuracy we may use
\begin{eqnarray}
\label{eq:expansion_soft_function}
\tilde{s}_a = 1+C_a\frac{\alpha_s(\mu_s^2)}{2\pi}\left(\zeta(2) + \ln^2\frac{Q^2}{\mu_s^2} + 2\ln\frac{Q^2}{\mu_s^2}\partial_{\eta_a} + \partial_{\eta_a}^2\right)\,.
\end{eqnarray}
Furthermore, the ratio of $\Gamma$-functions in eq.~\eqref{eq:LPalmostSCET} may be expanded in the large-$N$ limit, resulting in
\begin{eqnarray}
\label{eq:ratiogammaexp}
\frac{\Gamma(N-\eta_a)e^{-2\gamma_E\eta_a}}{\Gamma(N+\eta_a)} = {\rm e}^{-2\eta_a \ln\bar{N}}\left[1+\frac{\eta_a}{N}+\mathcal{O}\left(\frac{1}{N^2}\right)\right]  .
\end{eqnarray}
By using eq.~\eqref{eq:expansion_soft_function} and eq.~\eqref{eq:ratiogammaexp} in eq.~\eqref{eq:LPalmostSCET}, we obtain
\begin{eqnarray}
\bm{\widehat \Delta}^{\rm SCET,LP}(N,Q,\mu_s) &=& H(Q,Q)\,U(Q,\mu_s)\,{\rm e}^{-2\eta_a\ln\bar{N}} \Bigg\{\frac{C_a}{N}\frac{\alpha_s(\mu_s^2)}{\pi}\left(\ln\frac{Q^2}{\mu_s^2}-\ln\bar{N}^2\right) \\
&& \hspace{1cm}+ \left(1+\frac{\eta_a}{N}\right)\left[1+C_a\frac{\alpha_s(\mu_s^2)}{2\pi}\left(\zeta(2) + \ln^2\frac{Q^2}{\bar{N}^2\mu_s^2} \right)\right]\Bigg\} + \mathcal{O}\left(\frac{1}{N^2}\right).\,\,\, \nonumber
\label{eq:LPSCET}
\end{eqnarray}
Recognizing $\tilde{s}_a\left(\ln^2\frac{Q^2}{\bar{N}^2\mu_s^2},\mu_s\right)$ in the second line we can write this as 
\begin{eqnarray}
\bm{\widehat  \Delta}^{\rm SCET,LP}(N,Q,\mu_s) &=& H(Q,Q)\,\tilde{s}_a\left(\ln^2\frac{Q^2}{\bar{N}^2\mu_s^2},\mu_s\right)U(Q,\mu_s)\,{\rm e}^{-2\eta_a\ln\bar{N}}\left(1+\frac{\eta_a}{N}\right) \nonumber \\
&& \hspace{2cm} + \frac{C_a}{N}\frac{\alpha_s(\mu_s^2)}{\pi}H(Q,Q)\,U(Q,\mu_s)\,{\rm e}^{-2\eta_a\ln\bar{N}}\ln\frac{Q^2}{\bar{N}^2\mu_s^2}\,.\label{eq:scetstrictlyLP}
\end{eqnarray}
For the NLP SCET result in Mellin space we use eq.~\eqref{eq:NLPSCET} and take $\mu_s$ again independent of $z$, which results in
\begin{equation}
\bm{\Delta}^{\rm SCET,NLP}(N,Q,\mu_s) = -\frac{2C_a}{\pi b_0}\frac{1}{N}\ln\frac{\alpha_s(Q^2)}{\alpha_s(\mu_s^2)}{\rm exp}\left[-S_{a,{\rm LL}}(\mu_s^2,Q^2)\right].
\label{eq:NLPSCETN}
\end{equation}

We now compare eq.~\eqref{eq:scetstrictlyLP} and eq.~\eqref{eq:NLPSCETN} to $\bm{\Delta}^{\rm dQCD}$ as defined in eq.~\eqref{eq:resumgeneral} with $\bm{E}_a^{\rm LP+NLP}$ in eq.~\eqref{eq:Elpnlpll7} and $\bm{D}_{aa}^{\rm LP+NLP}$ in eq.~\eqref{eq:wideref}, starting with the former.
 By defining the derivative operator
\begin{eqnarray}
D_{\Gamma} \equiv \left(1+\frac{1}{2}\frac{\partial}{\partial N}\right){\rm e}^{-\gamma_E\frac{\partial}{\partial \ln N}}\Gamma\left(1-\frac{\partial}{\partial \ln N}\right),
\end{eqnarray}
eq.~\eqref{eq:Elpnlpll7} may be rewritten as
\begin{eqnarray}
\bm{E}^{\rm LP+NLP}_a &=& \frac{1}{2}D_{\Gamma}\int_{Q^2/\bar{N}^2}^{Q^2} \frac{{\rm d}\mu^2}{\mu^2} \int_{\alpha_s(Q^2)}^{\alpha_s(\mu^2)}\frac{{\rm d}\alpha_s}{\beta(\alpha_s)}A_a(\alpha_s)  \\
&=& \frac{1}{2}D_{\Gamma}\int_{\alpha_s(Q^2/\bar{N}^2)}^{\alpha_s(Q^2)}\frac{{\rm d}\alpha_s}{\beta(\alpha_s)}A_a(\alpha_s) \int_{\alpha_s(Q^2/\bar{N}^2)}^{\alpha_s}\frac{{\rm d}\alpha_s(\mu^2)}{\beta(\alpha_s(\mu^2))},\nonumber
\end{eqnarray}
where the order of integration of the two  integrals is interchanged on the second line. We now split up the integration domain of the two integrals on the second line by introducing an auxiliary scale $\mu_s$ for the first, and an intermediate scale $Q$ for the second integral. We obtain
\begin{eqnarray}
\bm{E}^{\rm LP+NLP}_a &=& \frac{1}{2}D_{\Gamma}\left(\int_{\alpha_s(Q^2/\bar{N}^2)}^{\alpha_s(\mu_s^2)}+\int_{\alpha_s(\mu_s^2)}^{\alpha_s(Q^2)}\right)\frac{{\rm d}\alpha_s}{\beta(\alpha_s)}A_a(\alpha_s) \nonumber\\
&& \hspace{5cm}\times \left(\int_{\alpha_s(Q^2)}^{\alpha_s}+\int_{\alpha_s(Q^2/\bar{N}^2)}^{\alpha_s(Q^2)}\right)\frac{{\rm d}\alpha_s(\mu^2)}{\beta(\alpha_s(\mu^2))} \nonumber \\
&\equiv& \bm{E}^{\rm Landau}_a + \bm{E}^{\rm SCET}_a, 
\end{eqnarray}
with 
\begin{eqnarray}
\bm{E}^{\rm SCET}_a = \frac{1}{2}D_{\Gamma}\int_{\alpha_s(\mu_s^2)}^{\alpha_s(Q^2)}\frac{{\rm d}\alpha_s}{\beta(\alpha_s)}A_a(\alpha_s)\int_{\alpha_s(Q^2)}^{\alpha_s}\frac{{\rm d}\alpha_s(\mu^2)}{\beta(\alpha_s(\mu^2))} = \frac{1}{2}D_{\Gamma}S_a(Q^2,\mu_s^2)\,,
\end{eqnarray}
where we recognize the SCET Sudakov exponent $S_a(Q,\mu_s)$ of eq.~\eqref{eq:softfuncSCET}, and 
\begin{eqnarray}
\bm{E}^{\rm Landau}_a &=& \frac{1}{2}D_{\Gamma}\Bigg[\int_{\alpha_s(Q^2/\bar{N}^2)}^{\alpha_s(\mu_s^2)}\frac{{\rm d}\alpha_s}{\beta(\alpha_s)}A_a(\alpha_s)\int_{\alpha_s(Q^2/\bar{N}^2)}^{\alpha_s}\frac{{\rm d}\alpha_s(\mu^2)}{\beta(\alpha_s(\mu^2))} \nonumber \\
&& \hspace{4cm}+\int_{\alpha_s(\mu_s^2)}^{\alpha_s(Q^2)}\frac{{\rm d}\alpha_s}{\beta(\alpha_s)}A_a(\alpha_s)\int_{\alpha_s(Q^2/\bar{N}^2)}^{\alpha_s(Q^2)}\frac{{\rm d}\alpha_s(\mu^2)}{\beta(\alpha_s(\mu^2))}\Bigg]. 
\end{eqnarray}
We call the latter function $\bm{E}^{\rm Landau}_a$ as it generates the branch cut starting at $\bar{N}_L = {\rm exp}\left[\frac{1}{2\alpha_s b_0}\right]{\rm exp}\left[\gamma_E\right]$ that is present in the dQCD result. This  contribution seems to be absent in eq.~\eqref{eq:LPSCET}, but actually it is contained by the ratio of $\Gamma$-functions (eq.~\eqref{eq:ratiogammaexp}). To see this, we set $\mu_s = Q/\bar{N}$, as motivated by ref.~\cite{Sterman:2013nya,Bonvini:2014qga}. We then find that 
\begin{eqnarray}
\label{eq:landaueasy}
\bm{E}^{\rm Landau}_a &=& -\frac{1}{2}D_{\Gamma}\Bigg[\eta_a\int_{\alpha_s(Q^2/\bar{N}^2)}^{\alpha_s(Q^2)}\frac{{\rm d}\alpha_s(\mu^2)}{\beta(\alpha_s(\mu^2))}\Bigg],
\end{eqnarray}
where we have used
\begin{eqnarray}
\int_{\alpha_s(Q^2/\bar{N}^2)}^{\alpha_s(Q^2)}\frac{{\rm d}\alpha_s}{\beta(\alpha_s)}A_a(\alpha_s) = \int_{Q/\bar{N}^2}^{Q^2}\frac{{\rm d}\mu^2}{\mu^2}\Gamma^{\rm cusp}_a(\alpha_s(\mu^2) = - \eta_a\,.
\end{eqnarray}
The integral in eq.~\eqref{eq:landaueasy} can be recognized to give
\begin{eqnarray}
\int_{\alpha_s(Q^2/\bar{N}^2)}^{\alpha_s(Q^2)}\frac{{\rm d}\alpha_s(\mu^2)}{\beta(\alpha_s(\mu^2))} = 2\ln\bar{N}.
\end{eqnarray}
Substituting this result in  eq.~\eqref{eq:landaueasy}, we obtain
\begin{eqnarray}
\label{eq:landau1}
\bm{E}^{\rm Landau}_a &=& -D_{\Gamma}\eta_a\ln\bar{N}.
\end{eqnarray}
With $\eta_a = \frac{A^{(1)}_a}{\pi b_0}\ln(1-2\lambda)$ at the first non-trivial order, we indeed observe that this factor gives rise to the branch cut starting at $\lambda = 1/2$. 
Given that we can write
\begin{eqnarray}
\label{eq:DgammaNLPNNLL}
D_{\Gamma} = \left(1+\frac{1}{2}\frac{\partial}{\partial N}\right)\left[1 + \frac{1}{2}\zeta(2)\left(\frac{\partial}{\partial \ln N}\right)^2+\dots\right],
\end{eqnarray}
where we set $D_{\Gamma} = 1$ at LP LL and NLL, we have now established that
\begin{eqnarray}
{\rm exp}\Bigg[2 \bm{E}^{\rm SCET}_a\Big|_{D_{\Gamma}=1} + 2 \bm{E}^{\rm Landau}_a\Big|_{D_{\Gamma}=1} \Bigg] = {\rm exp}\left[S_a(Q^2,Q^2/\bar{N}^2)\right]{\rm exp}\left[-2\eta_a\ln\bar{N}\right],
\end{eqnarray}
where the first factor is contained in $U(Q,\mu_s = Q/\bar{N})$, while the second one is generated by the ratio of $\Gamma$-functions (eq.~\eqref{eq:ratiogammaexp}). 

We can also relate $\bm{D}^{\rm LP+NLP}_{aa}$ (which is equal to $\bm{D}^{\rm LP}_{aa}$ up to NLP NNLL accuracy) to its SCET counterpart. From eq.~\eqref{eq:Uscet}, we see that this must be contained in $-a_{\gamma_S}(Q^2,\mu_s^2)+2a_{\gamma_a}(\mu_s^2,Q^2)$. We first write
\begin{eqnarray}
\label{eq:agammasum}
-a_{\gamma_S}(Q^2,\mu_s^2)+2a_{\gamma_a}(\mu_s^2,Q^2) = -\int_{\alpha_s(\mu_s^2)}^{\alpha_s(Q^2)}\frac{{\rm d}\alpha_s}{\beta(\alpha_s)}\left(\gamma_S+2\gamma_a\right). 
\end{eqnarray}
Let us first examine the NNLL DY case, for which
\begin{eqnarray}
\gamma_{V}^{(0)}+2\gamma_q^{(0)} &=& 0\,, \\
\gamma_{V}^{(1)}+2\gamma_q^{(1)} &=& 8D^{(2)}_{qq} - \pi^2\beta_0 C_F\,, \nonumber
\end{eqnarray}
which may derived by adding the explicit forms of these coefficients as given in appendix~\ref{app:resummationcoeff}.
We then find for eq.~\eqref{eq:agammasum}
\begin{eqnarray}
-a_{\gamma_S}(Q^2,Q^2/\bar{N}^2)+2a_{\gamma_a}(Q^2/\bar{N}^2,Q^2)
&=& \frac{8D^{(2)}_{qq} - \pi^2\beta_0 C_F}{(4\pi)^2b_0}\left(\alpha_s - \alpha_s(Q^2/\bar{N}^2)\right) \\
&=& -\alpha_s\left[\frac{D^{(2)}_{qq}}{\pi^2 b_0}\frac{\lambda}{1-2\lambda} - 3\zeta(2)\frac{C_F}{\pi}\frac{\lambda}{1-2\lambda}\right]\, \nonumber  \\
&=& \bm{D}_{qq}^{\rm LP} + 3\alpha_s\zeta(2)\frac{C_F}{\pi}\frac{\lambda}{1-2\lambda}\,.\nonumber
\end{eqnarray}
We shall return to the role of this last term, but we first turn our attention to the Higgs production case, where the story is somewhat more involved. For the Higgs case we have
\begin{eqnarray}
\gamma_{S}^{(0)}+2\gamma_g^{(0)} &=& 2\beta_0\,, \\
\gamma_{S}^{(1)}+2\gamma_g^{(1)} &=& 8D^{(2)}_{gg}  + 4 \beta_1 - \pi^2\beta_0 C_A\,, \nonumber
\end{eqnarray}
so that
\begin{eqnarray}
\label{eq:ahiggs}
-a_{\gamma_{S}}(Q^2,\mu_s^2)+2a_{\gamma_g}(\mu_s^2,Q^2)
&=& \\
&& \hspace{-1cm}2\ln\left(\frac{\alpha_s}{\alpha_s(\mu_s^2)}\right)+\frac{2\beta_1}{\beta_0}\frac{\alpha_s-\alpha_s(\mu_s^2)}{4\pi} + \left(8D^{(2)}_{gg}  - \pi^2\beta_0 C_A\right)\frac{\alpha_s-\alpha_s(\mu_s^2)}{4\pi\beta_0}\,.\nonumber
\end{eqnarray}
We have an apparent discrepancy with the DY case, but must remember that $U_{\rm h}$ (eq.~\eqref{eq:UscetHiggs}) has an additional prefactor that can be written as
\begin{eqnarray}
\frac{\alpha_s^2(\mu_s)^2}{\alpha_s^2}\left[\frac{\beta(\alpha_s(\mu_s^2))/\alpha_s(\mu_s^2)}{\beta(\alpha_s)/\alpha_s}\right]^2  &=& {\rm exp}\left[2\ln\left(\frac{\alpha_s(\mu_s^2)}{\alpha_s}\right)+2\ln\left(\frac{\beta(\alpha_s(\mu_s^2))/\alpha_s^2(\mu_s^2)}{\beta(\alpha_s)/\alpha_s^2}\right)\right]\\
&=& {\rm exp}\left[2\ln\left(\frac{\alpha_s(\mu_s^2)}{\alpha_s}\right)+2\frac{b_1}{b_0}\left(\alpha_s(\mu_s^2)-\alpha_s\right)+\mathcal{O}(\alpha_s^2)\right], \nonumber 
\end{eqnarray}
By substituting $\beta_1 = (4\pi)^2b_1$ and $\beta_0 = 4\pi b_0$, we observe that this factor cancels the first two terms of eq.~\eqref{eq:ahiggs}, such that up to NNLL and for $\mu_s = Q/\bar{N}$ we have
\begin{eqnarray}
\frac{\alpha_s^2(\mu_s)^2}{\alpha_s^2}\left[\frac{\beta(\alpha_s(\mu_s^2))/\alpha_s(\mu_s^2)}{\beta(\alpha_s)/\alpha_s}\right]^2{\rm exp}\left[-a_{\gamma_{S}}(Q^2,\mu_s^2)+2a_{\gamma_g}(\mu_s^2,Q^2)\right] && \\
&& \hspace{-4cm}= {\rm exp}\left[-\alpha_s\left(\frac{D^{(2)}_{gg}}{\pi^2 b_0}\frac{\lambda}{1-2\lambda} - 3\zeta(2)\frac{C_A}{\pi}\frac{\lambda}{1-2\lambda}\right)\right] \nonumber \\
&& \hspace{-4cm}=  {\rm exp}\left[\bm{D}_{gg}^{\rm LP} + 3\alpha_s\zeta(2)\frac{C_A}{\pi}\frac{\lambda}{1-2\lambda}\right]\,.\nonumber
\end{eqnarray}
After these first two steps, we have established that up to NLL accuracy, and for $\mu_s = Q/\bar{N}$ 
\begin{eqnarray}
\label{eq:ufuncresultLL}
U_{\mathrm{h/DY}}(Q,\mu_s){\rm e}^{-2\eta_a\ln\bar{N}} = {\rm exp}\left[\bm{D}^{\rm LP}_{aa}+ 2 \bm{E}^{\rm SCET}_a\Big|_{D_\Gamma = 1} + 2 \bm{E}^{\rm Landau}_a\Big|_{D_\Gamma = 1} + 3\alpha_s\zeta(2)\frac{C_a}{\pi}\frac{\lambda}{1-2\lambda}\right].\,\,\,\,\, \end{eqnarray}
By now truncating $D_{\Gamma}$ to LP NNLL instead (eq.~\eqref{eq:DgammaNLPNNLL} without the partial derivative towards $N$, which is of NLP), we obtain
\begin{eqnarray}
\label{eq:ufuncresult}
U_{\mathrm{h/DY}}(Q,\mu_s){\rm e}^{-2\eta_a\ln\bar{N}} = {\rm exp}\left[\bm{D}^{\rm LP}_{aa}+ 2 \bm{E}^{\rm SCET}_a\Big|_{\rm LP} + 2 \bm{E}^{\rm Landau}_a\Big|_{\rm LP} - \alpha_s\zeta(2)\frac{C_a}{\pi}\frac{\lambda}{1-2\lambda}\right],
\end{eqnarray}
Note that $D_{\Gamma}$ gives an additional factor of $\zeta(2)$ at NNLL (see also eq.~\eqref{eq:NNLLcontribution}), turning the $3\zeta(2)$ discrepancy into a $-\zeta(2)$ difference between the wide-angle contribution $D_{aa}$ for dQCD, and the sum of anomalous dimensions for SCET. We will see that this remaining (last) term is resolved after including the hard function of SCET and dQCD.

In eq.~\eqref{eq:ufuncresult}, we have achieved part of the LP result of  eq.~\eqref{eq:scetstrictlyLP}. What remains is to relate the hard function of dQCD, $g_0(\alpha_s)$, to the product $H(Q,Q)\,\tilde{s}_a(0,Q/\bar{N})$ in SCET (where we set $\mu_s = Q/\bar{N}$, such that the second line of eq.~\eqref{eq:scetstrictlyLP} is $0$). This comparison may be done easily. For DY we have (see eq.~\eqref{eq:hardfuncDY} of appendix~\ref{app:resummationcoeff})
\begin{eqnarray}
\label{eq:DYHdQCD}
g^{\rm DY}_0(\alpha_s) = 1+C_F\frac{\alpha_s}{\pi}\left(4\zeta(2)-4\right),
\end{eqnarray}
for dQCD, and for SCET (second line of eq.~\eqref{eq:hardfuncSCETdyhiggs})
\begin{eqnarray}
\label{eq:DYHSCET}
H_{\rm DY}(Q,Q)\tilde{s}_a(0,Q/\bar{N}) &=& 1 + C_F\frac{\alpha_s}{\pi}\left(4\zeta(2)-4\right) + C_F\frac{\alpha_s(Q^2/\bar{N}^2)-\alpha_s}{2\pi}\zeta(2) + \mathcal{O}(\alpha_s^2)\, \\
&=& \left(1 + C_F\frac{\alpha_s}{\pi}\left(4\zeta(2)-4\right)\right){\rm exp}\left[\alpha_s\zeta(2)\frac{C_F}{\pi}\frac{\lambda}{1-2\lambda}\right] + \mathcal{O}(\alpha_s^2)\,. \nonumber
\end{eqnarray}
For the Higgs case we have  (eq.~\eqref{eq:hardfunchiggs})
\begin{eqnarray}
\label{eq:higgsHdQCD}
g^{\rm h}_0(\alpha_s) = 1+C_A\frac{\alpha_s}{\pi}\left(\frac{11}{6}+4\zeta(2)\right),
\end{eqnarray}
for dQCD, and for SCET (first line of eq.~\eqref{eq:hardfuncSCETdyhiggs})
\begin{eqnarray}
\label{eq:higgsHSCET}
H_{\rm h}(Q,Q)\,\tilde{s}_{\rm h}(0,Q/\bar{N}) &=& \left(1 + C_A\frac{\alpha_s}{\pi}\left(\frac{11}{6}+4\zeta(2)\right)\right){\rm exp}\left[\alpha_s\zeta(2)\frac{C_A}{\pi}\frac{\lambda}{1-2\lambda}\right] + \mathcal{O}(\alpha_s^2)\,.\,\,\,\,\,\,\,
\end{eqnarray}
Note that in the SCET hard functions of eq.~\eqref{eq:DYHSCET} and~\eqref{eq:higgsHSCET} there is an additional exponent with respect to the dQCD hard functions of eq.~\eqref{eq:DYHdQCD} and~\eqref{eq:higgsHdQCD}. These contributions precisely cancel the last term in eq.~\eqref{eq:ufuncresult}. By now combining these results in eq.~\eqref{eq:LPSCET}, we find
\begin{eqnarray}
\hat{\bm{\Delta}}^{\rm SCET,LP}(N,Q,Q/\bar{N}) &=& g_0(\alpha_s)\,{\rm exp}\left[\bm{D}^{\rm LP}_{aa}+ 2 \bm{E}^{\rm SCET}_a\Big|_{\rm LP} + 2 \bm{E}^{\rm Landau}_a\Big|_{\rm LP} \right] \left(1+\frac{\eta_a}{N}\right) \\
&=& g_0(\alpha_s)\,{\rm exp}\left[\frac{2}{\alpha_s}g_a^{(1)}(\lambda) + 2g_a^{(2)}(\lambda) + 2\alpha_s g_a^{(3)}(\lambda)  \right] \left(1+\frac{\eta_a}{N}\right). \nonumber
\end{eqnarray}
By comparing this result to the dQCD result of eq.~\eqref{eq:resumgeneral} with $h^{(1)}$ set to zero, we see that the difference between dQCD and SCET at LP consists of the $\mathcal{O}(\eta_a/N)$-term. This term is included in the LP SCET result, but absent in the LP dQCD result. Therefore, we see that the difference between dQCD and SCET starts at NLP in Mellin space, as was noted before in ref.~\cite{Bonvini:2012az,Bonvini:2013td,Bonvini:2014qga}. 

We now study how the NLP LL contribution modifies this result. First we see that $\eta_a$ (eq.~\eqref{eq:etaa}) at the first logarithmic order is given by  
\begin{eqnarray}
\eta_a = \frac{A^{(1)}_a}{\pi b_0}\ln\frac{\alpha_s(Q^2)}{\alpha_s(Q^2/\bar{N}^2)} = \frac{A^{(1)}_a}{\pi b_0}\ln(1-2\lambda).
\end{eqnarray}
Moreover, we set $S_{a,{\rm LL}}(Q^2/\bar{N}^2,Q^2) = -\frac{2}{\alpha_s}g^{(1)}(\lambda)$ using eq.~\eqref{eq:sll}.
With this, we find that the NLP SCET result of eq.~\eqref{eq:NLPSCETN}  can be written in Mellin space as
\begin{eqnarray}
\bm{\Delta}^{\rm SCET,NLP}(N,Q,Q/\bar{N}) &=&  {\rm exp}\left[\frac{2}{\alpha_s}g^{(1)}_a(\lambda)\right]\left(1-\frac{2A^{(1)}_a}{\pi b_0}\frac{\ln(1-2\lambda)}{N}\right),
\end{eqnarray}
where we have used the result of eq.~\eqref{eq:sll}. 
Adding this to the LP result restricted to LL (where $H = \tilde{s} = 1$, and $a_{\gamma} = 0$), we find
\begin{eqnarray}
\bm{\Delta}^{\rm SCET,LP+NLP,LL}(N,Q,Q/\bar{N}) 
&=& {\rm exp}\left[\frac{2}{\alpha_s}g^{(1)}_a(\lambda)\right]\left(1-\frac{A^{(1)}_a}{\pi b_0}\frac{\ln(1-2\lambda)}{N}\right). \end{eqnarray}
The NLP correction in the second factor is precisely the first-order expansion of the NLP exponent $h_a^{(1)}(\lambda)$ of dQCD. Hence, at NLP LL, the contributions from SCET and dQCD are the same if one sets $\mu_s = Q/\bar{N}$. Differences between the two formalisms originate from i) truncations of higher-logarithmic terms, ii) truncations on higher-power terms (i.e.~a factor of $1/N^2$ is included in dQCD once the NLP exponent is expanded up to $\mathcal{O}(\alpha_s^2)$ or beyond, while it is missing in the SCET formalism), and iii) running-$\alpha_s$ effects.

We recall that we can obtain the NLP $z$-space contribution by deriving the LP LL result towards $\alpha_s(\mu_s)$ (eq.~\eqref{eq:derivSCET}). We now comment on how this compares to the dQCD result, where the NLP $N$-space result can be obtained via a derivative of the LP LL term towards $N$. The $N$-space form of eq.~\eqref{eq:derivSCET} is
\begin{eqnarray}
\label{eq:derivSCETN}
\mathbf{\Delta}^{\rm SCET,NLP}(z,Q,\mu_s,\mu) &=&  -\frac{1}{N}\beta(\alpha_s(\mu_s^2))\frac{\partial}{\partial\alpha_s(\mu_s^2)}U_{aa,{\rm LL}}(Q,\mu_s)\,,
\end{eqnarray}
as these functions do not depend on $z$. Using the definition of the $\beta$-function and setting $\mu_s = Q/\bar{N}$, the this equation can be rewritten as
\begin{eqnarray}
\mathbf{\Delta}^{\rm SCET,NLP}(z,Q,\mu_s,\mu) = \frac{1}{2}\frac{\partial}{\partial N} U_{aa,{\rm LL}}(Q,Q/\bar{N})\,.
\end{eqnarray}
Since we have established that $U_{aa,{\rm LL}}(Q,Q/\bar{N}) = {\rm exp}\left[\frac{2}{\alpha_s}g^{(1)}(\lambda)\right]$, we find that eq.~\eqref{eq:derivSCET} is equal to the derivative trick in dQCD (eq.~\eqref{eq:derivativetrick}) at the first-order expansion of the exponent. 

We have now established that the contributions from SCET and dQCD are the same at NLP LL accuracy if one sets $\mu_s = Q/\bar{N}$. Two important remarks are in order. Firstly, for our SCET starting point we have included a factor of $z^{-\eta_a}$ in eq.~\eqref{eq:scetlpwoBN}. If we would not include this factor as in eq.~\eqref{eq:scetlp}, the ratio of $\Gamma$-functions in eq.~\eqref{eq:ratiogammaexp} would instead be
\begin{eqnarray}
\label{eq:ratiogamma}
\frac{\Gamma(N)}{\Gamma(N+2\eta_a)}{\rm e}^{-2\gamma_E\eta_a} &=& {\rm e}^{-2\ln\bar{N}\eta_a}\left(1+\frac{\eta_a-2\eta_a^2}{N}\right).
\end{eqnarray}
The factor of $\eta_a^2$ is removed by including $z^{-\eta_a}$, which has a phase-space origin. In dQCD, we recognize this factor in the $\mathcal{J}^{\prime}_k$ contribution in eq.~\eqref{ELL2}, which also has a phase-space origin: it comes from the $\ln(z)$ expansion around $z=1$ as performed in eq.~\eqref{eq:expzdep}. This factor is needed in dQCD to obtain the NLP contribution directly from performing a derivative on the LP contribution, as shown in eq.~\eqref{derID}. However, even without performing the derivative trick, we truncate the dQCD NLP result at strictly NLP LL (resulting in the function $h^{(1)}_a$). If one would omit the factor of $z^{-\eta_a}$ in the SCET resummation formula, one would introduce an NLP NLL term via the factor of $\eta_a^2$. Therefore, by including this factor, the SCET and dQCD formalisms are put on equal grounds. We shall see that the numerical impact of this factor is sizeable. 

Secondly, although we have found analytical agreement at LP  + NLP LL accuracy between dQCD and SCET , we will see that there are numerical differences, which especially grow sizable when the LP part is evaluated at NNLL. This is caused by the fact that in dQCD the NLP contribution is \emph{multiplied} by the hard coefficient $g_0(\alpha_s)$, whereas in SCET, the NLP LL result is \emph{added} to the LP result. Indeed, in $\bm{\Delta}^{\rm SCET,NLP}$, we do not see a hard function, so for SCET the NLP result will not get enhanced by constant contributions. This is simply the result of a choice, i.e. of including strictly NLP LL logarithms in $z$-space in the SCET NLP resummation formula \eq{eq:NLPSCET}. As a remedy, we can upgrade the NLP SCET result for both DY and Higgs by using
\begin{equation}
\label{eq:NLPSCETwithH}
\bar{\Delta}^{\rm DY/h, SCET,NLP}(z,Q,\mu_s,\mu) = -\frac{2C_a}{\pi b_0}\ln\frac{\alpha_s(\mu^2)}{\alpha_s(\mu_s^2)}\,H_{\rm DY/h}(Q,Q){\rm exp}\,\left[S_{a,{\rm LL}}(Q^2,\mu^2)-S_{a,{\rm LL}}(\mu_s^2,\mu^2)\right].
\end{equation}
We will explore the numerical consequences of this in the next section. 

\begin{figure}[t]
\centering
\mbox{
    \centering
    \begin{subfigure}{.49\textwidth}
    \centering
        \includegraphics[width=\textwidth]{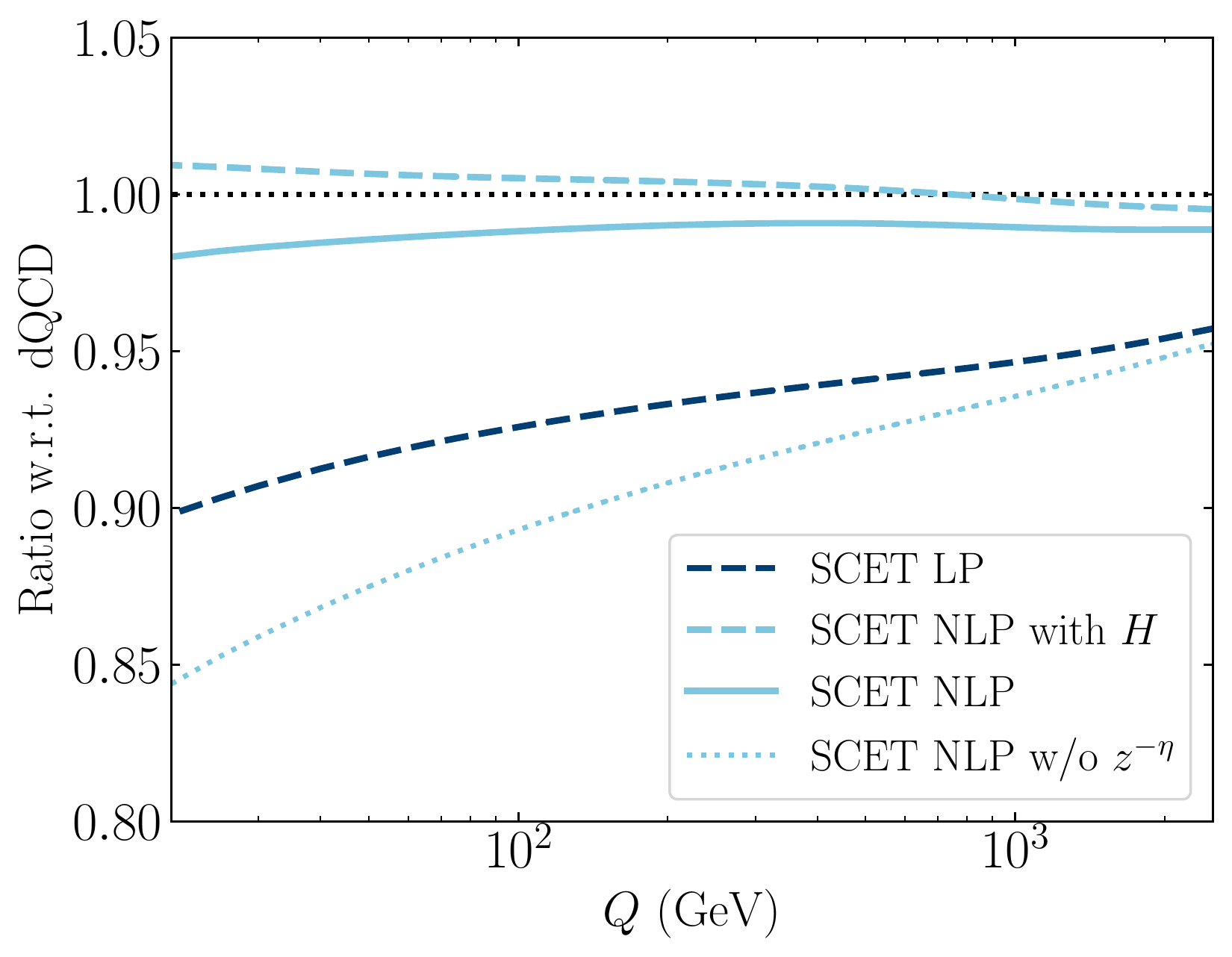}
  \label{fig:scetvspqcdDYNNLL}
    \end{subfigure}
    \hspace{0.2cm}
    \begin{subfigure}{.49\textwidth}
        \includegraphics[width=\textwidth]{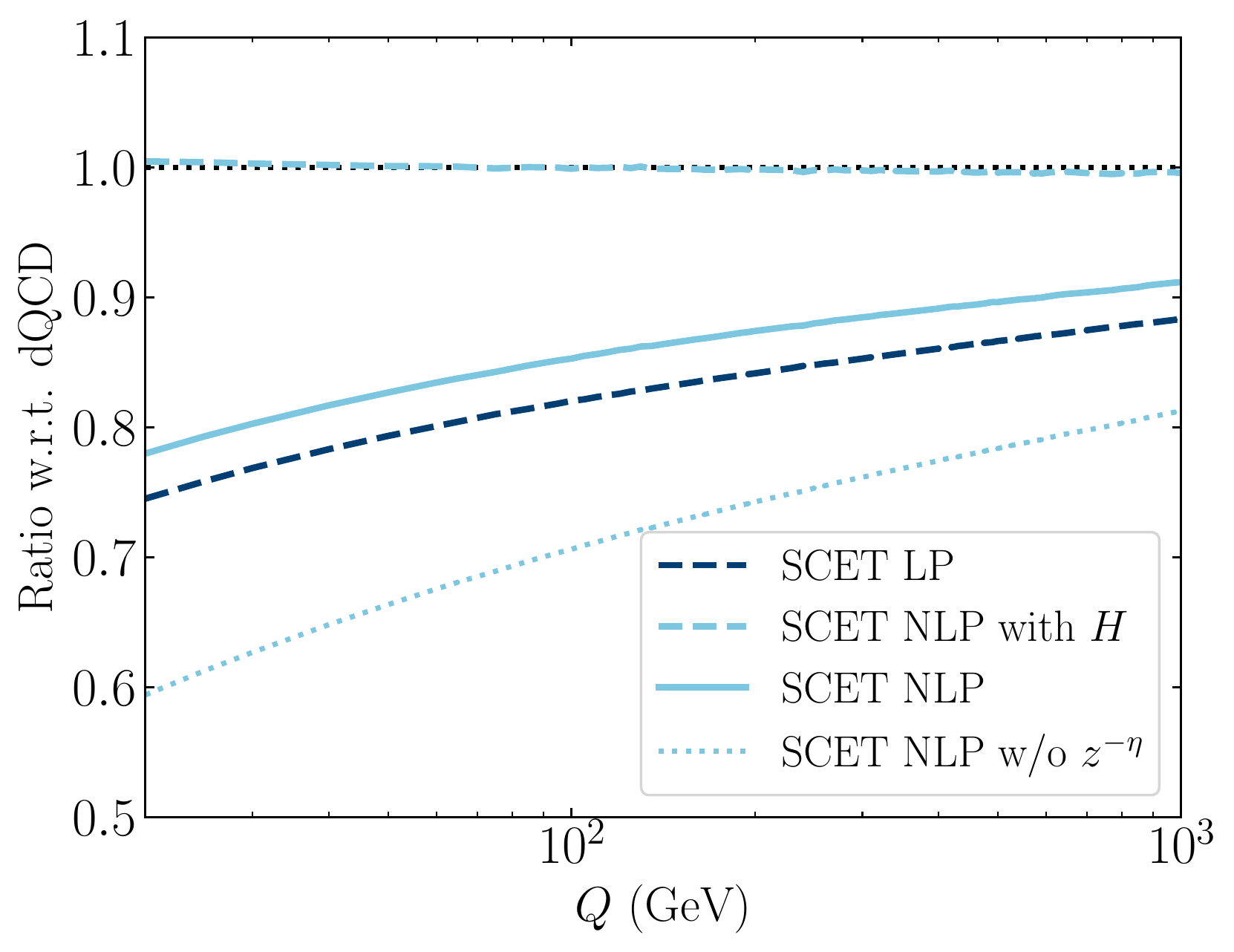}
  \label{fig:scetvspqcdhiggsNNLL}
    \end{subfigure} }
    \caption{Ratio plots for the NNLL results for DY (left) and Higgs (right) with respect to the dQCD distributions. The SCET LP result at NNLL (dark-blue dashed line) is divided by the dQCD result at LP NNLL. The SCET NLP results indicated by the light-blue lines are obtained via eq.~\eqref{eq:NLPSCETwithH} (dashed), eq.~\eqref{eq:NLPSCET} (dash-dotted) and eq.~\eqref{eq:NLPSCET} (dotted), and divided by the dQCD result at LP NNLL + NLP LL. }
  \label{fig:scetvspqcdNNLL}
\end{figure}
\subsection{Numerical comparison of dQCD and SCET resummation at NLP}

\begin{figure}[th!]
\mbox{
    \centering
    \begin{subfigure}{.49\textwidth}
    \centering
        \includegraphics[width=\textwidth]{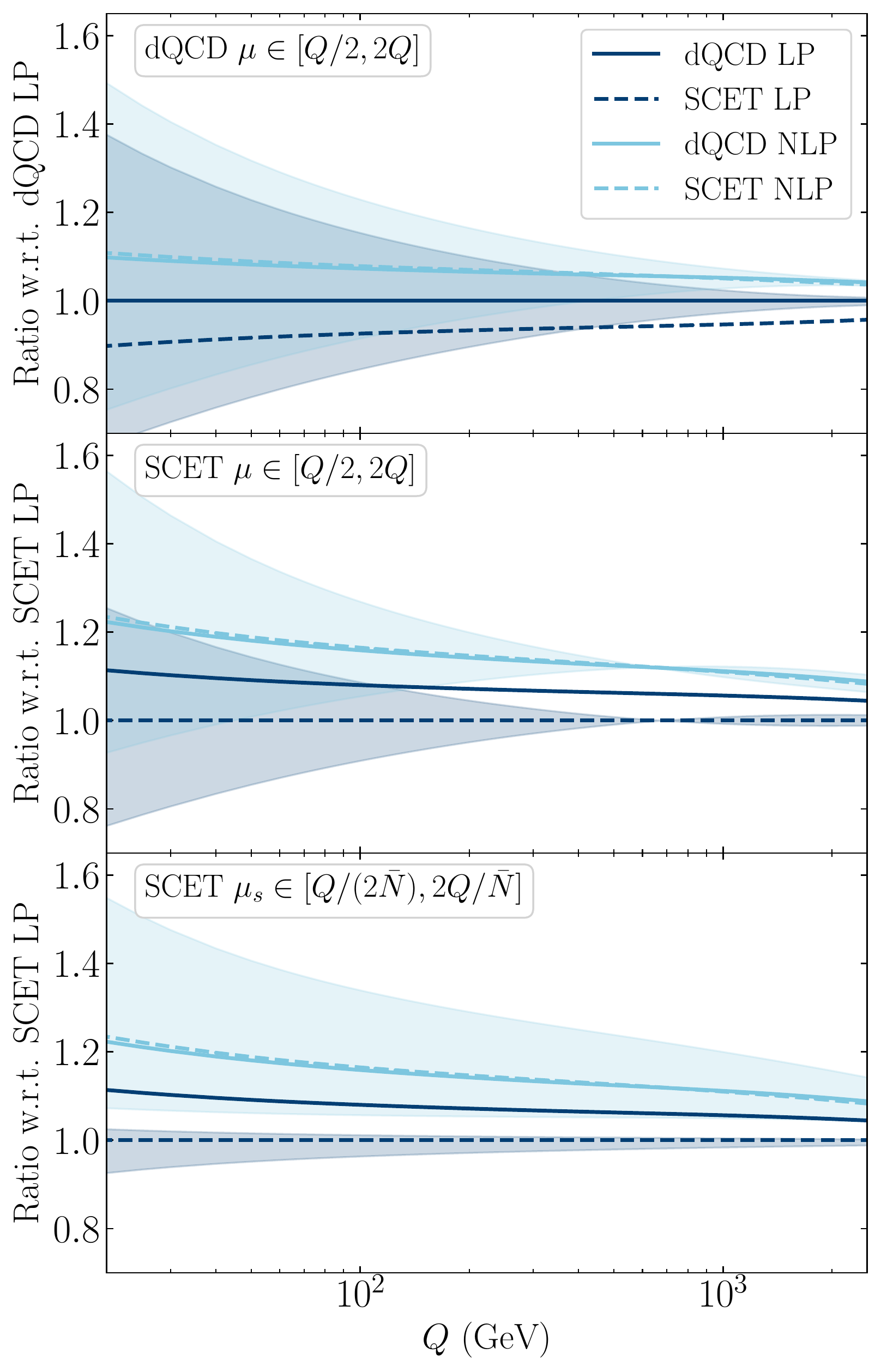}
    \end{subfigure}
    \hspace{0.1cm}
    \begin{subfigure}{.50\textwidth}
        \includegraphics[width=\textwidth]{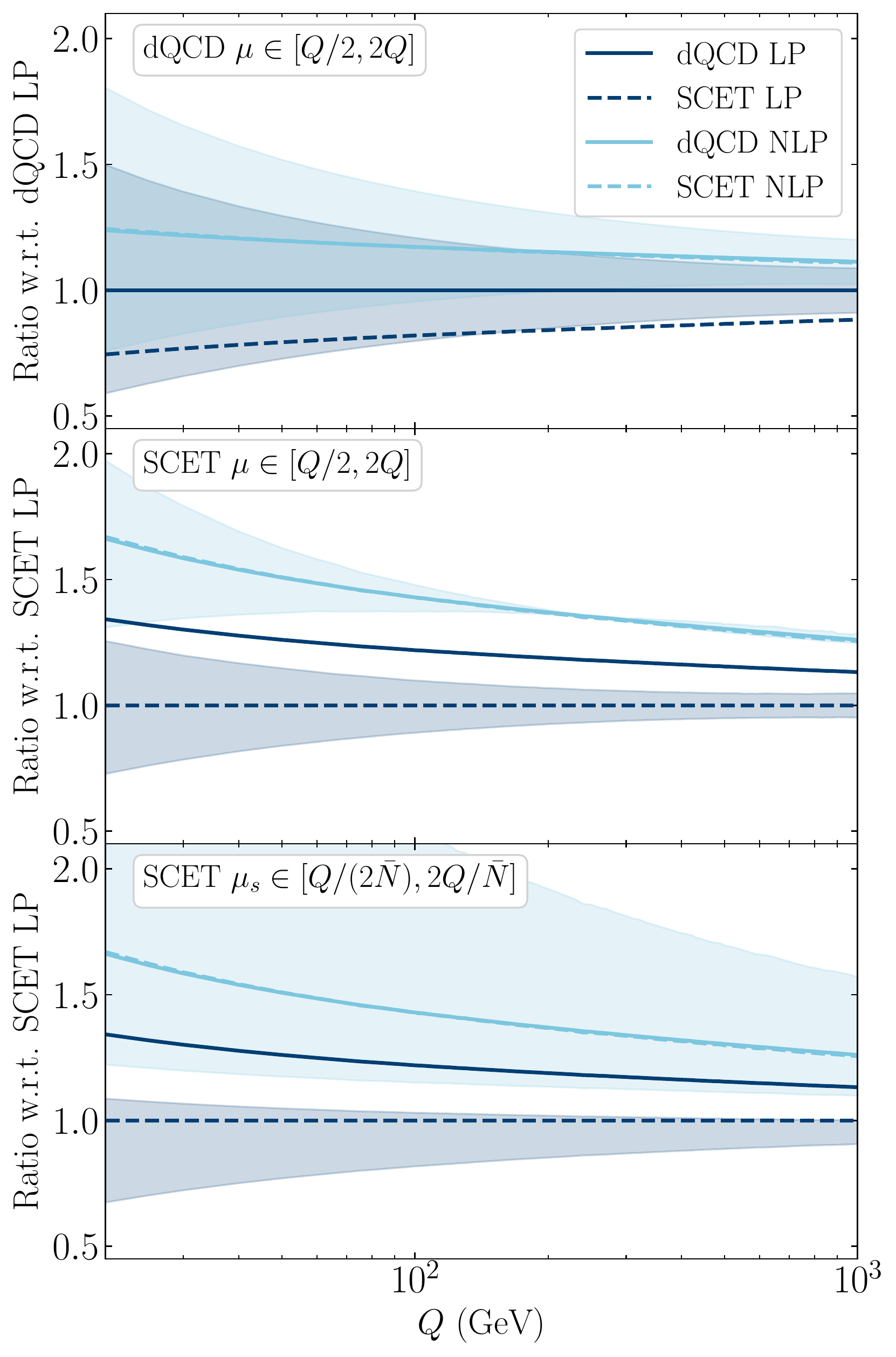}
    \end{subfigure} }
            \caption{Scale uncertainty of the LP and NLP dQCD and SCET results for DY (left) and Higgs (right). The LP (NLP) results are indicated in dark-blue (light-blue), and the dQCD (SCET) results are shown by the solid (dashed) lines. On the top panel, we normalize to the dQCD LP result showing the dQCD uncertainties obtained by varying $\mu \in [Q/2,2Q]$. The middle panel shows the curves normalized to the SCET LP result and with the scale variation of the SCET results with $\mu \in [Q/2,2Q]$. The bottom panel shows the SCET uncertainty by varying the soft scale $\mu_s \in [Q/(2\bar{N}),2Q/\bar{N}]$. }
        \label{fig:resumscalevar}
\end{figure}
Here we show the numerical results of the dQCD and SCET comparison. For the LP NNLL + NLP LL dQCD results, we use expression eq.~\eqref{eq:resumfunctions} in eq.~\eqref{eq:inversemel}, and we include $g_0(\alpha_s)$ up to $\mathcal{O}(\alpha_s)$. The SCET results are obtained by adding eq.~\eqref{eq:NLPSCET} (the NLP result) to eq.~\eqref{eq:scetlpwoBN} (the LP result with the factor of $z^{-\eta_a}$). We also show the results that are obtained excluding the factor of $z^{-\eta_a}$ from eq.~\eqref{eq:scetlpwoBN}, as in eq.~\eqref{eq:scetlp}, and those obtained by dressing the NLP SCET result by $H(Q,\mu_h)$ (eq.~\eqref{eq:NLPSCETwithH}). The results are shown in fig.~\ref{fig:scetvspqcdNNLL}.  Interestingly, we observe that near-perfect agreement is found between dQCD and SCET at NLP LL accuracy if one includes both the hard function and $z^{-\eta_a}$. If both these factors are included, we find that the difference between dQCD and SCET does not exceed $\mathcal{O}(0.5\%)$ ($\mathcal{O}(1\%)$) for the Higgs (DY) case. This may be contrasted to the LP case, where the differences are of $\mathcal{O}(25\%-12\%)$ ($\mathcal{O}(10\%-4\%)$).  Especially in the Higgs case, the inclusion of $H$ in the NLP SCET result is important, as without it, differences of $\mathcal{O}(10-20\%)$ are found between dQCD and SCET. The inclusion of $z^{-\eta_a}$ in the LP result also plays a significant role. Without it, the DY results differ between dQCD and SCET by $\mathcal{O}(15-5\%)$, while for Higgs, differences up to $40\%$ may be found for small values of the scale $Q$. The discrepancies decrease towards larger values of $Q$. 

In fig.~\ref{fig:resumscalevar} we examine how robust these results are under variation of the scale. We first examine the case where we vary $\mu$ between $\mu = 2Q$ and $\mu = Q/2$ (top and middle panel). For the SCET LP result, we again use eq.~\eqref{eq:scetlpwoBN} and for the NLP result, we use eq.~\eqref{eq:NLPSCETwithH}. In general, the dQCD uncertainty is larger than the SCET scale uncertainty. Note that the dQCD scale uncertainty band does not contain the LP SCET central value for large values of $Q$. The situation is worse for the LP SCET scale uncertainty, where especially for the Higgs production case the LP dQCD central value is not contained in the uncertainty band. This changes at NLP. At NLP, the dQCD uncertainty band does not change in size, but with the SCET result lying much closer to the dQCD result, it is now contained inside the uncertainty band. For SCET, the uncertainty band does shrink at NLP, which is especially pronounced for Higgs production. 

In dQCD, one implicitly varies the `soft scale' via the inverse Mellin transform. If we equate $\bar{N}$ to $Q/\mu_s$, as we have done above, we readily see that all values of $N$ (and hence of the soft scale) are accessible by integrating over $N$. In SCET one has to vary the soft scale explicitly, and we show the effect of this in the bottom panel of fig.~\ref{fig:resumscalevar}. There, one may observe that the uncertainty obtained by varying the soft scale grows very large (especially for the Higgs case) once the NLP corrections are included. The soft-scale variation of the LP SCET result is small, and the LP dQCD result is not contained in this uncertainty band, however, the LP dQCD is contained in the soft-scale uncertainty band of the NLP SCET result.

In this section, we have compared the NLP LL SCET and dQCD contribution both analytically and numerically. At the analytical level, differences between NLP LL SCET and dQCD start at $\mathcal{O}(\alpha_s^2)$ and $\mathcal{O}(1/N^2)$. However, we find that the numerical impact of these terms is small, as we see near-perfect numerical agreement between dQCD and SCET for both the Higgs and DY cases. However, this agreement can only be obtained if one includes the hard function in the NLP SCET contribution at the same order as the LP SCET contribution, and if the factor of $z^{-\eta_a}$ is included in the LP SCET contribution. Without this latter factor, the difference between the NLP dQCD and SCET result starts at $\mathcal{O}(\eta_a^2/N) = \mathcal{O}(\alpha_s^2 \ln^2\bar{N}/N)$, which is an NLP NLL contribution. This contribution turns out to be sizeable, as we have shown in fig.~\ref{fig:scetvspqcdNNLL}.

\section{Conclusion}
\label{sec:discussion}
In this paper we have studied the role and impact of NLP corrections in colour-singlet production processes, with a particular focus on DY and single Higgs production. In section~\ref{sec:fo_threshold} we assessed the quality of the threshold expansion for Higgs and DY,  for both the dominant ($q\bar{q}$ and $gg$) and the subdominant ($qg$) partonic channels.
The threshold expansion of the dominant Higgs production channel is less well-behaved than that of DY, due to a substantial part of the Higgs partonic cross section arising from the small $z$ region. The quality of the threshold expansion depends only marginally on the boson mass $Q$ for DY, while the convergence for Higgs noticeably improves as $Q$ increases.
The threshold expansion of the off-diagonal $qg$-channel in Higgs production convergences only very slowly, whereas for DY  convergence is already obtained after including the N$^{4}$LP contribution.

We reviewed the resummation of leading-logarithmic NLP corrections in dQCD in section~\ref{sec:dQCDresum}, and derived a slightly improved resummed expression involving a derivative with respect to the Mellin moment (eq.~\eqref{eq:Elpnlpll6}).
This we applied to the NLP LL + LP NNLL$^{\prime}$ DY and Higgs cross sections (as well as di-Higgs and di-vector boson production, albeit at lower logarithmic accuracy) and performed numerical studies (section~\ref{sec:dQCD_results}).
For the DY/Higgs diagonal channels we found that the NLP LL resummation has a notable numerical impact on the NNLL$^{\prime}$ resummed result matched to NNLO for both production processes, especially when compared to the effect of the N$^3$LL contribution. A similarly sizeable impact was seen in the di-boson processes. 
The off-diagonal $qg$ NLP LL contribution is more important for DY than for single-Higgs production given the large $\delta(1-z)$ contributions in the latter process at NLO and NNLO, which enter the resummation formula via the matching coefficient $g_0$ and inflate the resummation effects in the Higgs diagonal channel. 

In section~\ref{sec:comparison}, we performed an analytical and numerical comparison of the NLP LL resummed expressions in the dQCD and SCET frameworks. A derivative construction  to obtain the NLP LL contribution, similar to the Mellin derivative in dQCD, can be employed in SCET (eq.~\eqref{eq:derivSCET}), where the derivative is performed in the coupling evaluated at the SCET soft scale $\mu_s$. At NLP we found from our analytical analysis that differences between the two formalisms originate from truncations of higher-logarithmic or higher-power terms, and running-$\alpha_s$ effects. These differences resulted in very small numerical discrepancies once the NLP corrections were included; the two formalisms then agree numerically to $\mathcal{O}(0.5\%)$/$\mathcal{O}(1\%)$ for Higgs/DY production after setting hard scales equal to $Q$, and the SCET soft scale equal to $Q/\bar{N}$. To obtain this agreement, we argued that a factor of $z^{-\eta_a}$ is profitably included in the LP SCET expression, which originates from treating the kinematics of the scattering exact up to NLP. Indeed, sizeable numerical differences between the dQCD and SCET results of $\mathcal{O}(5-15\%)$ in the DY case and $\mathcal{O}(15-40\%)$ in the Higgs case are found if this factor is not included. Furthermore, we showed that the hard function of SCET that multiplies the NLP resummation should contain $\mathcal{O}(\alpha_s)$ corrections as well, certainly for the Higgs case. Large numerical differences are found for the Higgs case without including this factor, where the $\delta(1-z)$-contribution, and hence the hard function itself, is large. For the DY case the need is somewhat less compelling, as its $\delta(1-z)$-contribution is smaller.

As part of our SCET-dQCD comparison we examined the factorisation/renormalisation-scale variation of the SCET and dQCD results, varying $\mu$ between $Q/2$ and $2Q$, and the SCET soft scale $\mu_s$ between $Q/(2/\bar{N})$ and $2Q/\bar{N}$.
At LP, the dQCD and SCET central-value results do not lie in each other's scale-uncertainty bands. Conversely, at NLP, the central-value results show only a small numerical discrepancy and comfortably lie within each other’s scale-uncertainty bands. The uncertainty bands do increase upon including NLP effects for both dQCD (by varying $\mu$) and SCET (by varying $\mu_s$). We expect this to be reversed once NLP NLL contributions would be included.

Our study of the numerical effects of NLP corrections in hadronic collisions is, we believe, a valuable addition to  
an area where most effort has so far been on the analytical side. It moreover validates these efforts by showing that NLP 
threshold corrections can have a notable impact, and should motivate further development of the
understanding of NLP corrections.

\section*{Acknowledgments}
EL and MvB acknowledge support from the  NWO-I program 156, "Higgs as Probe and Portal", and MvB from the Science and Technology Facilities Council (grant number ST/T000864/1). JSD is supported by the D-ITP consortium, a program of NWO funded by the Netherlands Ministry of Education, Culture and Science (OCW). LV acknowledges support from the Fellini - Fellowship for Innovation at INFN, funded by the European Union's Horizon 2020 research programme under the Marie Sk\l{}odowska-Curie Cofund Action, grant agreement no. 754496. This paper is also based upon work from COST Action CA16201 PARTICLEFACE supported by COST (European Cooperation in Science and Technology).

\appendix
\section{Fitted parton distribution functions} \label{app:PDFs}
In this paper we rely on PDFs obtained from the LHAPDF library~\cite{Buckley:2014ana} and use the central member of the PDF4LHC15\_NNLO\_100 PDF set~\cite{Butterworth:2015oua}. For our purposes these PDFs need to be converted to $N$-space to perform the resummation. To this end, we expand the PDFs on a basis of polynomials whose Mellin transforms may be computed directly. The functional form of the PDFs that we use is inspired by that used by the MMHT~\cite{Harland-Lang:2014zoa} collaboration, and reads
\begin{align}
\label{eq:fit}
xf(x) &= A(1-x)^{a_1}x^{a_2}(1+b y + c (2 y^2-1))+B(1-x)^{a_3}x^{a_4}(1 + C x^{a_5})\hspace{5pt}\text{for}\hspace{5pt}  y= 1-2\sqrt{x}\,,
\end{align}
with $10$ (real) fit parameters. We require that the fitted function lies within the $1\sigma$ error as given by the LHAPDF grid implementation of the PDFs in the entire domain. We have checked our set by comparing the fixed-order results obtained with the $x$-space form of the fitted PDFs with those obtained using the grid directly, and found that differences are smaller than the numerical integration error. A tabulated form in \texttt{c++} format of the resulting fit parameters is available at~\cite{mbeekveldcode:2020}.

\section{Normalization of the 
partonic cross section}\label{AppNormalization}

In this appendix we discuss the normalization of the partonic cross section $\Delta_{ij}(z)$, with the particular aim to highlight the origin of the additional factor of $1/z$ appearing in Higgs production relative to DY. We start from the definition of the invariant mass distribution for DY, which according to the QCD factorisation theorem 
is given as 
\begin{equation}
\label{eqHadron1}
\frac{{\rm d}\sigma}{{\rm d} Q^2} = \sum_{a,b} \int  
{\rm d}x_a {\rm d}x_b \, f_{a/A}(x_a) \, f_{b/B}(x_b) \, 
\hat \sigma_{ab} \! \left(\frac{Q^2}{x_a x_b S}\right),
\end{equation}
where $\hat \sigma_{ab}(Q^2/(x_a x_b S))$ is the partonic 
cross section, and we drop the renormalisation/factorisation scale dependence for conciseness. For Higgs production an analogous equation holds, with the invariant mass distribution replaced by the total cross section. 

We first rewrite the partonic cross section in terms of $z = Q^2/(x_a x_b S) = Q^2/s$. This can be done by inserting
\begin{equation}\label{InsertDeltaz}
1= \int_0^1 {\rm d}z\, \delta\left(z-\frac{\tau}{x_a x_b}\right) 
= \int_0^1 {\rm d}z \, x_a x_b \, \delta\left(\tau - x_a x_b z\right)
= \int_0^1 {\rm d}z \, \frac{\tau}{z} \, \delta\left(\tau - x_a x_b z\right),
\end{equation}
where we recall that $\tau = Q^2/S$. With this, \Eq{eqHadron1} becomes
\begin{equation}\label{eqHadron4}
\frac{{\rm d}\sigma}{{\rm d} Q^2} = \sum_{a,b} 
\int {\rm d}z \, {\rm d} x_a {\rm d} x_b \, 
f_{a/A}(x_a) \, f_{b/B}(x_b) \,  
\delta\left(\tau - x_a x_b z\right) \, 
\frac{\tau}{z} \, \hat \sigma_{ab} (z)\,.
\end{equation}
This equation is then matched to 
\eq{eq:DYtotalxsec}, which we repeat here
\begin{equation}\label{eqHadron5}
\frac{{\rm d}\sigma}{{\rm d} Q^2} = \sigma_0(Q^2) 
\sum_{a,b} \int {\rm d}z\, {\rm d} x_a {\rm d} x_b \, 
f_{a/A}(x_a) \, f_{b/B}(x_b) \,  
\delta\left(\tau - x_a x_b z\right) 
 \Delta_{ab}(z)\,.
\end{equation}
The matching equation 
\begin{equation}\label{Matching}
\frac{\tau}{z} \, \hat \sigma_{ab} (z) 
= \sigma_0(Q^2)\,  \Delta_{ab}(z),
\end{equation}
is defined such that $\sigma_0(Q^2)$ contains the terms of the tree-level cross section which are $z$-independent, while $\Delta_{ab}(z)$ contains the $z$-dependent terms. 

We are now in a position to perform the matching in \eq{Matching} for the DY and Higgs tree-level
cross section. We start from the matrix element squared for the two processes, summed (averaged) over the final (initial) state partons
\begin{equation}
\label{TreeDYvsH}
|\mathcal{M}_{\rm DY}^{(0)}|^2 = \frac{4\alpha_{\rm EM}^2}{3 Q^2}\frac{s}{N_c}, 
\qquad \qquad 
|\mathcal{M}_{\rm h}^{(0)}|^2 = \bigg(\frac{\as}{3\pi v} \bigg)^2 \frac{s^2}{8(N_c^2 -1)}\,.
\end{equation}
The factor of $s^2$ in the Higgs matrix element squared is due to the derivative squared contained in the $G^a_{\mu\nu}G^{a,\mu\nu} H$ term of the effective Lagrangian. Given the one-particle phase space with the flux factor of $1/2s$
\begin{equation}
\frac{1}{2s}{\rm d}\Phi_1 = \int {\rm d}^4q \,2\pi\,
\delta^{(4)}(p_1 + p_2 - q) \delta^+\left(q^2-Q^2\right) =  \frac{\pi}{s^2} \delta(1-z)\,,
\end{equation}
the partonic cross section for the two processes reads 
\begin{equation}
\hat{\sigma}^{(0)}_{\rm DY}(z) = \frac{4\pi \alpha_{\rm EM}^2}{3 Q^2} \frac{1}{N_c} \frac{z}{Q^2} \delta(1-z)\,,
\qquad  
\hat{\sigma}^{(0)}_{\rm h}(z) = \frac{\as^2}{72\pi v^2} \frac{1}{N_c^2 -1}\delta(1-z)\,.
\end{equation}
We see that the DY cross section has an additional factor of $z$ compared to Higgs production, whose origin is  ultimately related to the dimensionful effective $ggH$ vertex versus the elementary $q\bar q \gamma$ vertex. The coefficient $\Delta_{aa}^{(0)}(z)$ is now obtained from \eq{Matching}, where $\sigma_0^{\rm h}$ and $\sigma_0^{\rm DY}$ are taken as in eqs.~(\ref{eq:higgsLO}) and~(\ref{eq:DYLOcoeff}), which implies
\begin{equation}
\Delta^{(0)}_{{\rm DY},q\bar q} = \delta(1-z)\,, 
\qquad \qquad 
\Delta^{(0)}_{{\rm h},gg} = 
\frac{1}{z}\delta(1-z)\,.
\end{equation}
At tree level the additional factor of $1/z$
is of course harmless, given the overall $\delta(1-z)$. However, this factor is general, and it is present also at higher orders in perturbation theory, explaining the origin of the factor $1/z$ in \eq{eq:etadef} (see also \eq{eq:DeltaHiggs1}).

\section{Singular contributions at threshold (LP)}
\label{app:singular}
In order to asses how the parton flux weighs the partonic cross section, it is useful to consider the point-by-point multiplication of the partonic cross section and the parton luminosity factor $\mathcal{L}_{ij}(\tau/z)/z$. For non-singular terms beyond LP this is trivial, but the singular contributions at LP consist of plus-distributions which have the non-local definition 
\begin{equation}
    \int_0^1\!{\rm d}z\, \left[g(z)\right]_+ f(z)  = \int_0^1\!{\rm d}z\,g(z)\left(f(z)-f(1)\right)\,.
\end{equation}
In this appendix we show explicitly how one obtains an equivalent point-by-point multiplication for the LP terms. Starting from the hadronic cross section
\begin{equation}
    \frac{{\rm d}\sigma_{ij}}{{\rm d}\tau} = \int_\tau^1\!{\rm d}z\,\frac{\mathcal{L}_{ij}(\tau/z)}{z}\frac{{\rm d}\hat{\sigma}_{ij}(z)}{{\rm d}z}\,,
\end{equation}
we express the partonic cross section, differential in $z$, as  \begin{equation}
    \frac{{\rm d}\hat{\sigma}(z)}{{\rm d}z} = -\frac{\partial}{\partial z} \int_z^1\! {\rm d}z^\prime \, \frac{{\rm d}\hat{\sigma}(z^\prime)}{{\rm d}z^\prime}\,.
\end{equation} 
Upon integration by parts we find
\begin{align}
    \frac{{\rm d}\sigma_{ij}}{{\rm d}\tau} &=  \left.-\frac{\mathcal{L}_{ij}(\tau/z)}{z}\int_z^1\!{\rm d}z^\prime\,\frac{{\rm d}\hat{\sigma}_{ij}(z^\prime)}{{\rm d}z^\prime}\, \right|_{\tau}^{1} + \int_\tau^1\! {\rm d}z\, \frac{\partial}{\partial z}\left[\frac{\mathcal{L}_{ij}(\tau/z)}{z}\right] \int_z^1\! {\rm d}z^\prime\,   \frac{{\rm d}\hat{\sigma}_{ij}(z^\prime)}{{\rm d}z^\prime}\nonumber \\ &= \int_\tau^1\! {\rm d}z\, \frac{\partial}{\partial z}\left[\frac{\mathcal{L}_{ij}(\tau/z)}{z}\right]\times \bigg( - \int_0^z\! {\rm d}z^\prime\,   \frac{{\rm d}\hat{\sigma}_{ij}(z^\prime)}{{\rm d}z^\prime} \bigg).\label{dertrick}
\end{align} 
The boundary term in the first line vanishes at $z=1$ by a vanishing domain for the integral over $z^\prime$, as well as at the lower boundary since $\lim_{z\to\tau }\mathcal{L}_{ij}(\tau/z)=0$. The plus-distributions in the remaining term on the first line of \eq{dertrick} are now separately integrated, with the trivial test function $f(z^\prime)=1$, such that we may use
\begin{equation}
    \int_z^1\!{\rm d}z'\,[g(z')]_+ = \cancelto{0}{\int_0^1\!{\rm d}z'\,[g(z')]_+} - \int_0^z\!{\rm d}z'\,g(z')\,.
\end{equation}
to obtain the second line of \eq{dertrick}. The integrand of \eq{dertrick} has a point-by-point multiplication form, consisting of the integrated plus-distributions weighted by the derivative of the parton luminosity function. 

\section{Resummation coefficients}
\label{app:resummationcoeff}
To evaluate the running of $\alpha_s$ we use the $\beta$-function as defined by
\begin{eqnarray}
\label{eq:beta1}
\frac{{\rm d}\alpha_s(\mu^2)}{{\rm d}\ln(\mu^2)} \equiv \beta(\alpha_s(\mu^2)) = -\alpha_s^2\sum_{n=0}b_n\alpha_s^n\,.
\end{eqnarray}
With this definition, we have the one- \cite{Gross:1973id,Politzer:1973fx}, two- \cite{Caswell:1974gg,Jones:1974mm,Egorian:1978zx} and three-loop \cite{Tarasov:1980au,Larin:1993tp} coefficients
\begin{eqnarray}
b_0 &=& \frac{11 C_A - 4 T_R n_f}{12 \pi}\,,\;\;\;\;  \;\;\;\;\;
b_1 \;=\; \frac{17 C_A^2-10 C_A T_R n_f-6 C_F T_R n_f}{24 \pi^2}\,,\;  \\
b_2 &=& \frac{1}{(4\pi)^3}\Bigg[\frac{2857}{54}C_A^3-\frac{1415}{27}C_A^2T_Rn_f -\frac{205}{9}C_AC_FT_Rn_f+2C_F^2T_Rn_f\nonumber \\
&&\hspace{1.28cm} +\frac{158}{27}C_AT_R^2n_f^2+\frac{44}{9}C_FT_R^2n_f^2\Bigg]\,,\nonumber
\end{eqnarray}
and $T_R = 1/2$, $C_A = 3$ and $C_F = \frac{4}{3}$. The number of active flavours is denoted by $n_f$ and is set equal to $5$ in this work. At $\mathcal{O}(\alpha_s^2)$, the solution to the $\beta$-function reads
\begin{eqnarray}
\alpha_s\left(k_T^2\right) = \frac{\alpha_s}{1+b_0\alpha_s \ln\frac{k_T^2}{\mu_R^2}} - \frac{\alpha_s^2}{\left(1+b_0\alpha_s \ln\frac{k_T^2}{\mu_R^2}\right)^2}\frac{b_1}{b_0}\ln\left(1+b_0\alpha_s \ln\frac{k_T^2}{\mu_R^2}\right) + \mathcal{O}\left(\alpha_s^3\right).\qquad
\end{eqnarray}
Alternatively, we occasionally use
\begin{eqnarray}
\label{eq:beta2}
\beta(\alpha_s(\mu^2)) = -\frac{\alpha_s^2}{4\pi}\sum_{n=0}\beta_n \left(\frac{\alpha_s}{4\pi}\right)^n,
\end{eqnarray}
where $\beta_0 = 4\pi b_0$, $\beta_1 = (4\pi)^2 b_1$, etc. 
\subsection{dQCD}
The initial state exponents for the LP LL ($g_a^{(1)}$), NLL ($g_a^{(2)}$) and NNLL resummations ($g_a^{(3)}$) are given by~\cite{Catani:1998tm,Catani:2003zt}
{\allowdisplaybreaks\begin{eqnarray}
\label{eq:g1}
g_a^{(1)} (\lambda) & =& \frac{A_a^{(1)}}{2\pi b_0^2}\Big[2\lambda + (1-2\lambda)\ln(1-2\lambda) \Big]\,, \\
\label{eq:g2}
g_a^{(2)} (\lambda,Q^2/\mu_F^2,Q^2/\mu_R^2)& =& \frac{1}{2\pi
  b_0}\left(-\frac{A^{(2)}_a}{\pi b_0} + A^{(1)}_a \ln\Big(\frac{Q^2}{\mu_R^2} \Big) \right)\Big[2\lambda + \ln(1-2\lambda) \Big] \qquad \\
            &&  + \frac{A^{(1)}_a b_1}{2\pi b_0^3}\Big[2\lambda + \ln(1-2\lambda)+\frac{1}{2}\ln^2(1-2\lambda) \Big]  - \frac{A^{(1)}_a}{\pi b_0}\lambda \ln\Big(\frac{Q^2}{\mu_F^2} \Big) \; ,\nonumber \\
\label{eq:g3}
g_a^{(3)}(\lambda,Q^2/\mu_F^2,Q^2/\mu_R^2) &=& \frac{2A^{(1)}_a}{\pi}\frac{\zeta(2)\lambda}{1-2\lambda}  +\frac{A^{(1)}_a b_2}{2\pi b_0^3}\left[2\lambda + \ln(1-2\lambda)+\frac{2\lambda^2}{1-2\lambda}\right] \\
&& +\frac{A^{(1)}_a b_1^2}{2\pi b_0^4(1-2\lambda)}\left[2\lambda^2+2\lambda\ln(1-2\lambda)+\frac{1}{2}\ln^2(1-2\lambda)\right] \nonumber \\
&& -\frac{A^{(2)}_a b_1}{2\pi^2b_0^3}\frac{1}{1-2\lambda}\left[2\lambda + \ln(1-2\lambda)+2\lambda^2\right]- \frac{A^{(2)}_a}{\pi^2b_0}\lambda\ln\frac{Q^2}{\mu_F^2}\nonumber \\
&&-\frac{A^{(1)}_a}{2\pi}\lambda \ln^2\frac{Q^2}{\mu_F^2}
 +\frac{A^{(1)}_a}{\pi}\lambda\ln\frac{Q^2}{\mu_R^2}\ln\frac{Q^2}{\mu_F^2}+\frac{A^{(1)}_a}{\pi}\frac{\lambda^2}{1-2\lambda}\ln^2\frac{Q^2}{\mu_R^2}  \nonumber \\
&& + \frac{1}{1-2\lambda}\left(\frac{A_a^{(1)}b_1}{2\pi b_0^2}\left[2\lambda + \ln(1-2\lambda)\right] -\frac{2A^{(2)}_a}{\pi^2b_0}\lambda^2\right)\ln \frac{Q^2}{\mu_R^2}\nonumber \\
&& +\frac{A^{(3)}_a}{\pi^3b_0^2}\frac{\lambda^2}{1-2\lambda} -\frac{D_{aa}^{(2)}}{2b_0\pi^2}\frac{\lambda}{1-2\lambda}\,, \nonumber
\end{eqnarray}}
where $\lambda = b_0 \alpha_s \ln \bar{N}$ and $\alpha_s\equiv\alpha_s(\mu_R^2)$. The N$^3$LL function $g_a^{(4)}(\lambda,Q^2/\mu_F^2,Q^2/\mu_R^2)$ is extracted from the TROLL~\cite{Bonvini:2014joa} code.
The function $h^{(1)}$ which is added to account for the NLP LL terms is
\begin{eqnarray}
\label{eq:hNLP}
h^{(1)}_a(\lambda,N) = -\frac{A^{(1)}_a}{2\pi b_0} \frac{\ln(1-2\lambda)}{N}\,.
\end{eqnarray}
The coefficients $A^{(n)}_a$ are given by \cite{Catani:1989ne,Vogt:2004mw,Moch:2004pa}
\begin{align}
\label{eq:37}
  A_a^{(1)} &= C_a\,, \qquad \qquad A_a^{(2)} = 
\frac{C_a}{2} \left[C_A\Bigg(\frac{67}{18}-\zeta(2)\Bigg)-\frac{10}{9}T_R n_f \right],  \\
    A_a^{(3)} &=  C_a\Bigg[\left(\frac{245}{96}-\frac{67}{36}\zeta(2)+\frac{11}{24}\zeta(3)+\frac{11}{8}\zeta(4)\right)C_A^2 -\frac{1}{108}n_f^2 \nonumber \\
    &\hspace{1.1cm}+\left(-\frac{209}{432}+\frac{5}{18}\zeta(2)-\frac{7}{12}\zeta(3)\right)C_A n_f + \left(-\frac{55}{96}+\frac{1}{2}\zeta(3)\right)C_Fn_f\Bigg]. \nonumber
\end{align}
The first order coefficient $D_{aa}^{(1)} = 0$, while the second order one is given by~\cite{Vogt:2000ci,Catani:2001ic}
\begin{eqnarray}
D_{aa}^{(2)} = C_a\left(C_A\left(-\frac{101}{27}+\frac{11}{18}\pi^2 +\frac{7}{2}\zeta_3\right)+n_f\left(\frac{14}{27}-\frac{1}{9}\pi^2\right)\right),
\end{eqnarray}
with $C_q = C_F$ and $C_g = C_A$. 
The perturbative (or hard) function $g_0(\alpha_s)$ reads
\begin{eqnarray}
g_0(\alpha_s) = \sigma_0\left[1+\alpha_s g_0^{1} + \alpha_s^2 g_0^{2} + \mathcal{O}\left(\alpha_s^3\right) \right].
\end{eqnarray}
While we extract the coefficients $g_0^i$ from TROLL~\cite{Bonvini:2014joa}, we give their explicit form below for convenience. For DY we have
\begin{eqnarray}
\label{eq:hardfuncDY}
\sigma_0 &=& \sigma^{\rm DY}\,,\quad g_0^{{\rm DY},1} = \frac{C_F}{\pi}\left(4\zeta_2 - 4 +\frac{3}{2}\ln\frac{Q^2}{\mu_F^2} \right),  \\
g_0^{{\rm DY},2}&=& -\frac{b_0 C_F}{\pi}\ln\frac{\mu_F^2}{\mu_R^2}\left(4\zeta(2)-4+\frac{3}{2}\ln\frac{Q^2}{\mu_F^2}\right) \nonumber \\ 
&& + \frac{C_F}{16 \pi^2}\Bigg[\ln^2\frac{Q^2}{\mu_F^2}(18 C_F-11C_A+2n_F)+C_F\left(\frac{511}{4}-198\zeta(2)-60\zeta(3)+\frac{552}{5}\zeta(2)^2\right) \nonumber\\
&& +C_A\left(-\frac{1535}{12}+\frac{376}{3}\zeta(2)+\frac{604}{9}\zeta(3)-\frac{92}{5}\zeta(2)^2\right) + n_f\left(\frac{127}{6}-\frac{64}{3}\zeta(2)+\frac{8}{9}\zeta(3)\right) \nonumber \\
&& +\ln\frac{Q^2}{\mu_F^2}\Bigg(C_F(48\zeta(3)+72\zeta(2)-93)+C_A\left(\frac{193}{3}-24\zeta(3)-\frac{88}{3}\zeta(2)\right) \nonumber \\
&&\hspace{1.5cm} +n_f\left(\frac{16}{3}\zeta(2)-\frac{34}{3}\right)\Bigg)\Bigg],\nonumber
\end{eqnarray}
with $\sigma^{\rm DY}$ given in eq.~\eqref{eq:DYLOcoeff}, 
and for Higgs
\begin{eqnarray}
\label{eq:hardfunchiggs}
\sigma_0 &=& \sigma^{\rm h}_0 \,,\quad g_0^{{\rm h},1} =  \frac{C_A}{\pi}\left(4\zeta_2+\frac{11}{6}+\frac{33-2n_f}{18}\ln\frac{\mu_R^2}{\mu_F^2}\right), \\
g_0^{{\rm h},2}&=& \frac{2 A^{(2)}_g\zeta(2)}{\pi^2} + 3 b_0^2\ln^2\frac{\mu_F^2}{\mu_R^2} + 
   \ln\frac{Q^2}{m_t^2}\left(\frac{2}{3}\frac{n_f}{\pi^2} + \frac{19}{8\pi^2}\right)  \nonumber \\
   && \frac{C_A}{\pi^2}\Bigg[ \ln\frac{Q^2}{\mu_F^2}\left(8 b_0 \pi\zeta(2) + \frac{11}{3} b_0 \pi+
        \frac{C_A}{6}(9\zeta(3) + 8) -\frac{4n_f}{9}\right)\nonumber \\
&& - \ln\frac{Q^2}{\mu_R^2}\left(\frac{11}{2}b_0\pi + \frac{2}{3}b_1+12 b_0 \pi\zeta(2)\right)  + 
   n_f \left(-\frac{5}{9}\zeta(2) + \frac{5}{18}\zeta(3) - \frac{1189}{432}\right)\nonumber \\
        && + 
 \frac{8}{3}\pi b_0\zeta(3) +
   2C_A\left(\zeta(2)^2 + 6\zeta(4)\right)-\frac{9}{60}\zeta^2(2) + \frac{199}{6}\zeta(2) - \frac{55}{4}\zeta(3) + \frac{11399}{432}\Bigg]
   \,, \nonumber 
\end{eqnarray}
where $\sigma^{h}_0$ is given in eq.~\eqref{eq:higgsLO}, and we have used the infinite-top mass limit for the effective $gg$-$h$ coupling. 
\subsection{SCET}
For SCET we have up to NNLL with $r \equiv \alpha_s(\mu^2)/\alpha_s(\nu^2)$
\begin{eqnarray}
S_{a,{\rm LL}}(\nu^2,\mu^2) &=& \frac{\Gamma^{(0)}_{\rm cusp,a}}{\beta_0^2}\frac{\pi}{\alpha_s(\nu^2)}\left(1-\frac{1}{r}-\ln r\right)\,, \\
S_{a,{\rm NLL}}(\nu^2,\mu^2) &=& \left(\frac{\Gamma^{(1)}_{\rm cusp,a}}{4\beta_0^2}-\frac{\Gamma^{(0)}_{\rm cusp,a}\beta_1}{4\beta_0^3}\right)\left(1-r+\ln r\right)+\frac{\Gamma^{(0)}_{\rm cusp,a}\beta_1}{8\beta_0^3}\ln^2 r\,,\nonumber \\
S_{a,{\rm NNLL}}(\nu^2,\mu^2) &=& \frac{\alpha_s(\nu^2)}{4\pi}\Bigg[\left(\frac{\Gamma^{(1)}_{\rm cusp,a}\beta_1}{4\beta_0^3}-\frac{\Gamma^{(0)}_{\rm cusp,a}\beta_2}{4\beta_0^3}\right)\left(1-r+r\ln r\right)\nonumber \\
&& \hspace{2cm} +\frac{\Gamma^{(0)}_{\rm cusp,a}}{4\beta_0^3}\left(\frac{\beta_1^2}{\beta_0}-\beta_2\right)(1-r)\ln r \nonumber \\
&& \hspace{2cm} -\, \frac{\Gamma^{(0)}_{\rm cusp,a}}{8\beta_0^2}\left(\frac{\beta_1^2}{\beta_0^2}-\frac{\beta_2}{\beta_0}-\frac{\beta_1}{\beta_0}\frac{\Gamma^{(1)}_{\rm cusp,a}}{\Gamma^{(0)}_{\rm cusp,a}}+\frac{\Gamma^{(2)}_{\rm cusp,a}}{\Gamma^{(0)}_{\rm cusp,a}}\right)\left(1-r\right)^2\Bigg]\,, \nonumber 
\end{eqnarray}
and
\begin{eqnarray}
a_{\gamma,{\rm LL}}(\nu^2,\mu^2) &=& \frac{\gamma^{(0)}}{\beta_0}\ln\frac{\alpha_s(\mu^2)}{\alpha_s(\nu^2)}\,,\\
a_{\gamma,{\rm NLL}}(\nu^2,\mu^2) &=& \frac{1}{4\pi}\left(\frac{\gamma^{(1)}}{\beta_0}-\frac{\beta_1\gamma^{(0)}}{\beta_0^2}\right)\left(\alpha_s(\mu^2)-\alpha_s(\nu^2)\right). \nonumber 
\end{eqnarray}
We expand the cusp anomalous dimension $\Gamma_{{\rm cusp}, a}$ and similarly the other anomalous dimensions $\gamma_S$, $\gamma_a$, as
\begin{eqnarray}
\gamma(\alpha_s) = \sum_{k=0}^{\infty}\gamma^{(k)}\left(\frac{\alpha_s}{4\pi}\right)^{k+1}.
\end{eqnarray}
The coefficients of the cusp anomalous dimension are  given by 
\begin{eqnarray}
\Gamma^{(0)}_{\rm cusp,a} = 4C_a\,,\qquad && \Gamma^{(1)}_{\rm cusp,a} = 4C_F\left[\left(\frac{67}{9}-\frac{\pi^2}{3}\right)C_A-\frac{20}{9}T_F n_f\right]. \nonumber
\end{eqnarray}
The other coefficients are given by 
\begin{eqnarray}
\gamma^{(0)}_{V} &=& -6 C_F\,, \qquad \gamma^{(0)}_{S} = 0\,,\\
\gamma^{(1)}_{V} &=& C_F^2(-3+4\pi^2-48\zeta(3))+C_FC_A\left(-\frac{961}{27}-\frac{11\pi^2}{3}+52\zeta(3)\right)+C_FT_Fn_f\left(\frac{260}{27}+\frac{4\pi^2}{3}\right), \nonumber\\
\gamma^{(1)}_{S} &=& C_A^2\left(-\frac{160}{27}+\frac{11\pi^2}{9}+4\zeta(3)\right)+C_AT_Fn_f\left(-\frac{208}{27}-\frac{4\pi^2}{9}\right)-8C_FT_Fn_f\,, \nonumber
\end{eqnarray}
and
\begin{eqnarray}
\gamma^{(0)}_q &=& 3 C_F\,, \qquad \gamma^{(0)}_g = \beta_0\,, \\
\gamma^{(1)}_q &=&C_F^2\left(\frac{3}{2}-2\pi^2+24\zeta(3)\right)+C_FC_A\left(\frac{17}{6}+\frac{22\pi^2}{9}-12\zeta(3)\right)-C_FT_Fn_f\left(\frac{2}{3}+\frac{8\pi^2}{9}\right),\nonumber\\
\gamma^{(1)}_g &=& 4C_A^2\left(\frac{8}{3}+3\zeta(3)\right)-\frac{16}{3}C_AT_Fn_f-4C_FT_Fn_f\,.\nonumber
\end{eqnarray}
The hard functions for DY and Higgs read
\begin{eqnarray}
\label{eq:hardfuncSCETdyhiggs}
H_{\rm DY}\left(Q,\mu_h\right) &=& |C_{S,{\rm DY}}(-Q^2-i\epsilon,\mu_h^2)|^2\,,\\
C_{S,{\rm DY}}(-Q^2-i\epsilon,\mu_h^2) &=& 1+C_F\frac{\alpha_s(\mu_h^2)}{4\pi}\left(-L^2+3L-8+\zeta(2)\right) + \mathcal{O}(\alpha_s^2)\,,\nonumber 
\end{eqnarray}
with $L = \ln\frac{Q^2}{\mu_h^2}-i\pi$, and
\begin{eqnarray}
H_{\rm h}(Q,\mu_h) &=& |C_{S,\rm h}(-Q^2,\mu_h^2)|^2\,, \\
C_{S,{\rm h}}(-Q^2-i\epsilon,\mu_h^2) &=& 1+C_A\frac{\alpha_s(\mu_h^2)}{4\pi}\left(-L^2+\zeta(2)\right) + \frac{\alpha_s(\mu_h^2)}{4\pi}\left(5C_A-3C_F\right)   \mathcal{O}(\alpha_s^2)\,. \nonumber 
\end{eqnarray}
Finally, the soft function reads
\begin{eqnarray}
\tilde{s}_{a}\left(L_{\eta_a}, \mu_s\right) &=& 1 + \frac{\alpha_s(\mu_s^2)}{2\pi}C_a\left(L_{\eta_a}^2+\zeta(2)\right) + \mathcal{O}(\alpha_s^2)\,,
\end{eqnarray}
with 
\begin{eqnarray}
L_{\eta_a} = \ln\frac{Q^2}{\mu_s^2}+\partial_{\eta_a}\,. 
\end{eqnarray}

\bibliographystyle{JHEP}
\bibliography{spire}

\providecommand{\href}[2]{#2}\begingroup\raggedright\begin{thebibliography}{100}

\bibitem{Parisi:1980xd}
G.~Parisi, {\it Summing large perturbative corrections in {QCD}},  {\em Phys.
  Lett.} {\bf B90} (1980) 295.

\bibitem{Curci:1979am}
G.~Curci and M.~Greco, {\it {Large infrared corrections in QCD processes}},
  {\em Phys. Lett.} {\bf B92} (1980) 175--178.

\bibitem{Sterman:1987aj}
G.~Sterman, {\it Summation of large corrections to short distance hadronic
  cross-sections},  {\em Nucl. Phys.} {\bf B281} (1987) 310.

\bibitem{Catani:1989ne}
S.~Catani and L.~Trentadue, {\it {Resummation of the QCD perturbative series
  for hard processes}},  {\em Nucl. Phys.} {\bf B327} (1989) 323.

\bibitem{Korchemsky:1993uz}
G.~P. Korchemsky and G.~Marchesini, {\it {Resummation of large infrared
  corrections using Wilson loops}},  {\em Phys. Lett.} {\bf B313} (1993)
  433--440.

\bibitem{Forte:2002ni}
S.~Forte and G.~Ridolfi, {\it Renormalization group approach to soft gluon
  resummation},  {\em Nucl. Phys.} {\bf B650} (2003) 229--270,
  [\href{http://arxiv.org/abs/hep-ph/0209154}{{\tt hep-ph/0209154}}].

\bibitem{Contopanagos:1997nh}
H.~Contopanagos, E.~Laenen, and G.~Sterman, {\it Sudakov factorization and
  resummation},  {\em Nucl. Phys.} {\bf B484} (1997) 303--330,
  [\href{http://arxiv.org/abs/hep-ph/9604313}{{\tt hep-ph/9604313}}].

\bibitem{Becher:2006nr}
T.~Becher and M.~Neubert, {\it {Threshold resummation in momentum space from
  effective field theory}},  {\em Phys. Rev. Lett.} {\bf 97} (2006) 082001,
  [\href{http://arxiv.org/abs/hep-ph/0605050}{{\tt hep-ph/0605050}}].

\bibitem{Kramer:1996iq}
M.~Kramer, E.~Laenen, and M.~Spira, {\it Soft gluon radiation in {H}iggs boson
  production at the {LHC}},  {\em Nucl. Phys.} {\bf B511} (1998) 523--549,
  [\href{http://arxiv.org/abs/hep-ph/9611272}{{\tt hep-ph/9611272}}].

\bibitem{Catani:2001ic}
S.~Catani, D.~de~Florian, and M.~Grazzini, {\it Higgs production in hadron
  collisions: Soft and virtual {QCD} corrections at {NNLO}},  {\em JHEP} {\bf
  0105} (2001) 025,
  [\href{http://arxiv.org/abs/http://arXiv.org/abs/hep-ph/0102227}{{\tt
  http://arXiv.org/abs/hep-ph/0102227}}].

\bibitem{Laenen:2008ux}
E.~Laenen, L.~Magnea, and G.~Stavenga, {\it {On next-to-eikonal corrections to
  threshold resummation for the Drell-Yan and DIS cross sections}},  {\em Phys.
  Lett.} {\bf B669} (2008) 173--179,
  [\href{http://arxiv.org/abs/0807.4412}{{\tt arXiv:0807.4412}}].

\bibitem{Moch:2009hr}
S.~Moch and A.~Vogt, {\it {On non-singlet physical evolution kernels and
  large-x coefficient functions in perturbative QCD}},  {\em JHEP} {\bf 0911}
  (2009) 099, [\href{http://arxiv.org/abs/0909.2124}{{\tt arXiv:0909.2124}}].

\bibitem{Soar:2009yh}
G.~Soar, S.~Moch, J.~Vermaseren, and A.~Vogt, {\it {On Higgs-exchange DIS,
  physical evolution kernels and fourth-order splitting functions at large x}},
   {\em Nucl. Phys.} {\bf B832} (2010) 152--227,
  [\href{http://arxiv.org/abs/0912.0369}{{\tt arXiv:0912.0369}}].

\bibitem{Grunberg:2009vs}
G.~Grunberg, {\it {Large-x structure of physical evolution kernels in Deep
  Inelastic Scattering}},  {\em Phys. Lett.} {\bf B687} (2010) 405--409,
  [\href{http://arxiv.org/abs/0911.4471}{{\tt arXiv:0911.4471}}].

\bibitem{deFlorian:2014vta}
D.~de~Florian, J.~Mazzitelli, S.~Moch, and A.~Vogt, {\it {Approximate N$^{3}$LO
  Higgs-boson production cross section using physical-kernel constraints}},
  {\em JHEP} {\bf 1410} (2014) 176, [\href{http://arxiv.org/abs/1408.6277}{{\tt
  arXiv:1408.6277}}].

\bibitem{Presti:2014lqa}
N.~Lo~Presti, A.~Almasy, and A.~Vogt, {\it {Leading large-x logarithms of the
  quark\textendash{}gluon contributions to inclusive Higgs-boson and
  lepton-pair production}},  {\em Phys. Lett. B} {\bf 737} (2014) 120--123,
  [\href{http://arxiv.org/abs/1407.1553}{{\tt arXiv:1407.1553}}].

\bibitem{Ajjath:2020ulr}
A.~Ajjath, P.~Mukherjee, and V.~Ravindran, {\it {On next to soft corrections to
  Drell-Yan and Higgs Boson productions}},
  \href{http://arxiv.org/abs/2006.06726}{{\tt arXiv:2006.06726}}.

\bibitem{Bahjat-Abbas:2019fqa}
N.~Bahjat-Abbas, D.~Bonocore, J.~Sinninghe~Damst\'e, E.~Laenen, L.~Magnea,
  L.~Vernazza, and C.~White, {\it {Diagrammatic resummation of
  leading-logarithmic threshold effects at next-to-leading power}},  {\em JHEP}
  {\bf 11} (2019) 002, [\href{http://arxiv.org/abs/1905.13710}{{\tt
  arXiv:1905.13710}}].

\bibitem{Beneke:2018gvs}
M.~Beneke, A.~Broggio, M.~Garny, S.~Jaskiewicz, R.~Szafron, L.~Vernazza, and
  J.~Wang, {\it {Leading-logarithmic threshold resummation of the Drell-Yan
  process at next-to-leading power}},  {\em JHEP} {\bf 1903} (2019) 043,
  [\href{http://arxiv.org/abs/1809.10631}{{\tt arXiv:1809.10631}}].

\bibitem{Beneke:2019oqx}
M.~Beneke, A.~Broggio, S.~Jaskiewicz, and L.~Vernazza, {\it {Threshold
  factorization of the Drell-Yan process at next-to-leading power}},  {\em
  JHEP} {\bf 07} (2020) 078, [\href{http://arxiv.org/abs/1912.01585}{{\tt
  arXiv:1912.01585}}].

\bibitem{Beneke:2019mua}
M.~Beneke, M.~Garny, S.~Jaskiewicz, R.~Szafron, L.~Vernazza, and J.~Wang, {\it
  {Leading-logarithmic threshold resummation of Higgs production in gluon
  fusion at next-to-leading power}},  {\em JHEP} {\bf 01} (2020) 094,
  [\href{http://arxiv.org/abs/1910.12685}{{\tt arXiv:1910.12685}}].

\bibitem{Moult:2018jjd}
I.~Moult, I.~W. Stewart, G.~Vita, and H.~X. Zhu, {\it {First Subleading Power
  Resummation for Event Shapes}},  {\em JHEP} {\bf 08} (2018) 013,
  [\href{http://arxiv.org/abs/1804.04665}{{\tt arXiv:1804.04665}}].

\bibitem{Moult:2016fqy}
I.~Moult, L.~Rothen, I.~W. Stewart, F.~J. Tackmann, and H.~X. Zhu, {\it
  {Subleading power corrections for N-jettiness subtractions}},  {\em Phys.
  Rev.} {\bf D95} (2017), no.~7 074023,
  [\href{http://arxiv.org/abs/1612.00450}{{\tt arXiv:1612.00450}}].

\bibitem{Boughezal:2016zws}
R.~Boughezal, X.~Liu, and F.~Petriello, {\it {Power corrections in the
  N-jettiness subtraction scheme}},  {\em JHEP} {\bf 1703} (2017) 160,
  [\href{http://arxiv.org/abs/1612.02911}{{\tt arXiv:1612.02911}}].

\bibitem{Moult:2017jsg}
I.~Moult, L.~Rothen, I.~W. Stewart, F.~J. Tackmann, and H.~X. Zhu, {\it {N
  -jettiness subtractions for $gg\to H$ at subleading power}},  {\em Phys.
  Rev.} {\bf D97} (2018), no.~1 014013,
  [\href{http://arxiv.org/abs/1710.03227}{{\tt arXiv:1710.03227}}].

\bibitem{Boughezal:2018mvf}
R.~Boughezal, A.~Isgr{\`o}, and F.~Petriello, {\it {Next-to-leading-logarithmic
  power corrections for $N$-jettiness subtraction in color-singlet
  production}},  {\em Phys. Rev.} {\bf D97} (2018), no.~7 076006,
  [\href{http://arxiv.org/abs/1802.00456}{{\tt arXiv:1802.00456}}].

\bibitem{Ebert:2018lzn}
M.~A. Ebert, I.~Moult, I.~W. Stewart, F.~J. Tackmann, G.~Vita, and H.~X. Zhu,
  {\it {Power Corrections for N-Jettiness Subtractions at ${\cal
  O}(\alpha_s)$}},  {\em JHEP} {\bf 12} (2018) 084,
  [\href{http://arxiv.org/abs/1807.10764}{{\tt arXiv:1807.10764}}].

\bibitem{Ebert:2018gsn}
M.~A. Ebert, I.~Moult, I.~W. Stewart, F.~J. Tackmann, G.~Vita, and H.~X. Zhu,
  {\it {Subleading power rapidity divergences and power corrections for
  q$_{T}$}},  {\em JHEP} {\bf 04} (2019) 123,
  [\href{http://arxiv.org/abs/1812.08189}{{\tt arXiv:1812.08189}}].

\bibitem{Cieri:2019tfv}
L.~Cieri, C.~Oleari, and M.~Rocco, {\it {Higher-order power corrections in a
  transverse-momentum cut for colour-singlet production at NLO}},  {\em Eur.
  Phys. J. C} {\bf 79} (2019), no.~10 852,
  [\href{http://arxiv.org/abs/1906.09044}{{\tt arXiv:1906.09044}}].

\bibitem{Moult:2019uhz}
I.~Moult, I.~W. Stewart, G.~Vita, and H.~X. Zhu, {\it {The Soft Quark
  Sudakov}},  {\em JHEP} {\bf 05} (2020) 089,
  [\href{http://arxiv.org/abs/1910.14038}{{\tt arXiv:1910.14038}}].

\bibitem{Liu:2020eqe}
Z.~L. Liu, B.~Mecaj, M.~Neubert, X.~Wang, and S.~Fleming, {\it {Renormalization
  and Scale Evolution of the Soft-Quark Soft Function}},  {\em JHEP} {\bf 07}
  (2020) 104, [\href{http://arxiv.org/abs/2005.03013}{{\tt arXiv:2005.03013}}].

\bibitem{Beneke:2020ibj}
M.~Beneke, M.~Garny, S.~Jaskiewicz, R.~Szafron, L.~Vernazza, and J.~Wang, {\it
  {Large-x resummation of off-diagonal deep-inelastic parton scattering from
  d-dimensional refactorization}},  {\em JHEP} {\bf 10} (2020) 196,
  [\href{http://arxiv.org/abs/2008.04943}{{\tt arXiv:2008.04943}}].

\bibitem{Appell:1988ie}
D.~Appell, G.~Sterman, and P.~Mackenzie, {\it Soft gluons and the normalization
  of the drell-yan cross- section},  {\em Nucl. Phys.} {\bf B309} (1988) 259.

\bibitem{Magnea:1990qg}
L.~Magnea, {\it {All Order Summation and Two Loop Results for the {Drell-Yan}
  Cross-section}},  {\em Nucl. Phys. B} {\bf 349} (1991) 703--713.

\bibitem{Anastasiou:2014lda}
C.~Anastasiou, C.~Duhr, F.~Dulat, E.~Furlan, T.~Gehrmann, F.~Herzog, and
  B.~Mistlberger, {\it {Higgs Boson Gluon-fusion Production Beyond Threshold in
  N$^{3}LO$ QCD}},  {\em JHEP} {\bf 03} (2015) 091,
  [\href{http://arxiv.org/abs/1411.3584}{{\tt arXiv:1411.3584}}].

\bibitem{Anastasiou:2016cez}
C.~Anastasiou, C.~Duhr, F.~Dulat, E.~Furlan, T.~Gehrmann, F.~Herzog,
  A.~Lazopoulos, and B.~Mistlberger, {\it {High precision determination of the
  gluon fusion Higgs boson cross-section at the LHC}},  {\em JHEP} {\bf 05}
  (2016) 058, [\href{http://arxiv.org/abs/1602.00695}{{\tt arXiv:1602.00695}}].

\bibitem{Anastasiou:2015vya}
C.~Anastasiou, C.~Duhr, F.~Dulat, F.~Herzog, and B.~Mistlberger, {\it {Higgs
  Boson Gluon-Fusion Production in QCD at Three Loops}},  {\em Phys. Rev.
  Lett.} {\bf 114} (2015) 212001, [\href{http://arxiv.org/abs/1503.06056}{{\tt
  arXiv:1503.06056}}].

\bibitem{Bonvini:2010tp}
M.~Bonvini, S.~Forte, and G.~Ridolfi, {\it {Soft gluon resummation of Drell-Yan
  rapidity distributions: Theory and phenomenology}},  {\em Nucl. Phys.} {\bf
  B847} (2011) 93--159, [\href{http://arxiv.org/abs/1009.5691}{{\tt
  arXiv:1009.5691}}].

\bibitem{Bonvini:2014qga}
M.~Bonvini, S.~Forte, G.~Ridolfi, and L.~Rottoli, {\it {Resummation
  prescriptions and ambiguities in SCET vs. direct QCD: Higgs production as a
  case study}},  {\em JHEP} {\bf 01} (2015) 046,
  [\href{http://arxiv.org/abs/1409.0864}{{\tt arXiv:1409.0864}}].

\bibitem{Bonvini:2014joa}
M.~Bonvini and S.~Marzani, {\it {Resummed Higgs cross section at N$^{3}$LL}},
  {\em JHEP} {\bf 09} (2014) 007, [\href{http://arxiv.org/abs/1405.3654}{{\tt
  arXiv:1405.3654}}].

\bibitem{Bonvini:2016frm}
M.~Bonvini, S.~Marzani, C.~Muselli, and L.~Rottoli, {\it {On the Higgs cross
  section at N$^{3}$LO+N$^{3}$LL and its uncertainty}},  {\em JHEP} {\bf 08}
  (2016) 105, [\href{http://arxiv.org/abs/1603.08000}{{\tt arXiv:1603.08000}}].

\bibitem{Basu:2007nu}
R.~Basu, E.~Laenen, A.~Misra, and P.~Motylinski, {\it {Soft-collinear effects
  in prompt photon production}},  {\em Phys. Rev. D} {\bf 76} (2007) 014010,
  [\href{http://arxiv.org/abs/0704.3180}{{\tt arXiv:0704.3180}}].

\bibitem{vanBeekveld:2019cks}
M.~van Beekveld, W.~Beenakker, R.~Basu, E.~Laenen, A.~Misra, and P.~Motylinski,
  {\it {Next-to-leading power threshold effects for resummed prompt photon
  production}},  {\em Phys. Rev. D} {\bf 100} (2019), no.~5 056009,
  [\href{http://arxiv.org/abs/1905.11771}{{\tt arXiv:1905.11771}}].

\bibitem{Herzog:2014wja}
F.~Herzog and B.~Mistlberger, {\it {The soft-virtual higgs cross-section at
  N$^3$LO and the convergence of the threshold expansion}},
  \href{http://arxiv.org/abs/1405.5685}{{\tt arXiv:1405.5685}}.

\bibitem{vanNeerven:1991gh}
W.~van Neerven and E.~Zijlstra, {\it {The $O(\alpha_s^2)$ corrected Drell-Yan
  $K$ factor in the DIS and MS scheme}},  {\em Nucl. Phys. B} {\bf 382} (1992)
  11--62. [Erratum: Nucl.Phys.B 680, 513--514 (2004)].

\bibitem{Contopanagos:1993xh}
H.~Contopanagos and G.~Sterman, {\it Normalization of the drell-yan
  cross-section in qcd},  {\em Nucl. Phys.} {\bf B400} (1993) 211.

\bibitem{Harlander:2002wh}
R.~V. Harlander and W.~B. Kilgore, {\it Next-to-next-to-leading order higgs
  production at hadron colliders},  {\em Phys. Rev. Lett.} {\bf 88} (2002)
  201801, [\href{http://arxiv.org/abs/hep-ph/0201206}{{\tt hep-ph/0201206}}].

\bibitem{Catani:2001cr}
S.~Catani, D.~de~Florian, and M.~Grazzini, {\it Direct higgs production and jet
  veto at the tevatron and the lhc in nnlo qcd},  {\em JHEP} {\bf 01} (2002)
  015, [\href{http://arxiv.org/abs/http://arXiv.org/abs/hep-ph/0111164}{{\tt
  http://arXiv.org/abs/hep-ph/0111164}}].

\bibitem{Dawson:1990zj}
S.~Dawson, {\it Radiative corrections to higgs boson production},  {\em Nucl.
  Phys.} {\bf B359} (1991) 283--300.

\bibitem{Djouadi:1991tka}
A.~Djouadi, M.~Spira, and P.~Zerwas, {\it {Production of Higgs bosons in proton
  colliders: QCD corrections}},  {\em Phys. Lett. B} {\bf 264} (1991) 440--446.

\bibitem{Spira:1995rr}
M.~Spira, A.~Djouadi, D.~Graudenz, and P.~M. Zerwas, {\it Higgs boson
  production at the {LHC}},  {\em Nucl. Phys.} {\bf B453} (1995) 17--82,
  [\href{http://arxiv.org/abs/hep-ph/9504378}{{\tt hep-ph/9504378}}].

\bibitem{Anastasiou:2002yz}
C.~Anastasiou and K.~Melnikov, {\it Higgs boson production at hadron colliders
  in nnlo qcd},  {\em Nucl. Phys.} {\bf B646} (2002) 220--256,
  [\href{http://arxiv.org/abs/hep-ph/0207004}{{\tt hep-ph/0207004}}].

\bibitem{Ravindran:2003um}
V.~Ravindran, J.~Smith, and W.~L. van Neerven, {\it Nnlo corrections to the
  total cross section for higgs boson production in hadron hadron collisions},
  {\em Nucl. Phys.} {\bf B665} (2003) 325--366,
  [\href{http://arxiv.org/abs/hep-ph/0302135}{{\tt hep-ph/0302135}}].

\bibitem{Butterworth:2015oua}
J.~Butterworth et~al., {\it {PDF4LHC recommendations for LHC Run II}},  {\em J.
  Phys.} {\bf G43} (2016) 023001, [\href{http://arxiv.org/abs/1510.03865}{{\tt
  arXiv:1510.03865}}].

\bibitem{DelDuca:2017twk}
V.~Del~Duca, E.~Laenen, L.~Magnea, L.~Vernazza, and C.~D. White, {\it
  {Universality of next-to-leading power threshold effects for colourless final
  states in hadronic collisions}},  {\em JHEP} {\bf 11} (2017) 057,
  [\href{http://arxiv.org/abs/1706.04018}{{\tt arXiv:1706.04018}}].

\bibitem{vanBeekveld:2019prq}
M.~van Beekveld, W.~Beenakker, E.~Laenen, and C.~D. White, {\it
  {Next-to-leading power threshold effects for inclusive and exclusive
  processes with final state jets}},  {\em JHEP} {\bf 03} (2020) 106,
  [\href{http://arxiv.org/abs/1905.08741}{{\tt arXiv:1905.08741}}].

\bibitem{Hamberg:1990np}
R.~Hamberg, W.~van Neerven, and T.~Matsuura, {\it {A Complete calculation of
  the order $\alpha_s^{2}$ correction to the Drell-Yan $K$ factor}},  {\em
  Nucl. Phys.} {\bf B359} (1991) 343--405.

\bibitem{Catani:1996yz}
S.~Catani, M.~L. Mangano, P.~Nason, and L.~Trentadue, {\it The resummation of
  soft gluons in hadronic collisions},  {\em Nucl. Phys.} {\bf B478} (1996)
  273--310, [\href{http://arxiv.org/abs/hep-ph/9604351}{{\tt hep-ph/9604351}}].

\bibitem{Vogt:2018miu}
A.~Vogt, F.~Herzog, S.~Moch, B.~Ruijl, T.~Ueda, and J.~A.~M. Vermaseren, {\it
  {Anomalous dimensions and splitting functions beyond the
  next-to-next-to-leading order}},  {\em PoS} {\bf LL2018} (2018) 050,
  [\href{http://arxiv.org/abs/1808.08981}{{\tt arXiv:1808.08981}}].

\bibitem{Moch:2004pa}
S.~Moch, J.~A.~M. Vermaseren, and A.~Vogt, {\it The three-loop splitting
  functions in qcd: The non-singlet case},  {\em Nucl. Phys.} {\bf B688} (2004)
  101--134, [\href{http://arxiv.org/abs/hep-ph/0403192}{{\tt hep-ph/0403192}}].

\bibitem{Vogt:2004mw}
A.~Vogt, S.~Moch, and J.~A.~M. Vermaseren, {\it The three-loop splitting
  functions in qcd: The singlet case},  {\em Nucl. Phys.} {\bf B691} (2004)
  129--181, [\href{http://arxiv.org/abs/hep-ph/0404111}{{\tt hep-ph/0404111}}].

\bibitem{Catani:2003zt}
S.~Catani, D.~de~Florian, M.~Grazzini, and P.~Nason, {\it Soft-gluon
  resummation for {H}iggs boson production at hadron colliders},  {\em JHEP}
  {\bf 0307} (2003) 028, [\href{http://arxiv.org/abs/hep-ph/0306211}{{\tt
  hep-ph/0306211}}].

\bibitem{Mistlberger:2018etf}
B.~Mistlberger, {\it {Higgs boson production at hadron colliders at N$^{3}$LO
  in QCD}},  {\em JHEP} {\bf 05} (2018) 028,
  [\href{http://arxiv.org/abs/1802.00833}{{\tt arXiv:1802.00833}}].

\bibitem{Dulat:2018rbf}
F.~Dulat, A.~Lazopoulos, and B.~Mistlberger, {\it {iHixs 2 \textemdash{}
  Inclusive Higgs cross sections}},  {\em Comput. Phys. Commun.} {\bf 233}
  (2018) 243--260, [\href{http://arxiv.org/abs/1802.00827}{{\tt
  arXiv:1802.00827}}].

\bibitem{Shao:2013bz}
D.~Y. Shao, C.~S. Li, H.~T. Li, and J.~Wang, {\it {Threshold resummation
  effects in Higgs boson pair production at the LHC}},  {\em JHEP} {\bf 07}
  (2013) 169, [\href{http://arxiv.org/abs/1301.1245}{{\tt arXiv:1301.1245}}].

\bibitem{deFlorian:2015moa}
D.~de~Florian and J.~Mazzitelli, {\it {Higgs pair production at
  next-to-next-to-leading logarithmic accuracy at the LHC}},  {\em JHEP} {\bf
  09} (2015) 053, [\href{http://arxiv.org/abs/1505.07122}{{\tt
  arXiv:1505.07122}}].

\bibitem{deFlorian:2018tah}
D.~De~Florian and J.~Mazzitelli, {\it {Soft gluon resummation for Higgs boson
  pair production including finite M$_{t}$ effects}},  {\em JHEP} {\bf 08}
  (2018) 156, [\href{http://arxiv.org/abs/1807.03704}{{\tt arXiv:1807.03704}}].

\bibitem{Plehn:1996wb}
T.~Plehn, M.~Spira, and P.~M. Zerwas, {\it {Pair production of neutral Higgs
  particles in gluon-gluon collisions}},  {\em Nucl. Phys.} {\bf B479} (1996)
  46--64, [\href{http://arxiv.org/abs/hep-ph/9603205}{{\tt hep-ph/9603205}}].
  [Erratum: Nucl. Phys.B531,655(1998)].

\bibitem{Catani:2014uta}
S.~Catani, L.~Cieri, D.~de~Florian, G.~Ferrera, and M.~Grazzini, {\it
  {Threshold resummation at N$^3$LL accuracy and soft-virtual cross sections at
  N$^3$LO}},  {\em Nucl. Phys. B} {\bf 888} (2014) 75--91,
  [\href{http://arxiv.org/abs/1405.4827}{{\tt arXiv:1405.4827}}].

\bibitem{DiMicco:2019ngk}
J.~Alison et~al., {\it {Higgs boson potential at colliders: Status and
  perspectives}},  {\em Rev. Phys.} {\bf 5} (2020) 100045,
  [\href{http://arxiv.org/abs/1910.00012}{{\tt arXiv:1910.00012}}].

\bibitem{Dawson:2013lya}
S.~Dawson, I.~M. Lewis, and M.~Zeng, {\it {Threshold resummed and approximate
  next-to-next-to-leading order results for $W^+W^-$ pair production at the
  LHC}},  {\em Phys. Rev. D} {\bf 88} (2013), no.~5 054028,
  [\href{http://arxiv.org/abs/1307.3249}{{\tt arXiv:1307.3249}}].

\bibitem{Wang:2014mqt}
Y.~Wang, C.~S. Li, Z.~L. Liu, and D.~Y. Shao, {\it {Threshold resummation for
  $W^{\pm}Z$ and $ZZ$ pair production at the LHC}},  {\em Phys. Rev. D} {\bf
  90} (2014), no.~3 034008, [\href{http://arxiv.org/abs/1406.1417}{{\tt
  arXiv:1406.1417}}].

\bibitem{Wang:2015mvz}
Y.~Wang, C.~S. Li, and Z.~L. Liu, {\it {Resummation prediction on gauge boson
  pair production with a jet veto}},  {\em Phys. Rev. D} {\bf 93} (2016), no.~9
  094020, [\href{http://arxiv.org/abs/1504.00509}{{\tt arXiv:1504.00509}}].

\bibitem{Grazzini:2015wpa}
M.~Grazzini, S.~Kallweit, D.~Rathlev, and M.~Wiesemann, {\it
  {Transverse-momentum resummation for vector-boson pair production at
  NNLL+NNLO}},  {\em JHEP} {\bf 08} (2015) 154,
  [\href{http://arxiv.org/abs/1507.02565}{{\tt arXiv:1507.02565}}].

\bibitem{Frixione:1993yp}
S.~Frixione, {\it A next-to-leading order calculation of the cross-section for
  the production of w+ w- pairs in hadronic collisions},  {\em Nucl. Phys.}
  {\bf B410} (1993) 280--324.

\bibitem{Brown:1978mq}
R.~W. Brown and K.~O. Mikaelian, {\it {$W^+W^-$ and $Z^0 Z^0$ Pair Production
  in $e^+ e^-, pp, p\bar{p}$ Colliding Beams}},  {\em Phys. Rev.} {\bf D19}
  (1979) 922.

\bibitem{Bonvini:2012an}
M.~Bonvini, S.~Forte, and G.~Ridolfi, {\it {The Threshold region for Higgs
  production in gluon fusion}},  {\em Phys. Rev. Lett.} {\bf 109} (2012)
  102002, [\href{http://arxiv.org/abs/1204.5473}{{\tt arXiv:1204.5473}}].

\bibitem{vanBeekveld:2020cat}
M.~van Beekveld and W.~Beenakker, {\it {The role of the threshold variable in
  soft-gluon resummation of the $t\bar{t}h$ production process}},
  \href{http://arxiv.org/abs/2012.09170}{{\tt arXiv:2012.09170}}.

\bibitem{Borowka:2016ehy}
S.~Borowka, N.~Greiner, G.~Heinrich, S.~Jones, M.~Kerner, J.~Schlenk,
  U.~Schubert, and T.~Zirke, {\it {Higgs Boson Pair Production in Gluon Fusion
  at Next-to-Leading Order with Full Top-Quark Mass Dependence}},  {\em Phys.
  Rev. Lett.} {\bf 117} (2016), no.~1 012001,
  [\href{http://arxiv.org/abs/1604.06447}{{\tt arXiv:1604.06447}}]. [Erratum:
  Phys.Rev.Lett. 117, 079901 (2016)].

\bibitem{Nandan:2016ohb}
D.~Nandan, J.~Plefka, and W.~Wormsbecher, {\it {Collinear limits beyond the
  leading order from the scattering equations}},  {\em JHEP} {\bf 1702} (2017)
  038, [\href{http://arxiv.org/abs/1608.04730}{{\tt arXiv:1608.04730}}].

\bibitem{Bonvini:2013td}
M.~Bonvini, S.~Forte, M.~Ghezzi, and G.~Ridolfi, {\it {The scale of soft
  resummation in SCET vs perturbative QCD}},  {\em Nucl. Phys. B Proc. Suppl.}
  {\bf 241-242} (2013) 121--126, [\href{http://arxiv.org/abs/1301.4502}{{\tt
  arXiv:1301.4502}}].

\bibitem{Abbate:2007qv}
R.~Abbate, S.~Forte, and G.~Ridolfi, {\it {A New prescription for soft gluon
  resummation}},  {\em Phys. Lett. B} {\bf 657} (2007) 55--63,
  [\href{http://arxiv.org/abs/0707.2452}{{\tt arXiv:0707.2452}}].

\bibitem{Forte:2006mi}
S.~Forte, G.~Ridolfi, J.~Rojo, and M.~Ubiali, {\it {Borel resummation of soft
  gluon radiation and higher twists}},  {\em Phys. Lett. B} {\bf 635} (2006)
  313--319, [\href{http://arxiv.org/abs/hep-ph/0601048}{{\tt hep-ph/0601048}}].

\bibitem{Ahrens:2008nc}
V.~Ahrens, T.~Becher, M.~Neubert, and L.~L. Yang, {\it {Renormalization-Group
  Improved Prediction for Higgs Production at Hadron Colliders}},  {\em Eur.
  Phys. J. C} {\bf 62} (2009) 333--353,
  [\href{http://arxiv.org/abs/0809.4283}{{\tt arXiv:0809.4283}}].

\bibitem{Bonvini:2012az}
M.~Bonvini, S.~Forte, M.~Ghezzi, and G.~Ridolfi, {\it {Threshold Resummation in
  SCET vs. Perturbative QCD: An Analytic Comparison}},  {\em Nucl. Phys. B}
  {\bf 861} (2012) 337--360, [\href{http://arxiv.org/abs/1201.6364}{{\tt
  arXiv:1201.6364}}].

\bibitem{Sterman:2013nya}
G.~Sterman and M.~Zeng, {\it {Quantifying Comparisons of Threshold
  Resummations}},  {\em JHEP} {\bf 05} (2014) 132,
  [\href{http://arxiv.org/abs/1312.5397}{{\tt arXiv:1312.5397}}].

\bibitem{Almeida:2014uva}
L.~G. Almeida, S.~D. Ellis, C.~Lee, G.~Sterman, I.~Sung, and J.~R. Walsh, {\it
  {Comparing and counting logs in direct and effective methods of QCD
  resummation}},  {\em JHEP} {\bf 04} (2014) 174,
  [\href{http://arxiv.org/abs/1401.4460}{{\tt arXiv:1401.4460}}].

\bibitem{Li:2011zp}
Y.~Li, S.~Mantry, and F.~Petriello, {\it {An Exclusive Soft Function for
  Drell-Yan at Next-to-Next-to-Leading Order}},  {\em Phys. Rev. D} {\bf 84}
  (2011) 094014, [\href{http://arxiv.org/abs/1105.5171}{{\tt
  arXiv:1105.5171}}].

\bibitem{Becher:2007ty}
T.~Becher, M.~Neubert, and G.~Xu, {\it {Dynamical Threshold Enhancement and
  Resummation in Drell-Yan Production}},  {\em JHEP} {\bf 07} (2008) 030,
  [\href{http://arxiv.org/abs/0710.0680}{{\tt arXiv:0710.0680}}].

\bibitem{Buckley:2014ana}
A.~Buckley, J.~Ferrando, S.~Lloyd, K.~Nordstr\"om, B.~Page, M.~R\"ufenacht,
  M.~Sch\"onherr, and G.~Watt, {\it {LHAPDF6: parton density access in the LHC
  precision era}},  {\em Eur. Phys. J. C} {\bf 75} (2015) 132,
  [\href{http://arxiv.org/abs/1412.7420}{{\tt arXiv:1412.7420}}].

\bibitem{Harland-Lang:2014zoa}
L.~Harland-Lang, A.~Martin, P.~Motylinski, and R.~Thorne, {\it {Parton
  distributions in the LHC era: MMHT 2014 PDFs}},  {\em Eur. Phys. J. C} {\bf
  75} (2015), no.~5 204, [\href{http://arxiv.org/abs/1412.3989}{{\tt
  arXiv:1412.3989}}].

\bibitem{mbeekveldcode:2020}
\url{https://github.com/melli1992/resummation/tree/master/src}.

\bibitem{Gross:1973id}
D.~J. Gross and F.~Wilczek, {\it {Ultraviolet Behavior of Nonabelian Gauge
  Theories}},  {\em Phys. Rev. Lett.} {\bf 30} (1973) 1343--1346.

\bibitem{Politzer:1973fx}
H.~Politzer, {\it {Reliable Perturbative Results for Strong Interactions?}},
  {\em Phys. Rev. Lett.} {\bf 30} (1973) 1346--1349.

\bibitem{Caswell:1974gg}
W.~E. Caswell, {\it {Asymptotic Behavior of Nonabelian Gauge Theories to Two
  Loop Order}},  {\em Phys. Rev. Lett.} {\bf 33} (1974) 244.

\bibitem{Jones:1974mm}
D.~Jones, {\it {Two Loop Diagrams in Yang-Mills Theory}},  {\em Nucl. Phys. B}
  {\bf 75} (1974) 531.

\bibitem{Egorian:1978zx}
E.~Egorian and O.~Tarasov, {\it {Two Loop Renormalization of the \{QCD\} in an
  Arbitrary Gauge}},  {\em Teor. Mat. Fiz.} {\bf 41} (1979) 26--32.

\bibitem{Tarasov:1980au}
O.~Tarasov, A.~Vladimirov, and A.~Zharkov, {\it {The Gell-Mann-Low Function of
  QCD in the Three Loop Approximation}},  {\em Phys. Lett. B} {\bf 93} (1980)
  429--432.

\bibitem{Larin:1993tp}
S.~A. Larin and J.~A.~M. Vermaseren, {\it The three loop qcd beta function and
  anomalous dimensions},  {\em Phys. Lett.} {\bf B303} (1993) 334--336,
  [\href{http://arxiv.org/abs/hep-ph/9302208}{{\tt hep-ph/9302208}}].

\bibitem{Catani:1998tm}
S.~Catani, M.~L. Mangano, and P.~Nason, {\it Sudakov resummation for prompt
  photon production in hadron collisions},  {\em JHEP} {\bf 9807} (1998) 024,
  [\href{http://arxiv.org/abs/hep-ph/9806484}{{\tt hep-ph/9806484}}].

\bibitem{Vogt:2000ci}
A.~Vogt, {\it Next-to-next-to-leading logarithmic threshold resummation for
  deep-inelastic scattering and the drell-yan process},  {\em Phys. Lett.} {\bf
  B497} (2001) 228--234, [\href{http://arxiv.org/abs/hep-ph/0010146}{{\tt
  hep-ph/0010146}}].

\end{thebibliography}\endgroup

\end{document}